\documentstyle[prb,aps,epsf,eqsecnum]{revtex}

\begin{document}
\draft

\begin{title}
{Ground-State Magnetization for Interacting Fermions in a Disordered
Potential : Kinetic Energy, Exchange Interaction and Off-Diagonal
Fluctuations}
\end{title}

\author{Philippe Jacquod$^{1}$ and A. Douglas Stone$^{2}$}

\address{$^{1}$ Instituut-Lorentz, Universiteit Leiden, P.O. Box 9506,
2300 RA Leiden, The Netherlands\\
$^{2}$ Department of Applied Physics, P.O.Box 208284,
Yale University, New Haven, CT 06520-8284 }

\date{ \today}

\maketitle

\begin{abstract}
We study a model of $n$ interacting fermions in a disordered potential,
which is assumed to generate uniformly fluctuating interaction matrix
elements. We show that the ground state magnetization is
systematically decreased by off-diagonal fluctuations of the
interaction matrix elements. This effect is  neglected in the Stoner
picture of itinerant ferromagnetism in which the ground-state
magnetization is simply determined by the balance between
ferromagnetic exchange and kinetic energy, and increasing the
interaction strength always favors ferromagnetism.  The physical
origin of the demagnetizing effect of interaction fluctuations is the
larger number $K$ of final states available for interaction-induced
scattering in the lower spin sectors of the Hilbert space. We analyze
the energetic role played by these fluctuations in the limits of
small and large interaction $U$. In the small $U$ limit we do
second-order perturbation theory and identify explicitly transitions
which are allowed for minimal spin and forbidden for
higher spin. These transitions then on average lower
the energy of the minimal spin ground state with respect to higher
spin; we analytically evaluate the size of this reduction and find it
to give a contribution $\Delta^s \propto n U^2/\Delta$ to the spin gap
between the two lowest spin ground-states. In term of an average effective
Hamiltonian, these contributions induce a $n U^2 S^2/\Delta $
term which decreases the strength of the ferromagnetic exchange, thereby
delaying the onset of Stoner ferromagnetism, and generate
a second, larger $S$ term $ \propto S^3 $
which results in a saturation
of the ground-state spin before full polarization is achieved,
in contrast to the Stoner scenario.
For large interactions $U$ we amplify on our earlier work
[Phys.\ Rev.\ Lett.\ {\bf 84}, 3938
(2000)] which showed that the broadening of the many-body density of states
is proportional to $\sqrt{K} U$ and hence favors minimal spin.
Numerical results are presented in both limits.
After evaluating the effect of fluctuations, we discuss the
competition between fluctuations plus kinetic energy and the exchange
energy. We finally present numerical
results for specific microscopic models and relate them to our
generic model of fluctuations. We discuss the different physical
situations to which such models may correspond, the importance of
interaction fluctuations, and hence the relevance of our results to
these situations and recall an experimental setup which we
proposed in an earlier work \cite{spings1} to measure the importance of 
interaction fluctuations on the ground-state spin of lateral quantum 
dots in the Coulomb blockade regime.
\end{abstract}
\vspace*{-0.05 truein}
\vspace*{-0.15 truein}

\section{Introduction}\label{intro}
\subsection{Stoner effect and disorder}

  More than fifty years ago Stoner proposed a simple route to
ferromagnetism in itinerant systems based on the competition between
one-body and interaction (exchange) energy \cite{stoner}.
The repulsive interaction energy can be minimized when the
fermionic antisymmetry requirement is satisfied by the spatial
wavefunction, as the overlap between different wavefunctions
is then minimal. This effect favors the alignment of spins and, if
the interaction is sufficiently strong, results in a large
ground-state spin
magnetization. This mechanism is the primary origin of
Hund's first rule in atomic physics. In contrast, when the
interaction is weak minimal spin is favored, since in order to align
spins electrons must be promoted from lower, doubly-occupied levels
to higher singly-occupied levels and the cost in one-body energy is
prohibitive.  Because the Pauli principle is essentially local,
ferromagnetism in metals has been studied within models such as the
Hubbard model \cite{hubbard,hubmag} which only retain the short-range part
of the electronic interaction; the long-range part of the interaction
being assumed to give spin-independent contributions to the
ground-state energy (the capacitance or charging energy). In the case
of a Hubbard interaction, only pairs of electrons of opposite spin
interact. The number of such pairs is a monotonically decreasing
function of the total magnetization
$\sim [(n/2)^2-\sigma^2]$ where $n$
is the number of electrons and $\sigma$ the total spin \cite{remspin}. 
On the other
hand, as just noted, flipping a spin requires the promotion of an
electron to a higher one-body level and in the case of a finite system
with a discrete spectrum of average spacing $\Delta$, a magnetization
$\sigma$ requires an energy $\sigma^2 \Delta$.
A simple first order perturbation treatment shows then that a
sufficiently strong interaction results in a finite magnetization,
when the corresponding reduction in interaction energy
counterbalances the increase in kinetic (one-body) energy

\begin{eqnarray}
(\Delta-V_c) \sigma^2 = 0
\end{eqnarray}

  This occurs when the typical
exchange interaction $V_c$ between two states close to the Fermi energy
is equal to the one-particle level spacing which for a Hubbard
interaction $U(\vec{r},\vec{r}')=U \delta(\vec{r}-\vec{r}')$
reads

\begin{equation}
V_c = U_c \int d\vec{r} \overline{|\psi_{\alpha}(\vec{r})|^2
|\psi_{\beta}(\vec{r})|^2} = \Delta
\end{equation}

  The upper bar indicates
an average over wavefunctions in the vicinity of the Fermi level.
In a clean system this gives
$U_c=\Delta$ and this threshold is known as the {\it Stoner
instability}. As both the kinetic energy and the interaction energy
have the same parametric dependence on the magnetization $\sigma$,
reaching this threshold results in a second order phase transition
to a ferromagnetic phase, the divergence of the magnetic susceptibility,
and a macroscopic magnetization.

  Quite naturally one may wonder in what way does the presence
of a disordered potential modify this Stoner picture, and this question has
recently attracted a lot of attention, both in the context of bulk
metals (i.e. infinitely extended systems with diffusive eigenstates)
and
in finite-sized metallic systems such as quantum dots and
metallic nanoparticles.
Two types of questions have been considered: 1) The effect
of a disordered potential on the {\it average} threshold for the Stoner
instability. 2)  The statistical properties of the threshold in an
ensemble of mesoscopic metallic samples. Both aspects have been
recently investigated
theoretically. For the bulk case, it has been known for some time
\cite{altshuler} that within perturbation theory disorder enhances
the exchange effect in the susceptibility; recently
Andreev and Kamenev constructed a mean-field theory which they argue
describes the Stoner transition
\cite{andreev} and found a significant reduction of the Stoner
threshold in low-dimensional disordered systems due to correlations in
diffusive wavefunctions which enhance the average exchange term.
In the framework of the same
mean-field approach which neglects the fluctuations of the
interactions, but takes into account those of the one-body
spectrum, Kurland, Aleiner and Altshuler proposed that below, but
in the immediate vicinity of the Stoner instability, there is a broad
distribution of magnetization and that each sample's free
energy is characterized by a large number of local magnetization minima
\cite{newaleiner1}. Brouwer, Oreg and Halperin \cite{brouwer} considered
the effect of
mesoscopic wavefunction fluctuations on the exchange interaction
and found that their effect was to increase substantially the probability
of non-zero spin magnetization in the ground state before the Stoner
threshold is reached.
Baranger, Ullmo and Glazman \cite{baranger} suggested that the
observed "kinks" in the parametric variations of Coulomb blockade
peak positions (e.g. as one varies an external magnetic field) could
reflect changes in the ground state spin of the quantum dot. It was
noted that the statistical
occurence of nonzero ground-state magnetizations can account for the
absence of bimodality of the conductance
peak spacings distribution for tunneling experiments with quantum dots
in the Coulomb blockade regime \cite{marcus,ensslin,berko}.
Another aspect of large disordered metallic samples
is that the Stoner threshold can be locally exceeded, while
the exchange averaged over the full sample has a value well below
the threshold. In this case one may expect that
localized regions with nonzero magnetization will
be formed even though the full system is nonmagnetic. This
scenario has been investigated by
Narozhny, Aleiner and Larkin \cite{newaleiner2} who also considered
the effect of such {\it local spin droplets} on dephasing. They found
that the probability to form a local spin droplet though
exponentially small, does not rigorously vanish as it would in a clean
system, and that neither
this probability, nor the corresponding spin depends on the droplet's size.
In a different approach focusing on the large interaction regime
close to half-filling, Eisenberg and Berkovits numerically found that
the presence of disorder may stabilize
Nagaoka-like ferromagnetic phases at larger
number of holes ($\ge 2$) \cite{eberko}.
Finally, Stopa has suggested
that scarring of one-body wavefunctions in a chaotic
confining potential may lead to strong enhancements of the exchange interaction
and to the occurence of few-electron polarization in finite-sized systems
\cite{stopa}. Thus the general message of these works is that disorder tends to
favor novel magnetic states over paramagnetic states.

\subsection{Overview and outline}

  In a recent letter \cite{spings1}, we pointed out a
competing effect of interactions in disordered systems
which {\it reduces} the probability of ground-state magnetization and
hence favors paramagnetism. This effect had not (to our knowledge)
been treated in any of the previous works on itinerant magnetization
of
disordered systems.  The works cited above neglect the effect of
disorder in inducing fluctuations in the {\it off-diagonal}
interaction matrix elements \cite{andreev,newaleiner1}.
However it is well-known from studies of complex few-body systems
like nuclei and atoms \cite{brody,kota} that the
band-width of the many-body density of states in finite interacting fermi
systems is strongly modified by the fluctuations of these off-diagonal
matrix elements already at moderate strength of the interactions.
Such studies did not directly address the effect of this broadening
on the ground-state spin of the system.  However our extension of
these models immediately revealed that these fluctuations are largest
for the states
of minimal spin, due to the larger number of final states (non-zero
interaction matrix elements) for interaction-induced transitions (we
will review this argument below). This effect then significantly
increases the probability that
the extremal (low-energy) states in the band are those of minimal spin and
opposes the exchange effect. In our earlier works
\cite{spings1,spings2} we focused on the regime of large fluctuations
to deduce the scaling properties of the ground-state energy as a
function of spin, and verified these scaling laws with numerical
tests. In the present work we will review and extend these results
for large fluctuations, but we will focus mostly on the perturbative
regime of small $U$.  While in this regime the correction to the
ground-state energy due to fluctuations is small by assumption, one
is able to evaluate these  corrections analytically and show that
they favor minimal spin for an arbitrary number of particles.
Specifically, the larger number of interaction-induced transitions
for lower spin leads to more and larger interaction contributions to
the (negative) second-order correction to the ground-state energy in
each spin block.
This illustrates explicitly the "phase-space"
argument introduced in \cite{spings1,spings2} which implies that
fluctuations generically suppress magnetization.  We expect this
effect to be significant in
quantum dots where it will reduce the probability of high spin ground states.
We recall that the ground state spin of lateral quantum dots can be
experimentally determined by following the motion
of Coulomb blockade conductance peaks as an in-plane magnetic field
is applied \cite{spings1}. Therefore the strength of the demagnetizing 
effect of fluctuations of interaction is experimentally accessible.

The paper is organized as follows. In section \ref{model} we start
by an explicit derivation of our model and describe its main
features. In section \ref{tip} we begin for pedagogical reasons with
an analytical treatment of the model for the case of only two
particles, in both the perturbative regime of weak off-diagonal
fluctuations and the asymptotic regime where they dominate.
Section \ref{pert} will be devoted to a second order
perturbative treatment of the model for an arbitrary number of particles;
this will be followed in section \ref{tbrim} by a discussion of the
magnetization properties of the system's ground state in the asymptotic
regime. As noted above, some of the results presented there have
already been presented in \cite{spings1,spings2} but are nevertheless
included to make the article self-contained. In the next section \ref{six}
we consider the competition between exchange and fluctuations in more
details, both from the point of view of average Stoner threshold and in terms
of probability of finding a polarized ground-state. We will see in particular
that the off-diagonal fluctuations induce a term $\sim \sigma^2$
in the Hamiltonian
which delays the Stoner instability and a second term $\sim \sigma^3$
which strongly
suppresses the occurence of large ground-state spins even above the Stoner
instability. In section \ref{realsp}
we consider more standard microscopic models for disordered
interacting fermions and relate their properties to our generic model
of fluctuating interactions.
We determine the conditions to be satisfied
in order for the results obtained from our random interaction model to
be relevant in different physical situations.
Finally we summarize our findings, put them
in perspective and discuss possible extensions of this work
in the final section \ref{concl}.

\section{Derivation of the Model}\label{model}

  Our starting point is a lattice model for fermions in a
disordered potential coupled by a two-body, spin-independent
interaction of arbitrary range. We make a unitary transformation to
the basis of single-particle eigenstates of the disordered potential
and introduce the assumption that the single-particle states are
random and uncorrelated.  Upon averaging over disorder
we arrive at a completely generic model describing both the nonvanishing
average interactions (exchange,charging and BCS) and the statistical
fluctuations in both the one and two-body terms. Finally we
introduce the assumption that all interaction matrix elements have
the same statistical variance. Hence our construction excludes both
one-body integrable systems and strongly-localized systems. We also
note that with this assumption geometric or commensurability effects
(such as spin waves or antiferromagnetic instabilities) cannot be
captured by our treatment, as the statistical character of the
construction erases most real-space details of the model.

  We consider the following tight-binding Hamiltonian
for $n$ spin-$1/2$ particles

\begin{eqnarray}\label{hamsite}
{\cal H} & = & {\cal H}_0 + {\cal U} \nonumber =
\sum_{i,j;s} {\cal H}_0^{i,j} d^{\dagger}_{i,s} d_{j,s}
+ \sum_{i \ne j} {\cal U}(i-j) n_i n_j + \sum_{i} {\cal U}(0)
n_{i,\uparrow} n_{i,\downarrow}
\end{eqnarray}

\noindent $s = \uparrow,\downarrow$ is a spin index, $d^{\dagger}_{i,s}$
($d_{i,s}$) creates (destroys) a fermion on the i$^{\rm th}$ site of a
$D$-dimensional lattice of linear dimension $L a$ and volume
   $\Omega = (L a)^D$.
This latter quantity defines the number $m/2=\Omega/a^D$ of
spin-degenerate one-body eigenenergies which we will refer to as {\it orbitals}
in what follows, $a \equiv 1$ is the lattice constant.
${\cal H}_0$ is a one-body, spin independent, disordered
Hamiltonian with
eigenvalues $\epsilon_{\alpha}$ and eigenvectors $\psi_{\alpha}$, i.e. one has

\begin{equation}\label{hamnot}
{\cal H}_0 |\psi_{\alpha} \rangle = \epsilon_{\alpha} |\psi_{\alpha} \rangle
= \epsilon_{\alpha} \sum_i \psi_{\alpha}(i) | i \rangle
\end{equation}

\noindent where $| i \rangle$ refers to a lattice site ket.

  We assume that $ {\cal H}_0 $
has no degeneracy besides twofold spin-degeneracy and
distribute the $m/2$ different one-body energies as
$\epsilon_{\alpha} \in [0;m/2]$ so as to fix $\Delta \equiv 1$ without
spin degeneracy. Below we will discuss three different eigenvalue
distributions : constant spacing distribution \cite{caveat}
($\epsilon_{\alpha}=(\alpha-1) \Delta$, note that due to the
level degeneracy the single-particle level spacing is $\Delta$,
whereas $\Delta/2$ is the mean level spacing),
randomly distributed $\epsilon_{\alpha}$ with a Poisson spacing
distribution, or with
a Wigner-Dyson spacing distribution.
Finally ${\cal U}(i-j)$ is the electron-electron interaction potential and
$n_i = \sum_s n_{i,s} = \sum_s d^{\dagger}_{i,s} d_{i,s}$. The
Hamiltonian is Spin Rotational Symmetric (SRS)
so that both the total spin $|\vec{S}|$ and its projection
$S_z$ commute with the Hamiltonian and the corresponding eigenvalues
$\sigma$ and $\sigma_z$ are good quantum numbers.
This results in a block structure of the Hamiltonian
${\cal H}$ which will be described in detail below. Performing the unitary
transformation defined by

\begin{eqnarray}\label{unitary}
\sum_{\alpha} \psi_{\alpha}(i) c^{\dagger}_{\alpha,s} & = & d^{\dagger}_{i,s},
\end{eqnarray}

\noindent we rewrite the Hamiltonian as

\begin{eqnarray}\label{hamran}
{\cal H} & = & \sum \epsilon_{\alpha}
c^{\dagger}_{\alpha,s} c_{\alpha,s}
+\sum
    U_{\alpha,\beta}^{\gamma,\delta}
c^{\dagger}_{\alpha,s} c^{\dagger}_{\beta,s'} c_{\delta,s'}
c_{\gamma,s}
\end{eqnarray}

\noindent where the Interaction Matrix Elements (IME) are given by

\begin{eqnarray}\label{ime}
U_{\alpha,\beta}^{\gamma,\delta} & = & \sum_{i,j}
{\cal U}(i-j) \psi_{\alpha}^*(i)  \psi_{\beta}^*(j)
      \psi_{\delta}(j) \psi_{\gamma}(i)
\end{eqnarray}

These IME
induce transitions between many-body states
differing by at most two one-body occupation numbers.
The distribution  and properties of the IME depend on both the range of
the interaction potential and the one-particle dynamics.  If there
are conserved quantities other than energy in the one-particle
dynamics (and hence good quantum numbers describing the one-body
states) this will lead to selection rules in the IME; the extreme
case of this would be an integrable one-particle hamiltonian for
which a complete set of quantum numbers exists. Selection rules
greatly reduce
the number of allowed interaction-induced transitions, and lead to a
very singular distribution of IME (this is most easily seen by considering
a clean hypercubic lattice model with Hubbard interaction).
Perturbing a clean lattice with a
disordered potential destroys translational symmetry and
these selection rules disappear, which induces a
crossover of the distribution of IME from a set of $\delta-$functions
to a smooth distribution. In Fig.\ref{fig:puodt1t2} we illustrate
this by plotting
the distribution of IME for a one-dimensional lattice model with
on-site disorder, nearest and next-nearest-neighbor hopping and a
Hubbard
interaction as described e.g. in reference \cite{daul}.

  The key assumption of our model is that such a smooth
distribution of interaction matrix elements exists and that in fact
all matrix elements which preserve SRS
{\it have the same nonzero variance}
(these matrix elements may vanish on average of course).
This assumption rules out both the case of
integrable one-body dynamics as discussed above, and the case of
strongly-localized wavefunctions for which interaction matrix
elements between states separated spatially by more than a
localization length will have different (and much smaller) variance
than those in the same localization volume.  Our assumption is
reasonable for metallic disordered states with a
randomness generated by either impurities or chaotic boundary
scattering.

  With this motivation, we assume that the fluctuations of the
off-diagonal $U_{\alpha,\beta}^{\gamma,\delta} $
are random with a zero-centered gaussian distribution of width $U$.
Only matrix elements $U_{\alpha,\beta}^{\alpha,\beta}$,
$U_{\alpha,\beta}^{\beta,\alpha}$ and $U_{\alpha,\alpha}^{\beta,\beta}$
have nonzero averages
$\langle U_{\alpha,\beta}^{\alpha,\beta} \rangle$,
$\langle U_{\alpha,\beta}^{\beta,\alpha} \rangle$ and
$\langle U_{\alpha,\alpha}^{\beta,\beta} \rangle$
which
lead (respectively) to mean-field charge-charge, spin-spin and 
BCS-like interaction terms.
Note that the average of both the exchange and BCS terms
is dominated by the short-range part of the interaction, and that
${\cal U}_{BCS}$ vanishes as time-reversal symmetry is broken.
Consequently, the electronic interactions give us four contributions.
The first three are
the average charge-charge, ferromagnetic spin-spin and BCS terms that we
just discussed and which can be written as
($n_{\alpha}=\sum_{\alpha,s} c^{\dagger}_{\alpha,s} c_{\alpha,s}$ and
$n = \sum_\alpha n_{\alpha}$.)

\begin{eqnarray}\label{cc}
{\cal U}_{avg} & = & {\cal U}_{cc} + {\cal U}_{ss} + {\cal U}_{BCS} 
= \left[ \langle U_{\alpha,\beta}^{\alpha,\beta} \rangle
-\frac{1}{2} \langle U_{\alpha,\beta}^{\beta,\alpha} \rangle \right]
n (n+1)/2 - \lambda U \vec{S} \cdot \vec{S}
+ \sum_{\alpha,\beta} \langle U_{\alpha,\alpha}^{\beta,\beta}
\rangle c^{\dagger}_{\alpha,\uparrow} c^{\dagger}_{\alpha,\downarrow}
c_{\beta,\downarrow} c_{\beta,\uparrow}
\end{eqnarray}

\noindent where we have introduced spin operators
$\vec{S}_{\alpha} \equiv (1/2)\sum_{s,t}
c^{\dagger}_{\alpha,s} \vec{\sigma}_{s,t} c_{\alpha,t}$ and
$\vec{S} = \sum_{\alpha} \vec{S}_{\alpha}$.
Note that the strength of the average ferromagnetic exchange term has
been written
in units of the rms fluctuation $U$, i.e. we have introduced a
parameter $\lambda$ which is the ratio of the average exchange to the
fluctuations

\begin{equation}
\lambda \equiv 2 \langle U_{\alpha,\beta}^{\beta,\alpha}\rangle/U
\end{equation}

\noindent much in the same spirit as the usual Stoner picture where another
energy ratio $\langle U_{\alpha,\beta}^{\beta,\alpha}\rangle/\Delta$,
between the exchange energy and the one-body energy spacing
at the Fermi level, is the relevant parameter.

  The fourth interaction
contribution to our model hamiltonian goes beyond the mean-field
approximation and contains the off-diagonal fluctuations of the
electronic interactions

\begin{eqnarray}\label{od}
{\cal U}_{f} & = & \sum_{\alpha,\beta;\gamma,\delta}\sum_{s,s'}
    \bar{U}_{\alpha,\beta}^{\gamma,\delta}
c^{\dagger}_{\alpha,s} c^{\dagger}_{\beta,s'} c_{\delta,s'}
c_{\gamma,s}
\end{eqnarray}

Having removed the average interactions, we now assume that
both the diagonal and off-diagonal IME
$\bar{U}_{\alpha,\beta}^{\gamma,\delta}$ have zero-centered uncorrelated
gaussian distributions $ P(\bar{U}_{\alpha,\beta}^{\gamma,\delta}) \propto
e^{-(\bar{U}_{\alpha,\beta}^{\gamma,\delta})^2/2U^2}$ of width $U$.
We stress that in general, not all IME have the same variance
but being interested in generic features of the interaction,
we will neglect these variance discrepancies.
${\cal U}_f$ contains three kind of
matrix elements, the variances of which depend on the number of
transferred one-body occupancies between the connected Slater determinants.
Diagonal matrix elements ${\cal U}_f^{I,I}
=\langle I| {\cal U}_f |I \rangle $ ($|I \rangle$ denotes a Slater determinant)
have a variance $\sim n(n-1)U^2/2$,
one-body off-diagonal elements that change only one occupancy
have a variance $\sim (n-1) U^2$ whereas generic two-body off-diagonal
matrix elements inducing transitions between Slater
determinants differing by exactly two occupancies, have the generic
variance $U^2$. In diagrammatic language, these discrepancies occur due
to the presence of up to two closed loops in the diagram corresponding
to these matrix elements, each loop corresponding to 
a sum over $O(n)$ uncorrelated IME.

Our full Hamiltonian then reads

\begin{eqnarray}\label{ham}
{\cal H} & = & {\cal H}_0 + {\cal U}_{avg} + {\cal U}_{f}
\end{eqnarray}

The mean-field Hamiltonian proposed in
\cite{newaleiner1} was constructed along similar lines but
neglects the fluctuations of interaction ${\cal U}_{f}$ and
is thus embedded in 
the above Hamiltonian (\ref{ham}). Consequently, all results
derived there can be obtained from the treatment to be presented
below after setting the strength of fluctuations $U \rightarrow 0$.
In a condensed matter context this is justified in the limit of
large conductance $g \rightarrow \infty$. As recent experiments in
quantum dots seem to be consistant with a conductance $g \approx 6-8$
\cite{ensslin}, it is a priori not obvious that ${\cal U}_f$ can
be neglected. We also stress that both the Random Matrix Theory (RMT)
symmetry under orthogonal (or unitary) basis transformation in the one-body
Hilbert space (which in metallic samples is satisfied for energy scales
smaller than the Thouless energy $E_c=g \Delta$ \cite{altshklo}), 
and the $SU(2)$ symmetry under rotation in spin space are
satisfied by each of the three terms in the above Hamiltonian.

The charge-charge mean-field
contribution results in a constant energy shift of the full spectrum
and has thus no influence on the
ground-state spin; we therefore neglect it henceforth.
This must however be kept in mind, as it is for instance
well known that including self-consistently
the mean-field charge-charge contribution of the interactions
(e.g. in a Hartree-Fock approach) leads to significant
corrections to the one-particle density of
states at the Fermi level \cite{altshuler,kopietz,blanter00}.
The BCS term gives rise to
superconducting fluctuations for a negative
effective interaction in the Cooper
channel $\langle U_{\alpha,\alpha}^{\beta,\beta} \rangle < 0$. We 
shall only consider disordered
metallic samples which have $\langle U_{\alpha,\alpha}^{\beta,\beta} 
\rangle > 0$. In this case the renormalization group flow brings the 
BCS coupling
to zero \cite{abrikosov}. We thus also
neglect this term and set ${\cal U}_{avg} = {\cal U}_{ss} $.
Note however that the presence of a nonzero
(repulsive or attractive) BCS coupling may stabilize a paramagnetic phase.

After these considerations we reach our model Hamiltonian

\begin{eqnarray}\label{hamfin}
{\cal H} & = & \sum_{\alpha} \epsilon_{\alpha}
n_{\alpha}
-\lambda U \vec{S} \cdot \vec{S}
+ \sum_{\alpha,\beta;\gamma,\delta} \sum_{s,s'}
    \bar{U}_{\alpha,\beta}^{\gamma,\delta}
c^{\dagger}_{\alpha,s} c^{\dagger}_{\beta,s'} c_{\delta,s'}
c_{\gamma,s}
\end{eqnarray}

  Due to the SRS that we imposed on
the original Hamiltonian (\ref{hamsite}), the interaction
commutes with the total magnetization $|\vec{S}|^2$ and its projection $S_z$
so that the Hamiltonian acquires a block structure where blocks are
labelled by a quantum number
$\sigma_z$ and subblocks of given $\sigma \ge \left| \sigma_z \right|$
appear within each of these blocks. Each block's size is given in term of
binomial coefficients as $N(\sigma_z) =
\left(_{n/2-\sigma_z}^{m/2}\right)
\left(_{n/2+\sigma_z}^{m/2}\right)$, while the size of a subblock of given
$\sigma$ is given by $N(\sigma) = N(\sigma_z=\sigma)-N(\sigma_z=\sigma+1)$.
Due to SRS it is sufficient to study the block with
lowest projection $\sigma_z=0$ $(1/2)$ for even (odd) number of
particles, as all values of $\sigma$ will be included in this block. For
simplicity, we will consider an even number $n$ of particles in the initial
discussion presented below, and will generalize the discussion later on to
include odd
$n$, highlighting the main differences between the two
cases. It is important to remark
that both $\sigma$ and $\sigma_z$ are not only good quantum numbers
for the full Hamiltonian, but also individually for ${\cal H}_0$,
${\cal U}_{avg}$ and ${\cal U}_{f}$. This allows us to consider each of these
terms separately and in the next two chapters we will make use of this
property, first neglecting ${\cal U}_{ss}$ : as it only
generates constant energy shifts within each sector,
it can be added after the restricted problem ${\cal H}_0+{\cal U}_{f}$
has been solved.

  In (\ref{od}) the sums in both the spin and orbital indices
are not restricted, i.e $s,s'=\downarrow,\uparrow$
and $\alpha,\beta,\gamma,\delta=1,2,...m/2$. It
is both convenient and instructive to rewrite it as

\begin{eqnarray}\label{usrs}
{\cal U}_{f} & = & \sum_{\alpha>\beta;\gamma<\delta} \sum_{s_z=0,\pm 1}
    V_{\alpha,\beta}^{\gamma,\delta} T^{\dagger}_{\alpha,\beta}(s_z)
T_{\gamma,\delta}(s_z) 
+ \frac{1}{2} \sum_{\alpha \ge \beta;\gamma \le \delta}
    W_{\alpha,\beta}^{\gamma,\delta} S^{\dagger}_{\alpha,\beta}
S_{\gamma,\delta}
(1-\frac{1}{2} \delta_{\alpha,\beta}) (1-\frac{1}{2} \delta_{\gamma,\delta})
\end{eqnarray}

\noindent where we have introduced totally symmetric and
antisymmetric matrix elements

\begin{eqnarray}\label{symasym}
W_{\alpha,\beta}^{\gamma,\delta} & = &\bar{U}_{\alpha,\beta}^{\gamma,\delta}+
\bar{U}_{\beta,\alpha}^{\delta,\gamma}+\bar{U}_{\alpha,\beta}^{\delta,\gamma}+
\bar{U}_{\beta,\alpha}^{\gamma,\delta} \nonumber \\
V_{\alpha,\beta}^{\gamma,\delta} & = &\bar{U}_{\alpha,\beta}^{\gamma,\delta}+
\bar{U}_{\beta,\alpha}^{\delta,\gamma}-\bar{U}_{\alpha,\beta}^{\delta,\gamma}-
\bar{U}_{\beta,\alpha}^{\gamma,\delta}
\end{eqnarray}

\noindent as well as two-body creation and destruction
operators for either singlet-paired

\begin{equation}\label{singlet}
S^{\dagger}_{\alpha,\beta} = (c^{\dagger}_{\alpha,\uparrow}
c^{\dagger}_{\beta,\downarrow}-
c^{\dagger}_{\alpha,\downarrow} c^{\dagger}_{\beta,\uparrow})/\sqrt{2}
\ \ \  ; \ \ \
S^{\dagger}_{\alpha,\alpha} = c^{\dagger}_{\alpha,\uparrow}
c^{\dagger}_{\alpha,\downarrow}
\end{equation}

\noindent or triplet-paired fermions 

\begin{equation}\label{triplet0}
T^{\dagger}_{\alpha,\beta}(0) = (c^{\dagger}_{\alpha,\uparrow}
c^{\dagger}_{\beta,\downarrow}+
c^{\dagger}_{\alpha,\downarrow} c^{\dagger}_{\beta,\uparrow})/\sqrt{2}
\ \ \  ; \ \ \ T^{\dagger}_{\alpha,\beta}(s) =
c^{\dagger}_{\alpha,s} c^{\dagger}_{\beta,s} \ \ , \ \ s= \uparrow,\downarrow
\end{equation}

\noindent As we consider fully uncorrelated IME
$U_{\alpha,\beta}^{\gamma,\delta}$, both the symmetrized and
antisymmetrized matrix elements have the same variance which for no
doubly appearing indices reads

\begin{eqnarray}\label{vwvariance}
\sigma^2 \left(W_{\alpha,\beta}^{\gamma,\delta} \right)
 = \sigma^2 \left(V_{\alpha,\beta}^{\gamma,\delta} \right) = 4 U^2
\end{eqnarray}

\noindent In principle, the ratio of the variances strongly depends on 
microscopic details, in particular the range of the
interaction. For instance, it can easily be seen that 
$\sigma^2 \left(V_{\alpha,\beta}^{\gamma,\delta} \right)/\sigma^2 
\left(W_{\alpha,\beta}^{\gamma,\delta} \right) \in [0,1]$ and that
the ratio vanishes for a Hubbard interaction. We will neglect 
this discrepancy however, but note that an increased variance of the 
symmetrized IME with respect to the antisymmetrized ones favors
a low spin ground-state \cite{lev}.

  The hamiltonian can now be regarded as acting on singlet
or triplet bonds between levels. SRS is then reflected in
the simple statement that the destruction of a bond between
two fermions  must be followed by the re-creation of a bond of the same nature.
We note that the triplet operators
(\ref{triplet0}) create
either a $\sigma_z=0$, $\sigma=1$ or a $\sigma_z=\pm 1$, $\sigma=1$
two-fermion state in a fixed spin basis.
A rotation in spin space would bring the operators in
(\ref{triplet0}) into one another and the first three terms in the
brackets in (\ref{usrs}) are not individually SRS but must be
considered as one single spin-conserving operator.
We illustrate this point in Appendix \ref{appendix1},
where we evaluate the effect
of this operator acting on a four-particle state with two double
occupancies. Note also that from (\ref{ime}) and (\ref{symasym}),
a purely on-site interaction influences only the singlet channel
as in this case the antisymmetrized IME vanish identically.

  The procedure leading to (\ref{usrs}) amounts to a projection of the
interaction operator
onto the two irreducible representations of the two-fermion
symmetry group. In this way the two-body singlet
matrix elements are explicitly separated from their triplet
counterparts, and the rewriting leading to (\ref{usrs}) allows
us to formulate the many-body problem in term of two-particles bonds
of different nature in a similar way as the authors of \cite{lev}.
Any even $n$-fermion state is represented as a $n/2$-boson
state where each boson has either spin $\sigma=0$ or 1. These
bosons can be constructed by acting on
the vacuum $|0\rangle$ with an $S$ or a $T$ operator respectively, and
the spin of these composite bosonic states depends on the
bond between the two fermions, i.e. whether the fermionic antisymmetry
is supported by the spin or the spatial degrees of freedom.
Alternatively, this means that for a $n$-body state of total spin
$\sigma$, the number of triplet bonds is given by $\sigma$ \cite{remark5}. Also
double orbital occupancies result in singlet bonds, so that their
number is restricted to $[0,n/2-\sigma]$. This construction
leads however to an overcomplete basis for $n \ge 6$. We were
unable to propose a systematic reduction to an orthonormal set of states
nor are we aware of any such systematic construction in the literature. 
For the computations
to be carried below it will however be sufficient to know that such
a basis can in principle
be constructed (via e.g. reduction and orthogonalization
of the constructed overcomplete basis), and how to construct it for the
special case of four particles above the filled fermi sea, as those are the
only states one encounters when doing second order perturbation theory for
the levels of lowest energy in the $\sigma=0$ and 1 sectors.

  Equation (\ref{usrs}) helps us see the key qualitative
point of our work. In second order perturbation theory ${\cal U}_f$ will
generate transitions in each spin
sub-block between the ground state 
and excited states differing by two occupation numbers (or
less). Both the triplet and singlet terms will generate transitions,
but there are certain types of transitions which can be generated by
the singlet term which {\it cannot} be generated by the triplet term.
For instance the triplet operator cannot generate transitions to final
states with additional double occupancies
nor is it possible to scatter a triplet bonded pair into
a double occupancy (see Fig. \ref{fig:sketch}).
As the magnetization increases, the number of singlet transitions
decreases accordingly as the
number of singlet two-particle bonds in a many-body state
obviously decreases with its total magnetization. Eventually, when $\sigma$
is maximal, only triplet transitions survive and we can readily
conclude that the number of two-body transitions is a monotonically decreasing
function of the magnetization as is therefore the number of (energy-decreasing)
second order contributions.
We will see below that this condition on
the available volume for phase-space scattering is crucial for the
ground-state magnetization properties, both in the perturbative regime
($U/\Delta \ll 1$) and in the asymptotic limit of dominant fluctuations
($U/\Delta \gg 1$).
It is important to understand that the relevant variable
here is the number of transitions and not the size of the
Hilbert space;
the block size $N(\sigma)$ is in general (for a sufficient number of
particles) a nonmonotonic function of $\sigma$, as on one hand
$N(\sigma=n/2) < N(\sigma=0)$ (or $N(\sigma=n/2) < N(\sigma=1/2)$
for odd number of particles), whereas on the other hand, and in the limit
$\sigma \ll n/2 \ll m/2$, it can be shown using Stirling's
formula that $\partial N(\sigma)/\partial \sigma > 0$. Except for very
few particles, $N(\sigma)$ has its maximum at a finite magnetization,
whereas the number of transitions is always maximum for $\sigma=0$.

We close this introductory chapter with a brief historical
survey of random interaction models similar to (\ref{hamfin}-\ref{usrs}).
These models originated in nuclear physics and are
based on similar principles as those which led Wigner to
propose the gaussian ensembles of random matrices, with the additional
requirement that they represent particles interacting via a $k$-body
interaction. Only when the rank $k$ of the interaction
is equal to the number $n$ of particles does one recover the Wigner
gaussian ensembles. Physically, interactions are in principle not random
{\it per se}, however once one postulates the invariance of the
one-body Hamiltonian matrix ensemble under unitary (i.e. basis)
transformation,
a postulate motivated e.g. by a chaotic one-body dynamics,
random IME naturally appear (see (\ref{ime})), and this
results in a similar invariance
for the many-body Hamiltonian ensemble and the associated probability
distribution $P({\cal H}) \propto \exp(-{\rm Tr} {\cal H}^2/2)$.
The first proposed model with random interactions was the fermionic
Two-Body Random Interaction Model (TBRIM) for spinless fermions
which was introduced independently by French and
Wong and Bohigas and Flores \cite{french}. This model is essentially
a spinless version of ${\cal U}_{f}$. While significant
deviations from the usual gaussian ensemble of random matrices 
were found in the
tails of the spectrum - in particular the MBDOS for $n \gg 2$
has a gaussian, not a semicircular shape -
these authors found no significant differences in the spectral
properties at high excitation energy. (This latter finding has been however
challenged very recently \cite{weiden} and may be due to the smallness of the
systems considered.)
More recently this spinless TBRIM was
extended with a one-body part and it has been
discovered that the critical interaction strength $U_c$ at which
WD statistics sets in is governed by the energy spacings $\Delta_c$
between directly coupled states \cite{m2body}.
This model and similar
ones have also been studied in the framework of quantum chaos in
atomic physics \cite{flam}, in particular the thermalization of
few-body isolated systems has attracted a significant
attention \cite{flam2,zel2}, and more recently
in solid-state physics to study quasiparticle
lifetime \cite{m2body,mbchaos1,mbchaos2,mbchaos3p,mbchaos4,mbchaos5}
and fluctuations of Coulomb blockade conductance peak spacings and
heights in quantum dots \cite{yoram,yoramrev}.
In a solid-state context however
the invariance of the one-body Hamiltonian under basis transformation
is satisfied only in an energy interval of the order of the Thouless
energy $E_c = g \Delta$
around the Fermi energy, where $g$ is the conductance \cite{altshklo}. 
Wavefunctions correlations become stronger and stronger
beyond the Thouless energy where IME
start to decay algebraically as a function of the energy.
It is thus reasonable to consider our random interaction model
as an effective truncated Hamiltonian in an energy window given by the
Thouless energy $E \in [E_f-E_c,E_F+E_c]$ \cite{mbchaos1}, so that
the number of particles and orbitals behave as $m$,$n \sim g$.
Nuclear shell models may also be represented by randomly
interacting models, differing from the original TBRIM in
the presence of additional
quantum numbers like spin, isospin, parity and so forth\cite{zel2}.
Most of those models consider
the limit of dominant fluctuations $U/\Delta \gg 1$ and quite
unexpectedly, it has been found that even in this regime,
random interactions may
result in an orderly behavior \cite{nuphys}, in particular a strong
statistical bias toward a low angular momentum ground-state.
In particular, for the special
case of an angular momentum restricted to $j_z=\pm 1/2$, the probability
of finding a zero angular momentum ground-state for an even number of nucleons
reaches almost 100\% \cite{spings1,lev}.
While the reasons for this behavior
in the asymptotic regime are still not clear \cite{dmitri},
we will see below that a strong
bias toward low angular momentum ground-state results from a stronger
broadening of the Many-Body Density of States (MBDOS) in the low spin sector,
associated with a larger number of off-diagonal transitions.
The same phenomenon
with qualitatively the same origin will be shown to influence the
ground-state magnetization in the perturbative limit.

\section{The Case of $\lowercase{n}=2$ Fermions}\label{tip}

For $n=2$ particles, only
the sectors $\sigma=0$ and $1$ exist whose size is
given by $N(\sigma)=m/2(m/2+1-2 \sigma)/2$. In each sector,
the interaction matrix ${\cal U}_{f}$
is a GOE matrix (the number of
particles is equal to the rank of the interaction) and all
Hamiltonian matrix elements are non-zero and have the same variance.
For simplest case of two orbitals ($m/2=2$) one can demonstrate the
magnetization reducing effect of interaction
Fluctuations by an argument which is {\it exact} for all values of the
off-diagonal fluctuations $U$. The two orbitals are spin-degenerate and
have energies $\epsilon_1=0$ and $\epsilon_2=\Delta > 0$. In the absence
of interaction fluctuations, the three eigenvalues in the $\sigma=0$
sector are $0$, $\Delta$ and $2 \Delta$. Switching on the interaction,
the determinant of the Hamiltonian matrix in the time-reversal
symmetric case can be written

\begin{eqnarray}
{\rm Det} {\cal H} & = & \frac{1}{2} W_{1,1}^{1,1} (\Delta +
\frac{1}{2} W_{1,2}^{1,2}) (2 \Delta + \frac{1}{2} W_{2,2}^{2,2})
+ \frac{1}{4} W_{1,2}^{1,1} W_{2,2}^{1,2} W_{2,2}^{1,1} \nonumber \\
& - & \frac{1}{4} \left[ \frac{1}{2} W_{1,1}^{1,1} (W_{2,2}^{1,2})^2 +
(W_{1,2}^{1,1})^2 (2 \Delta + \frac{1}{2} W_{2,2}^{2,2})+
(W_{2,2}^{1,1})^2 (\Delta + \frac{1}{2} W_{1,2}^{1,2})\right]
\end{eqnarray}

  Every single term in this expression has a symmetric
distribution, i.e. an equal probability of being positive or negative,
except for a term $-((W_{2,2}^{1,1})^2+2 (W_{1,2}^{1,1})^2 ) \Delta/4$
which is always negative. It results that the determinant
has a higher probability of being negative which in its turn means
that the lowest eigenvalue (which vanishes at $U/\Delta=0$)
is statistically more often negative than positive when $U$ is switched
on - it is more likely to be reduced than increased by the off-diagonal
fluctuations. Simultaneously, and in absence of exchange,
the energy of the only $\sigma=1$
level is given by $\Delta + V_{1,2}^{1,2}/2$, so that
the fluctuations lower or increase it with equal probability. Hence
fluctuations always increase the average spin gap in this case.

We next consider an arbitrary number of orbitals, $m$.
First consider the limit of dominant fluctuations $U/\Delta \gg 1$.
${\cal H} \approx {\cal U}_f$ is then a GOE matrix and
its MBDOS is well approximated by a semicircle law ($E^2 \le E_0^2$)

\begin{eqnarray}\label{halfcircle}
\rho_{GOE} = \frac{2}{\pi E_0^2(\sigma)} \sqrt{E_0^2(\sigma)-E^2}
\end{eqnarray}

\noindent where $E_0(\sigma) \approx \sqrt{2 N(\sigma)} U$.
This expression is not exact however
as there are corrections in the tail of the distribution \cite{mehta} as
one can see on Fig. \ref{fig:dostip}. These corrections behave as
$O(N^{-1/6})$ while the level density there is $O(N^{1/6})$ \cite{mehta},
i.e. the number of
levels outside the semicircle is independent of N (and hence of $m$)
and for simplicity we will neglect these corrections in what follows.

Henceforth we shall be focusing attention on the ground state in each
spin sector and the gaps between these ground states, so it is useful
to adopt the standard term in nuclear physics for the lowest levels, of a
given spin or angular momentum, {\it yrast} levels. 
In the current model, in the asymptotic regime of
large fluctuations $U/\Delta \gg 1$,
(and neglecting the exchange interaction)
we can approximate the energy of the yrast states
by $E_0(\sigma)$ and hence readily
predict that the average $\sigma=0$ yrast energy
will be lower than its $\sigma=1$ counterpart by an amount

\begin{equation}\label{gapinftip}
\Delta^s \equiv \overline{{\cal E}_{0,\sigma=1}-{\cal E}_{0,\sigma=0}}
\approx
U (\sqrt{m (m/2+1)/2}-\sqrt{m (m/2-1)/2}) = U + O(\frac{1}{m})
\end{equation}

\noindent i.e. on average there is a spin gap for $U/\Delta \gg 1$ in the
large
$m$ limit. Next we can calculate the average energy of the first excited
$\sigma=0$ level ${\cal E}_{1,\sigma=0}$ via integration of the average
MBDOS (\ref{halfcircle}) as

\begin{equation}
\overline{{\cal E}_{1,\sigma=0}-{\cal E}_{0,\sigma=0}} =
O(N(\sigma)^{-1/6}) = O(m^{-1/3})
\end{equation}

In the relevant limit $ m \gg 1$ the splitting
between this first excited $\sigma=0$ level and the $\sigma=0$ yrast
level is negligible and both states are below the $\sigma=1$ yrast
level by a gap of order $U$, independent of $m$.
This calculation can be extended to higher $\sigma=0$
excited states and the result suggests that on average there is a large
number
$O(m^{1/3})$ of $\sigma=0$ levels which have a lower energy than the
first
spin excited state. Remember however that we have neglected
corrections to the tails of the density of states, and it turns out that
these corrections result in an $m$-independent number $p \approx 3$
of $\sigma=0$ levels in the spin gap as shown by the numerical
data presented in Fig. \ref{fig:numgaptip}. In Fig. \ref{fig:gapinftip}
we show a numerical check which confirm the validity
of (\ref{gapinftip}) up to prefactors which are
due to additional correlations between the considered levels and cannot
be captured by the simple arguments presented here. We will come back to
this point in section \ref{tbrim}. Note however that the distance between
${\cal E}_{0,\sigma=0}$ and ${\cal E}_{1,\sigma=0}$ seem to remain
constant as $m$ increases which is a manifestation of the presence of the
tail correction to the semicircle law (\ref{halfcircle}) and is beyond
the
reach of the simplified reasoning we have presented.

We can next calculate perturbatively
the energy of the yrast state in each sector
up to the second order in $U/\Delta$. These states
can be written as (the singlet and triplet creation operators
$S_{1,1}^{\dagger}$ and $T_{1,2}^{\dagger}(0)$
have been defined in (\ref{singlet}) and (\ref{triplet0}))

\begin{eqnarray}
|\Psi_0^{(\sigma=0)} \rangle & = &
S_{1,1}^{\dagger} | 0 \rangle \nonumber \\
|\Psi_0^{(\sigma=1)} \rangle & = &
T_{1,2}^{\dagger}(0) | 0 \rangle
\end{eqnarray}

\noindent Up to the first order their energies are given by

\begin{eqnarray}\label{pertgstip}
{\cal E}_{0,0} & = & 2 \epsilon_1 +
W_{1,1}^{1,1}/2          \nonumber \\
{\cal E}_{0,1} & = & \epsilon_1 + \epsilon_2 + V_{1,2}^{1,2}/2 -2 \lambda
U
\end{eqnarray}

\noindent and the second order corrections read (using the
constant spacing model for the one-particle levels)

\begin{eqnarray}\label{2nd}
\overline{\Delta{\cal E}_{0,0}^{(2)}}
& = & -\frac{1}{2} \sum_{\alpha \ge \beta > 2}
\frac{\overline{(W_{\alpha,\beta}^{1,1})^2}(1-\delta_{\alpha,\beta}/2)}
{(\alpha+\beta-2) \Delta}
-\frac{1}{2} \sum_{\alpha \ge 2}
\left[\frac{\overline{(W_{1,\alpha}^{1,1})^2}}{(\alpha-1) \Delta}+
\frac{\overline{(W_{2,\alpha}^{1,1})^2}(1-\delta_{\alpha,2}/2)}
{(\alpha-2) \Delta}\right]\nonumber \\
\overline{\Delta{\cal E}_{0,1}^{(2)}} & = & -\sum_{\alpha > \beta>2}
\frac{\overline{(V_{\alpha,\beta}^{1,2})^2}}{(\alpha+\beta-3) \Delta}
-\sum_{\alpha > 2}
\left[\frac{\overline{(V_{2,\alpha}^{1,2})^2}}{(\alpha-1) \Delta}+
\frac{\overline{(V_{1,\alpha}^{1,2})^2}}{(\alpha-2) \Delta}\right]
\end{eqnarray}

\noindent Note that any double occupancy in either the initial or the
final state, results
in a $1/\sqrt{2}$ reduction of the transition amplitude, hence the
factors $1/2$ appearing on the right-hand side of the first and second
lines of (\ref{2nd}). These factors are however exactly counterbalanced
by the IME averages, since one has (see (\ref{vwvariance})

\begin{eqnarray}
\overline{(W_{\alpha,\alpha}^{1,1})^2} & = & \overline{(
4 U_{\alpha,\alpha}^{1,1})^2} = 16 U^2 \nonumber \\
\overline{(W_{\alpha,\beta}^{1,1})^2} & = & \overline{(
2 U_{\alpha,\beta}^{1,1}+2 U_{\beta,\alpha}^{1,1})^2} = 8 U^2 \\
\overline{(V_{\alpha,\beta}^{1,2})^2} & = & \overline{(
U_{\alpha,\beta}^{1,2}+U_{\beta,\alpha}^{2,1}-U_{\beta,\alpha}^{1,2}
-U_{\alpha,\beta}^{2,1})^2} = 4 U^2 \nonumber
\end{eqnarray}

\noindent The second order contributions for the energies of the
lowest levels in each spin sector is therefore given by

\begin{eqnarray}\label{2ndf}
\overline{\Delta{\cal E}_{0,0}^{(2)}}
& = &
-\frac{4 U^2}{\Delta} \sum_{\alpha \ge \beta > 2}
\frac{1}{\alpha+\beta-2} - \frac{8 U^2}{\Delta} \sum_{\alpha \ge 2}
\left[ \frac{1}{\alpha-1}+\frac{1}{\alpha-2}\right]\nonumber \\
\overline{\Delta{\cal E}_{0,1}^{(2)}} & = &
-\frac{4 U^2}{\Delta} \sum_{\alpha > \beta>2}
\frac{1}{\alpha+\beta-3} - \frac{8 U^2}{\Delta} \sum_{\alpha>2}
\left[ \frac{1}{\alpha-1}+\frac{1}{\alpha-2}\right]
\end{eqnarray}

\noindent The expressions given in equation (\ref{2ndf})
are in very good agreement with numerical
data obtained from exact diagonalization
as we show on Figs. \ref{fig:e0tip} and \ref{fig:e1tip}.
It is clearly seen from (\ref{2ndf}) that the singlet and triplet second
order corrections differ only by a restriction
in the sums which arises in the triplet case because transitions to
doubly occupied states are not allowed; it is straightforward to show
that there are exactly $m/2$ such transitions.
As each contribution
in second order perturbation theory
reduces the energy of the lowest energy state in each sector,
these $m/2$ additional transitions will therefore
favor a singlet ground-state in the perturbative regime.

All other transitions give on average the
same contribution to ${\cal E}_{0,0}$ as to ${\cal E}_{0,1}$
as symmetric
and antisymmetric matrix elements have the same variance. As
the first order corrections do not survive disorder averaging,
we can write the average energy difference between those
two levels in second order perturbation theory as

\begin{eqnarray}\label{tip2nd}
\Delta^S & \approx & \Delta -2 \lambda U + A \frac{U^2}{\Delta} \ln
(m/2)
\end{eqnarray}

\noindent where $A>0$ is a numerical prefactor that can be extracted
from (\ref{2ndf}) and the above expression is
valid in the large $m$ limit.
It follows from (\ref{tip2nd}) that in order to align spins,
the exchange has to overcome more than just one level spacing.
Equivalently, (\ref{tip2nd}) states that off-diagonal fluctuations
increase the energy spacing between the lowest
energy states of each sector. Equation (\ref{tip2nd}) has been
checked numerically and the result is shown in Fig. \ref{fig:gaptip}.

  One can also compute perturbatively the splitting induced
by the off-diagonal fluctuations between the first $\sigma=0$
excited state and the $\sigma=1$ yrast. As a matter of fact, except for
the
exchange interaction,
all corrections in the first two orders in perturbation theory give the
same average contributions up to second order contributions which exist
only
for $|\Psi_1^{(0)} \rangle$ and correspond to scattering onto a double
occupancy. In second order perturbation theory, this splitting reads

\begin{eqnarray}\label{splitting}
\overline{{\cal E}_{0,1}-{\cal E}_{1,0}} & = & -2 \lambda U +
\sum_{\alpha \ge 3}
\frac{\overline{(W_{\alpha,\alpha}^{1,2})^2}}{2 (2 \alpha-3) \Delta}
\approx -2 \lambda U + A' \frac{U^2}{\Delta} \ln (m/2)
\end{eqnarray}

\noindent In particular we see that the splitting induced by the
interaction fluctuations favors the
spatially {\it symmetric} singlet state and opposes
the exchange term ($A'>0$). Note also that for $n=2$, both the splitting
(\ref{splitting}) and the spin gap (\ref{tip2nd}) have a similar
magnitude.
We will see below that this is no longer the case for larger $n$.
Replacing the sum by an integral one finds $A' \approx 2$ in
(\ref{splitting}).

Some remarks are in order here as the case of two particles
is somehow special. For $n=2$,
${\cal U}_f$ is a GOE matrix for which
the number of transitions in each sector is equal
to its size. However,
as one adds particles, the matrix
becomes sparser and sparser as the Hilbert space
size grows exponentially with the number of particles,
whereas the number of transitions is a polynomial in $n$. It is however
clear from the perturbative treatment presented above that
what matters is the number of {\it transitions} not the sector
size. Generically and for sufficient number
of particles, the sector
with largest number of states has finite (nonzero) magnetization, whereas
it is always for $\sigma=0$ that one has the most transitions and
hence the largest probability to find the ground-state.
Simultaneously, for increasing number of particles,
the MBDOS undergoes a crossover
to a gaussian shape in the limit $n \gg 2$ \cite{brody,kota}. It is
understood that the sparsity of the resulting matrices alone does not
invalidate the semicircle law, sparse matrices with uncorrelated
matrix elements may have a semicircle law \cite{fyo,greek}. However,
as noted already, the IME in the TBRIM are highly correlated and this
apparently drives the MBDOS to the gaussian form.
For a very recent and interesting
analytical study of this crossover, we refer the reader to \cite{weiden}.
Of importance for us is that even
for $n \gg 2$ one still has a reliable expression for the MBDOS
in term of $n$ and $m$ that one may use to extract the average
energy difference between yrast states
in the regime of large fluctuations. We will implement
this procedure for $n>2$ in chapter \ref{tbrim}.

\section{Perturbative Treatment for $\lowercase{n}>2$}\label{pert}

We now discuss the perturbation theory for the yrast states for arbitrary
$n$.  These results are of particular interest since numerical results
for large $U$ are necessarily restricted to small $n$ and one may worry
that the large $n$ behavior is qualitatively different.  In this case,
within the perturbative regime, we can show analytically that
fluctuations reduce the probability of a magnetized ground state for
arbitary $n$.  To estimate the size of $U$ one must consider the
disorder averaged typical amplitude of fluctuations of the IME
(\ref{ime}), which has been computed for diffusive metallic samples
\cite{blanter,mirlin}. In this case the effective
static electronic interaction
is strongly screened and can therefore be well approximated by
a Hubbard interaction. Then, the variance of the IME
(\ref{ime}) is given by

\begin{eqnarray}\label{sigmaU}
\sigma^2(U_{\alpha,\beta}^{\gamma,\delta}) & = & {\cal U}^2
\sum_{i} \sum_{j} \overline{
\psi_{\alpha}^*(i)  \psi_{\beta}^*(i)
   \psi_{\gamma}(i)  \psi_{\delta}(i)
\psi_{\alpha}(j)  \psi_{\beta}(j)
   \psi_{\gamma}^*(j)  \psi_{\delta}^*(j)}
\end{eqnarray}

  In diffusive systems for which $l_e \ll L$ holds ($l_e$ is the elastic
mean free path), the wavefunctions can be
estimated using classical return probabilities as
extracted from the diffusion equation and one gets
$\sigma(U_{\alpha,\beta}^{\gamma,\delta})
\equiv U \sim \Delta/g$ \cite{blanter}.
In metallic samples
the conductance $g$ is very large and even in small quantum dots
it is typically
of the order of ten. It is therefore of interest to start with
a perturbative treatment up to second order
in the small parameter $U/\Delta$.
Each contribution in second order perturbation theory
is always negative for each yrast state
and we will see, as for the case $n=2$, that the number of such
contributions is larger in
the lowest spin sector, thereby favoring the absence
of magnetization; however additional and more subtle interference effects
in the transition matrix elements also appear and favor
$\sigma =0$. Here and
if not stated otherwise in the rest of the paper, we will make use of SRS
and consider each $\sigma$ sector in the $\sigma_z=0$ block.
This means that there are as many particles with up as with down spins,
and states with different $\sigma$'s but the same occupancies
will differ only in the nature of two-particle bonds between pairs of
fermions on partly occupied orbitals (see Fig. \ref{fig:sketch} and the
discussion in section \ref{model}).
We will also focus most of our discussion on the case of an even number
of particles,
but will eventually generalize our results to an odd number of particles.
To simplify numerical checks of the
perturbation theory we will consider
only the case of equidistant one-body orbitals
$\epsilon_{\alpha}=(\alpha-1)\Delta$ in this section
and will discuss generic spectra later on.

For $\sigma_z=0$, there are an equal number of
spin up and spin down fermions and  $N(0) =
\left(_{n/2}^{m/2}\right)^2$ Slater determinants. At $U=0$ the
ground-state
can be written as

\begin{eqnarray}\label{fermisea}
|F_n \rangle & = &
\prod_{\alpha=1,n/2}
c_{\alpha,\uparrow}^{\dagger}
c_{\alpha,\downarrow}^{\dagger} |0 \rangle
\end{eqnarray}

  Obviously this state has $\sigma=0$, as doubly occupied
orbitals form a singlet two-particle state. Acting
on $|F_n \rangle$ with the $S^{\dagger}$ and $T^{\dagger}(0)$ operators 
(see ((\ref{singlet}-\ref{triplet0})) $Q$
and $P$ times respectively

\begin{eqnarray}\label{basis}
\prod_{\alpha,\beta}^{P} S^{\dagger}_{\alpha,\beta}
\prod_{\gamma,\delta}^{Q} T^{\dagger}_{\gamma,\delta}(0) |F_n \rangle
\end{eqnarray}

\noindent allows to construct a $\sigma_z=0$ state which is in general
a linear combination of Slater determinants of total spin
$\sigma = 1,2,...Q-1,Q$. One can in principle
represent a complete basis with good quantum numbers $\sigma$,
$\sigma_z$ and one-particle occupations from the states
(\ref{basis}) following the rules :

$\bullet$ Fermions on the same orbital are singlet paired

$\bullet$ Fermions on singly occupied orbitals are arbitrarily
bonded in pairs, $\sigma$ of the latter being triplet, the
rest being singlet bonded

$\bullet$ The triplet bonds combine to maximize the total spin

  While the first rule is imposed by the Pauli principle,
the second and third rules are a matter of convention. This set
of rules is similar to the one employed by Kaplan, Papenbrock
and Johnson \cite{lev} for the case of $n=4$ particles. 
As noted above, the generalization to more particles
is not trivial: following the above prescription,
one obtains an overcomplete basis and
one should construct a proper orthogonalization procedure
to reduce this basis. In what follows however we will compute
perturbative corrections up to the second
order for only three different states : the $\sigma=0$ and
$\sigma=1$ yrast states ($|\Psi_0^{(\sigma=0)} \rangle$ and
$|\Psi_0^{(\sigma=1)} \rangle$), and the first $\sigma=0$ excited state
($|\Psi_1^{(\sigma=0)} \rangle$) For comparison of these states the
construction of a
basis for $n=4$ is sufficient. We can write these three states as

\begin{eqnarray}\label{3states}
|\Psi_0^{(0)} \rangle & = & |F_n \rangle \nonumber \\
|\Psi_1^{(0)} \rangle & = &
S_{n/2,n/2+1}^{\dagger} |F_{n-2} \rangle \nonumber \\
|\Psi_0^{(1)} \rangle & = &
T_{n/2,n/2+1}^{\dagger}(0) |F_{n-2} \rangle
\end{eqnarray}

  The difference between the
$\sigma=1$ yrast state and the first $\sigma=0$ excited state
lies exclusively in the bond
between the last two particles : it is a triplet in the first case
and a singlet in the second. Up to first order, the energies of
the states (\ref{3states}) are given by

\begin{eqnarray}
\overline{{\cal E}^{(1)}_{0,0}} & \equiv &
\overline{\langle \Psi_0^{(0)}|{\cal H}|\Psi_0^{(0)} \rangle} =
\frac{n}{2}(\frac{n}{2} -1)\Delta \nonumber \\
\overline{{\cal E}^{(1)}_{1,0}} & \equiv &
\overline{\langle \Psi_1^{(0)}|{\cal H}|\Psi_1^{(0)} \rangle} =
\overline{{\cal E}^{(1)}_{0,0}}+\Delta  \\
\overline{{\cal E}^{(1)}_{0,1}} & \equiv &
\overline{\langle \Psi_0^{(1)}|{\cal H}|\Psi_0^{(1)} \rangle} =
\overline{{\cal E}^{(1)}_{0,0}}+\Delta-2 \lambda U
\nonumber
\end{eqnarray}

\noindent Without interactions, the latter two levels are degenerate and in
first order they are on
average splitted only by the exchange interaction favoring as usual the
spatially antisymmetric
triplet state. To calculate the average second order corrections, we
need to know the number of direct interaction induced
transitions which we will call the {\it connectivity}
$K$ and which is calculated in detail in Appendix B. $K$
is a monotonously decreasing
function of the total spin and in particular the difference
between its values at $\sigma=0$ and $\sigma=1$
is always $m/2$, independently on the number of particles.
This decrease of $K$ as a function of $\sigma$ results in a smaller
number of second order contributions for states in higher
$\sigma$ sectors and thus a smaller reduction of the energy
of the corresponding yrast state. We will identify
below the transitions which give the major contributions
to the difference in second order shift between the two lowest yrast
states. The second order correction
to the energy of the $\sigma=0$ yrast reads

\begin{eqnarray}\label{pertnip0}
\overline{\Delta{\cal E}_{0,0}^{(2)}} & = &
  \sum_{\alpha \ge \beta \ge n/2+1}
\sum_{\gamma \le \delta \le n/2}
\frac{\overline{(W_{\alpha,\beta}^{\gamma,\delta})^2}
(1-\delta_{\gamma,\delta}/2)
(1-\delta_{\alpha,\beta}/2)}
{(\alpha+\beta-\gamma-\delta) \Delta} \nonumber \\
& + & \sum_{\alpha > \beta \ge n/2+1}
\sum_{\gamma < \delta \le n/2}
\frac{3 \overline{(V_{\alpha,\beta}^{\gamma,\delta})^2}}
{(\alpha+\beta-\gamma-\delta) \Delta} \approx A \frac{U^2}{\Delta}
n^2 m \ln(m)
\end{eqnarray}

\noindent Note that the singlet and triplet contributions add
incoherently and that triplet transition acquire a factor 3 reflecting
the corresponding number of channels ($\sigma_z=0$, $\pm 1$).
In order to estimate (\ref{pertnip0}),
the sums can be replaced by four-fold integral
which gives
an homogeneous polynomial of order three in $n$ and $m$,
each term being multiplied by a logarithmic correction. In the dilute
limit $1 \ll n \ll m$ the $m^3$ and $m^2n$ terms drop out exactly
and this gives the dominant $n^2m$ dependence expressed in
(\ref{pertnip0}).
This estimate is also confirmed by numerical evaluation of the
sum in (\ref{pertnip0}).

The above formula is found to be
in good agreement with numerical data as shown on Fig. \ref{fig:e0nip}.
Note that at larger number of particles, the dependence of the energies
of the yrast states starts to have a linear dependence in $U/\Delta$
much earlier, signaling an earlier breakdown of perturbation theory than
for small number of particles. We will discuss this point below.
The correction $\overline{\Delta{\cal E}_{0,1}^{(2)}}$ for
the $\sigma=1$ yrast can be calculated in the same way and one can show
that differences between
$\overline{\Delta{\cal E}_{0,0}^{(2)}}$ and
$\overline{\Delta{\cal E}_{0,1}^{(2)}}$ occur first due to
denominators differing by $\pm \Delta$
as transitions involving the two uppermost particles start from the
orbitals $(n/2,n/2)$ and $(n/2,n/2+1)$ for $\sigma=0$ and 1 respectively
and second due to transitions either increasing or decreasing
the number of double occupancies (which only occur for $\sigma = 0$). As
noted, the number of such transitions
is $m/2$ and they give a contribution to the spin gap which
can be written ($\nu \equiv n/m$ is the filling factor)

\begin{eqnarray}\label{pertngap}
& & 2A\frac{U^2}{\Delta} \ln(\nu)
\end{eqnarray}

\noindent where $A$ is a numerical factor. This is exactly analogous to
the energy difference we found in (\ref{tip2nd}),(\ref{splitting}) 
except that $\ln(m/2)$ has been replaced by $\ln(\nu)$.

While the term just calculated is easiest to identify, a more important
contribution to the spin energy gap comes from a more subtle source.
There is a certain class of transitions starting from the $\sigma=1$
ground-state which have exactly the same energy denominator as the
corresponding class in $\sigma = 0$ case (see Fig. \ref{fig:sketchgap}) 
but the $\sigma=1$ transitions have a reduced {\it amplitude} in comparison to the
$\sigma =0$ transitions.
The corresponding (negative) second order contributions will therefore
reduce more strongly the energy of the $\sigma=0$ ground-state.
The $\sigma=1$ transitions with this property are of the following kind.
The $\sigma=1$ non-interacting ground state has two partially occupied
levels at the top of the fermi sea which are triplet-bonded. The relevant
transition cause one of these partially occupied states to become
doubly-occupied while creating a hole in the Fermi sea and a particle above
the Fermi sea. For this kind of scattering process the number of double
occupancies does not change and one can show that the singlet and triplet
terms in the hamiltonian induce transitions onto
the same final state. Correspondingly, the two transition amplitudes
must be added coherently, and it turns out that this results in a reduced
transition probability
from $16 U^2$ down to $12 U^2$ (a detailed calculation of the amplitude
of these transitions is given in Appendix A). The
corresponding contribution to the spin gap can be estimated as

\begin{equation}\label{gapevenn}
\overline{\Delta{\cal E}_{0,0}^{(2)} -
\Delta{\cal E}_{0,1}^{(2)}} \approx - 4 \frac{U^2}{\Delta}
\int_{0}^{n/2} dx \int_{n/2}^{m/2} dy \sum_{z=n/2}^{n/2+1}
\frac{1}{y+z-x-n/2} \approx -4 \frac{n U^2}{\Delta} \ln(\nu)
\end{equation}

\noindent This result is valid in the dilute limit
$1 \ll n \ll m$ and this contribution dominates the spin
gap as soon as the number of particles is sufficiently large, i.e. for $n
\ge 4$.

  As for $n=2$ it is straightforward to calculate the splitting
induced by off-diagonal fluctuations between the $\sigma=1$ yrast and
the first excited $\sigma=0$ state :
there is no difference in the energy denominators and there is a
one-to-one
correspondence between all second order contributions for these two
states, except for the transitions which
don't exist for $\sigma=1$.
The latter correspond to scattering onto a double occupancy on the
$(n/2+1)^{\rm th}$ orbital, or from the $(n/2)^{\rm th}$ onto a double
occupancy on a previously empty orbital. These then are the only
contributions to the average splitting, which takes the form

\begin{eqnarray}\label{nsplit}
\overline{{\cal E}^{(2)}_{0,\sigma=1}-{\cal E}^{(2)}_{1,\sigma=0}} & =  &
\frac{1}{2}
\sum_{\gamma < n/2}
\frac{\overline{(W_{n/2,n/2+1}^{\gamma,\gamma})^2}}
{(n+1-2\gamma) \Delta} + \frac{1}{2} \sum_{\alpha > n/2+1}
\frac{\overline{(W_{\alpha,\alpha}^{n/2,n/2+1})^2}}
{(2\alpha-n-1) \Delta} \approx \frac{U^2}{\Delta}(\ln(m-n)+\ln(n))
\end{eqnarray}

\noindent and is thus positive.

We now briefly discuss the case of odd number
of particles. The lowest possible magnetization is $\sigma=1/2$, and at
$U/\Delta=0$, the yrast
corresponds to a singly occupied $(n/2+1)^{\rm th}$ orbital
above a filled Fermi sea (see Fig. \ref{fig:sketchodd} (a)).
The next magnetization is $\sigma=3/2$ and the corresponding
$U/\Delta=0$ yrast state is represented on Fig. \ref{fig:sketchodd} (b).
It has three single occupancies above
the Fermi sea and one of the two bonds between the corresponding
particles
must be a triplet (the choice of the bond is arbitrary).
We identified above the dominant second
order contributions to the spin gap for even $n$ as those which have an
amplitude reduction due to partial occupancies in both the initial and
final states. An example of such a transition
for $\sigma=3/2$ is depicted on Fig. \ref{fig:sketchgap}. From the
presented
data one sees that the expression corresponding to (\ref{gapevenn})
for the case of odd $n$ reads

\begin{equation}\label{gapoddn}
4 \frac{U^2}{\Delta}
\int_{0}^{n/2} dx \int_{n/2}^{m/2} dy \sum_{z>t=n/2}^{n/2+2}
\frac{1}{y+z-x-t} \approx -4 \frac{3}{2} \frac{nU^2}{\Delta} \ln(\nu)
\end{equation}

\noindent and differs from (\ref{gapevenn}) by the boundary values for
the
sums over $z$ and $t$. Correspondingly the contribution to the spin
gap picks up a factor $3/2$ and this results in an even odd
effect where the gap asymptotically behaves as $\Delta^s \approx -4 B n
\ln (\nu) U^2/\Delta$
where $B = 1$ for even and $B = 1.5$ for odd number of particles.
In particular, it is
more difficult to magnetize a system of odd number of fermions
\cite{remark6} which is in
agreement with the experimental results presented
in references \cite{berko,marcus2}. The above expression (\ref{gapevenn})
and (\ref{gapoddn}) have been checked numerically and the results are
shown
on Fig. \ref{fig:gapnip}. Both the even-odd dependence and the
$n$-dependence
of the gap are confirmed for larger number of particles $n>3$.
Note that the processes mentioned above and leading to
the scaling expression (\ref{gapevenn})
and (\ref{gapoddn}) do not exist for $n=2$ and 3 in agreement with the
data of
Fig. \ref{fig:gapnip}.

From the second order corrections to the yrast levels in each sector, 
it is possible
to construct an effective Hamiltonian which takes into
account the average effect of the off-diagonal fluctuations of interaction.
The number and strength of second order
contributions decreases with increasing magnetization and
the relevant contributions are those emphasized in this section
corresponding to one partially occupied orbital in both initial and final
two-particle state. For large magnetization $\sigma \gg 1$ the second order
contributions corresponding to the energy difference $\Delta^{(\sigma)}$
between the lowest spin yrast and the yrast 
level of spin $\sigma$ can
be approximated by the four-dimensional integral

\begin{eqnarray}\label{gapsigma}
\Delta^{(\sigma)} & \equiv & \overline{{\cal E}_{0,\sigma}^{(2)}}-
\overline{{\cal E}_{0,0}^{(2)}} = 4 \frac{U^2}{\Delta}
\int_{0}^{n/2-\sigma} dx \int_{n/2+\sigma}^{m/2} dy
\int_{n/2-\sigma}^{n/2+ \sigma}
\frac{dz dt}{y+z-x-t} \nonumber \\
& \approx & \frac{8 U^2}{\Delta}
(\sigma^2(\frac{n}{2}-\sigma)(\frac{5}{3}-\ln(\nu))
+8 \sigma^3 (\ln(\frac{2 \sigma}{n})+\frac{7}{3} \ln(2)))
\end{eqnarray}

We first note that
$\Delta^{(\sigma)}$ is a monotonically increasing function of $\sigma$.
The first term on the right-hand side of (\ref{gapsigma})
dominates the low-$\sigma$ behavior. This term is a generalization of the
$nU^2/\Delta$ term giving rise to the spin gap between the $\sigma=0$ and
$\sigma=1$ yrasts. Its $\sigma^2$ parametric dependence results in a
delay of the Stoner instability, equivalently in a reduction of the
strength of the spin-spin exchange coupling 

\begin{equation}
-\lambda U \vec{S} \cdot \vec{S} \rightarrow 
-(\lambda U - A_{\nu} \frac{nU^2}{\Delta}) \vec{S} \cdot \vec{S}
\end{equation}

\noindent where $A_{\nu}$ is a prefactor of order one, weakly depending
on the filling factor $\nu$.

For larger magnetization, i.e. when the polarization ratio
becomes finite (roughly at $\sigma \approx n/8$), $\Delta^{(\sigma)}$
starts to be dominated by the second term in (\ref{gapsigma}) which has
a larger $\sigma^3$ dependence and hence a stronger effect, beyond
the simple shift of the Stoner instability just mentioned : it results in
a saturation of the ground-state magnetization for exchange couplings
not much stronger than the critical Stoner value. Its $\sigma$-dependence 
suggests an higher-order effective spin coupling

\begin{equation}\label{effham}
{\cal H}_S \propto \frac{nU^2}{\Delta} S^3
\end{equation}

\noindent which is switched on roughly at a polarization ratio
$2 \sigma/n > 1/5$ (for which $\ln(2 \sigma/n) + 7/3 \ln(2)$ becomes 
positive). 
The higher $S^3$-dependence of this effective coupling can
also be obtained from a dimensional analysis. The number of second order 
contributions to the ground-state energy in each sector 
decreases as $\sigma^4$ for large enough $\sigma$ (see Appendix B). 
When summing over all of these
contributions, we must take into account their
energy denominator, which 
leads to a $\sim \sigma^3 \ln(\sigma)$ parametric dependence
for the second order contributions, in agreement with (\ref{gapsigma}).
Neglecting the logarithmic correction we finally get (\ref{effham}).
It is important to note 
that this latter effective Hamiltonian term is left invariant by both 
$SU(2)$ rotation in spin space and
rotation in the one-body Hilbert space. 

The above treatment illustrates the {\it average}
magnetization decreasing effect of the interaction fluctuations which results
in a shift of the Stoner threshold to higher exchange strength.
Simultaneously, contributions to
the {\it fluctuations} of the ground-state energy around
this average in a finite-sized system (quantum dot) can be of the same order
of magnitude as the average itself, possibly resulting in large fluctuations
of the ground-state spin around its average value.
We therefore close this section with a calculation of 
the contributions to the variance of the ground-state energy 
arising from the fluctuations of interaction. In first order we get
a contribution given by the variance of diagonal Hamiltonian matrix 
elements

\begin{equation}
\sigma^2(\langle \Psi_0^{(\sigma)}| {\cal U}_f | \Psi_0^{(\sigma)} \rangle)
= O(n^2 U^2)
\end{equation}

\noindent while the variance of the second order is given by
the square of the average contribution (\ref{pertnip0}) and is therefore
of order $O(n^4m^2U^4/\Delta^2)$. Consequently, these
fluctuations are dominated by the second order for $U>\Delta/(mn)$. Of 
physical relevance however are {\it relative} fluctuations between 
ground-states in different spin sectors or at different number of particles.
Both these quantities influence for instance the distribution of conductance
peak spacings for quantum dots in the Coulomb blockade 
regime \cite{marcus,ensslin}. In first order, the relative fluctuations 
between the ground-states with $n$ and $n+1$ particles are given
by 

\begin{equation}
\sigma \left(\sum_{\alpha} U_{\alpha,n+1}^{\alpha,n+1} 
\right) \propto
\sqrt{n} U
\end{equation}

\noindent and the relative fluctuations 
between consecutive (i.e. $\sigma$ and $\sigma+1$) yrasts can be written

\begin{equation}
\sigma \left(\sum_{\alpha} 
(U_{\alpha,n+\sigma+1}^{\alpha,n+\sigma+1}-
U_{\alpha,n-\sigma}^{\alpha,n-\sigma})\right) \propto
\sqrt{n} U
\end{equation}

\noindent where the sums run over occupied
orbitals. Both these last two expressions have the same parametric
dependence on $n$ as they both depend only on the change of one 
orbital occupancy in the immediate vicinity of the Fermi level.
In second order, the relative fluctuations between consecutive yrasts
can be estimated to
have the same order of magnitude as the spin gap 
(\ref{gapevenn})-(\ref{gapoddn}), and 
this also gives the contributions to 
the relative fluctuations between ground-states of consecutive number
of particles, i.e. $O(n U^2/\Delta)$. Consequently, the relative
fluctuations will be dominated by the first (second) 
order for $U<\Delta/\sqrt{n}$ ($U>\Delta/\sqrt{n}$). These estimates
neglect however the spectral fluctuations and are thus valid in 
the case of a rigid equidistant spectrum only. The variance of the gap
distribution is however dominated by these spectral fluctuations 
(which are proportional to the average level spacing $\Delta$)
for both
Wigner-Dyson and Poisson statistics as long as $U < \Delta/\sqrt{n}$.

\section{Asymptotic Regime}\label{tbrim}

  In the regime of dominant fluctuations $U/\Delta \gg 1$,
all energy scales
(Width of the MBDOS, ground-state energy, gaps and splitting between
eigenvalues...) become linear in the fluctuation strength $U$. Most of the
properties of the Hamiltonian can then be obtained by assuming
${\cal H} \approx {\cal U}_{f}$, and for random interaction models
of this form, the shape and width of the MBDOS can be extracted from
a computation of its variance and higher moments. 
We begin this chapter with a short overview of this method mostly
developed in \cite{brody}.

  When ${\cal U}_{f}$ dominates,
${\cal H}_0$ and ${\cal U}_{avg}$ may introduce
a constant shift of the full MBDOS due to the mean
field charge-charge interaction, a shift of
each sector's MBDOS by an amount $-\lambda U \sigma(\sigma+1)$
from the mean-field spin-spin exchange
and a subdominant ($O(\Delta/U)$)
nonhomogeneous modification of the MBDOS due to ${\cal H}_0$ which is
negligible in the limit considered here.
We thus first consider the MBDOS
corresponding to ${\cal U}_{f}$ and will introduce later on
the only relevant mean-field contributions :
the $\sigma$-dependent shifts due to the exchange interaction.
The average shape and width of the
MBDOS of ${\cal U}_f$ can be extracted from its moments

\begin{eqnarray}
{\cal M}^{(j)}(\sigma) & = & \frac{1}{N(\sigma)} \sum_I \left(E_I(\sigma)
\right)^{j}
= \frac{1}{N(\sigma)}{\rm Tr} \left({\cal U}_f\right)^{j}
=  \frac{1}{N(\sigma)} \sum_{I_i} {\cal U}_f^{I_1,I_2}
{\cal U}_f^{I_2,I_3} ... {\cal U}_f^{I_{2j-1},I_1}
\end{eqnarray}

\noindent where $ {\cal U}_f^{I,J} = \langle I| {\cal U}_f |J \rangle $ and
$|I \rangle$ refers to a Slater determinant.
Taking the average of this expression,
we easily see that only the even moments of the average MBDOS
do not vanish. In the last sum furthermore, only terms with pairs
of indices occuring twice
give a non-zero average contribution, i.e. to compute the
moments, one needs to perform contractions over the Hamiltonian
operators such that

\begin{equation}\label{contraction}
(I_k,I_{k+1}) = (I_l,I_{l+1})
\end{equation}

\noindent for a pair of indices $(k,l)$. 
We can readily calculate the second moment, i.e. the variance of the MBDOS

\begin{eqnarray}\label{variance}
{\cal M}^{(2)}(\sigma) & = &\frac{1}{N(\sigma)} {\rm Tr} {\cal H}^{2} (\sigma)
= (\frac{n (n-1)}{2} K_0(\sigma) + (n-1) K_1(\sigma) + K_2(\sigma)) 4 U^2
\end{eqnarray}

\noindent where we have taken care of the number of matrix
elements and different variances of the three classes of Hamiltonian
matrix elements mentioned
in chapter \ref{model}. In the limit of large number of particles,
equation (\ref{variance}) explicitly expresses the dominance of generic
two-body IME : since their number is given by $K_2(\sigma)
\sim n^2 m^2$, their contribution to ${\cal M}^{(2)}(\sigma)$
goes parametrically like $n^2 m^2 U^2$, whereas the contribution
from one-body IME and diagonal matrix elements is
$n^2 m U^2$ and $n^2 U^2$ respectively. This motivates us to
neglect the subdominant contributions to ${\cal M}^{(2)}(\sigma)$ and to
use the approximation

\begin{eqnarray}
{\cal M}^{(2)}\approx K(\sigma) U^2
\end{eqnarray}

A calculation of the connectivity $K(\sigma)$ is given in 
Appendix \ref{appendix2}.

  Higher moments are also easily estimated in the dilute
limit $1 \ll n \ll m$. In this case, the contractions (\ref{contraction})
can be performed
independently as the probability to create (or destroy) the same fermion
on the same orbital is vanishingly small. This remains the case
as long as the number of creation (destruction) operators in ${\cal U}^j$
is smaller than the number of particles, i.e. for
$2 j \ll n $. A second condition
$n \ll m$ must also be satisfied, for which creations and destructions
statistically occur on different orbitals. In this case, higher moments
are simply multiple of the second moment, with a combinatorial factor
reflecting the number of different possible contractions

\begin{eqnarray}
{\cal M}^{(2j)}(\sigma) & = & (2j-1)!!\left({\cal M}^{(2)}(\sigma)\right)^{j}
\end{eqnarray}

  This relation defines a Gaussian MBDOS, and corrections occur
only due to higher moments ($2j > n$), mainly affect the tails of the
distribution, and vanish in the large $n$ limit. It is remarkable that
the order of the moments which fail to behave like those of a Gaussian
distribution depends almost exclusively on the number of particles, at
least as long as one restricts oneself to the lowest magnetization
blocks away from full polarization.
Therefore, corrections affect each partial (i.e. $\sigma$-dependent,
or block-) MBDOS in the same way, and we will assume
that the relative parametric dependence
of the bulk of the MBDOS at different $\sigma$ can be extrapolated
to the tails. This means that, as in the case of a gaussian distribution,
the knowledge of its variance fully determines the MBDOS .
For more details on the shape of the MBDOS for random interaction
models similar to (\ref{hamran}) we refer the reader to \cite{brody} and
the more precise, very recent treatment given in \cite{weiden}.

Based on these previous works \cite{brody} 
establishing the quasi-Gaussian shape of the MBDOS, we can now derive a
simple parametric expression for the average energy difference
between yrast states in each magnetization block. Indeed, the MBDOS
satisfy a scaling law

\begin{eqnarray}
E \rightarrow \bar{E} & = & \frac{E}{\sqrt{K(\sigma)}}
\end{eqnarray}

\noindent which allows to rescale all of them approximately on top of
each other. This behavior is illustrated in Fig. \ref{fig:mbdos}
which shows both the multiple gaussian structure of the MBDOS
and the scaling with
$\sqrt{K(\sigma)}$ obtained from numerical
calculations for $\lambda =0$.

  The yrast states are distributed in the low energy tail
of the partial MBDOS, where the corrections due to higher moments
are the largest. We have nevertheless seen above that these corrections
affect each block's MBDOS in the same way
(this is true only for not too large magnetizations).
Thus the tails undergo the same modification, say for $\sigma=0$ and
$\sigma=1$. If we then make the (a priori not justified) assumption
that the yrast levels are uncorrelated, i.e. that for a given
realization of ${\cal U}_{f}$ their positions around their
respective average value are not correlated, then we can conclude that
the average distance between two yrast states is parametrically given
by the difference of the width of the corresponding MBDOS.
Assuming, as just discussed, that the tails of the
distribution
scale with the variance with a factor $\beta$ and
neglecting contributions arising from $H_0$, the typical
spin gap can be estimated (for $\lambda = 0$) as

\begin{equation}\label{spingap}
\Delta_s^U \approx \beta U[ \sqrt{K(\sigma_{min})} -
\sqrt{K(\sigma_{min} + 1)}]
\end{equation}

  In Fig. \ref{fig:gapinf} we show the computed spin gap
$\Delta_s^{U}$ between the minimally magnetized ground-state and the
first spin excited level for $\lambda=0$
in the limit of dominant interaction, i.e. for ${\cal U}_f$.
One of the main features emerging from the presented numerical data
is a strong even-odd effect which is reminiscent of a similar behavior
in the limit of vanishing interactions.
However the origin here is the fluctuating interaction and the
energy differences scale as $U$ instead of $\Delta$.
As in the perturbative regime discussed in the previous chapter, the
occurence of this even-odd effect is due to the connectivity
$K$ and from Fig. \ref{fig:gapinf} we see
that the probability for a magnetic ground-state is more strongly
reduced for odd
than for even number of particles also in the asymptotic regime.
We next note that the gap first increases with increasing
number of particles before it seems to stabilize above $n=6$. We
have checked (dashed and dotted-dashed lined in Fig. \ref{fig:gapinf})
that this behavior, which is not captured by the dilute
estimate (\ref{spingap}), is partly due to the
neglect in (\ref{variance}) of nongeneric matrix elements with enhanced
variance mentioned above. However,
even though the exact variance gives a much better estimate, it
still underestimates the gap at larger $n$ and we have numerically
determined that this is due to a strong positive correlation of the
ground state energies in adjoining spin blocks which is larger at large
$n$. Qualitatively, these correlations are due to the fact that
the different block hamiltonians are not statistically independent, but 
are constructed out of the same set of
two-body matrix elements. More precisely, for a given realization 
of ${\cal U}_f$, all blocks have
$K(\sigma) N(\sigma) = O(\exp(n),\exp(m))$ nonzero matrix elements
which are constructed out of
the same set of only $O(m^4)$ different two-body interaction matrix
elements. Yrast levels are then due to special realizations of
the latter inside the blocks. These realizations are presumably
not very different in blocks with consecutive magnetization which
results in strong eigenvalues correlations. The above
estimate (\ref{spingap}) which relies only on distribution averages
completely neglects these correlations. 
This is the reason why it underestimates the gap at larger $n$
where they are largest.

  The arguments presented in this section are based on estimates
for the average yrast energy in each sector extracted from the shape
and width of the corresponding MBDOS. We have seen in particular that
the MBDOS in low spin sectors and for a sufficent number of particles
are almost gaussian with a width given by the square root of
the corresponding connectivity (\ref{connec}) $\sqrt{K} \sim n m$.
It follows that the ground-state energy in each sector roughly satisfies
${\cal E}_{0,\sigma} \sim n m U$ in the asymptotic regime, whereas
in the perturbative regime we found
${\cal E}_{0,\sigma} \sim n^2 m U^2 \ln(m)/\Delta$ (see 
Equation (\ref{pertnip0})).
Neglecting logarithmic corrections (which arise due to the denominators
in the second order of perturbation theory) we arrive at the critical
border between perturbative and asymptotic regime (radius of convergence
of the perturbation theory)

\begin{eqnarray}\label{bord}
U_c \sim \Delta/n
\end{eqnarray}

\noindent Equation (\ref{bord}) indicates the
breakdown of perturbation theory at a much smaller strength of the
fluctuations of interaction than previously expected. This is due to the
coherent addition of many small second order contributions for the
perturbation expansion in the
immediate vicinity of the ground-state. A more detailed
study of this breakdown has been presented in reference \cite{jack}.

\section{Spin polarization threshold : discrepancies
from Stoner's scenario}\label{six}

  Having established the demagnetizing effect of the off-diagonal
fluctuations both in the perturbative and asymptotic regimes at $\lambda=0$,
we now switch on the mean-field spin-spin interaction $\lambda >0$.
The competition between one-body energy, exchange interaction
and off-diagonal fluctuations will determine both
the average threshold at which the ground-state starts to
be polarized and the probability of finding a magnetized ground-state
at a given set of parameters $(\lambda U/\Delta,U/\Delta)$. The theory
presented in the previous chapters focused essentially on the first aspect
and we already know that the average threshold for magnetization is
increased by non-zero interaction fluctuations. The exchange
induces energy shifts of $-\lambda U\sigma(\sigma+1)$ of each sector's
MBDOS but has no effect whatsoever on the width of the MBDOS.
Considering first the asymptotic regime,
the average spin gap becomes $\Delta_s = \Delta_s^{U} -
\bar{\lambda}U$, where $\bar{\lambda}=(5-(-1)^n)\lambda/2$. In particular,
the relative shift between the two lowest magnetized blocks
is larger for odd number of particles, as is the spin gap (See
Fig. \ref{fig:gapinf}). From (\ref{spingap})
the average threshold becomes parametrically

\begin{equation}
\lambda_c \sim \sqrt{K(\sigma_{min})} - \sqrt{K(\sigma_{min} + 1)}
\end{equation}

  From Fig. \ref{fig:gapinf}
$\lambda_c \approx 2.5$ (3.5) for even (odd) $n$.
Note that as both the spin gap and the exchange are
linear in $U$ in the asymptotic
regime, this average threshold is $U$-independent.
This is no longer the case in the perturbative regime.
As shown in section \ref{pert}, the perturbative spin gap
can be approximated by
$\Delta^s(U) -\Delta \sim B n U^2/\Delta$,
where we recall that $B=1$ (1.5) for even (odd) $n$.
We then get

\begin{equation}\label{lambdac}
\lambda_c(U)-\langle U_{\alpha,\beta}^{\beta,\alpha} \rangle_0 
\sim B n U/\Delta
\end{equation}

\noindent where we used the critical (Stoner) exchange strength
$\langle U_{\alpha,\beta}^{\beta,\alpha} \rangle_0 \equiv \Delta/2$.
Once this threshold is reached, the spins start to align, but in contrast
to the Stoner scenario, full polarization is not achieved at once,
because of a parametric decrease of the second-order contributions
from off-diagonal
fluctuations as $\sigma$ is increased. From the perturbative
treatment presented in section \ref{pert} a $S^3$ term takes over at
large spin which induces saturation of the ground-state spin. The mechanism
for the appearance of that term is a reduction of the
probability for transitions from or onto partially occupied orbitals
with respect to transitions from doubly occupied orbitals onto
empty orbitals. Off-diagonal fluctuations result in two
effective Hamiltonian terms $\sim \vec{S} \cdot \vec{S}$ and
$\sim S^3$ and the second term influences the
system's magnetization properties at large spin, but before full polarization.
Neglecting logarithmic corrections in $n$, $m$ and $\sigma$
and for a given $\lambda = \lambda_c(U)+\delta\lambda$ (i.e. $\lambda$
measures the distance to the Stoner threshold), $\Delta$ and $U$, the
magnetization will saturate at a value

\begin{equation}\label{sigmax}
\sigma_{max} \approx \delta\lambda \frac{\Delta}{U}
\end{equation}

  This is a major modification of the Stoner scenario for which
once the magnetization threshold is reached,
full polarization of the electrons is achieved at once.
The presence of off-diagonal
fluctuations, no matter how weak, induces this saturation, as their
relative weakness will eventually be counterbalanced by the larger
parametric dependence in $\sigma$ of the number of second order contributions
at large $\sigma$. We stress that this saturation is entirely induced by
the off-diagonal fluctuations and does not depend on any modification of the
one-body density of states at larger spin.

  We next show on Fig. \ref{fig:gapxover} the behavior of the spin gap
between the two lowest yrasts
as a function of $U/\Delta$ and for different values of $\lambda$.
The variance of the gap distribution is of course unaffected by the exchange
and we already know that the
probability $P(\sigma>0)$ \cite{remark2} of finding a magnetized ground-state
is reduced by the off-diagonal matrix elements. This probability
will eventually saturate above a finite value of $U/\Delta$,
since the width of the gap distribution
is proportional to its average $\sim U/\Delta$ \cite{remark}.
This is shown on Fig. \ref{fig:gapxover} where the error bars reflect
the width of the gap distribution. Their linear increase with $U$ means that
the fraction of negative ``gaps'' (contributing to the
probability of being magnetized) is constant with $U$.
The same behavior is characteristic of gaps between higher consecutive yrast,
which results in a $U$-independent behavior of $P(\sigma>0)$ at large $U$.

  Finally, $P(\sigma>0)$ is shown on
Fig. \ref{fig:probpf} as a function
of the exchange strength $\lambda U/\Delta$ for different values of $U/\Delta$
and different distributions of one-particle orbitals. This figure
shows a clear demagnetizing effect of the fluctuations of interaction
except below
the Stoner threshold in the case of an equidistant spectrum. We recall that
the demagnetizing effect is in fact only an average effect, and that for
an equidistant spectrum, $U$ may for particular realizations
reduce the level density at the Fermi level, thereby favoring the
appearance of a higher spin ground-state as can be seen on
Fig. \ref{fig:probpf} for small interaction fluctuations $U/\Delta = 0.1$ and
small exchange strength $\lambda U/\Delta < 0.5$.
In the two other cases of a randomly
distributed and Wigner-Dyson one-body spectrum, fluctuations of interaction
always reduce $P(\sigma>0)$. At larger $U$, the dependence
on the orbital distribution is rather weak,
as shown in Fig. \ref{fig:probcomp}. Note in Fig. \ref{fig:probpf}
the bending of $P(\sigma>0)$ 
above the onset
of magnetization which is a clear difference from the Stoner behavior : even
at quite large exchange, $P(\sigma>0)$ remains smaller than one.
 From these data, we define an average
magnetization threshold $\lambda_c(U)$ for which
$P(\sigma>0)=0.01$ and extract from Fig. \ref{fig:probpf} 
the additional exchange strength $\delta \lambda$ necessary to
achieve $P(\sigma>0)=0.5$. The results are shown on Fig. \ref{fig:dlambda}
and indicate a linear increase of $\delta \lambda$ with $U$ which illustrate
the demagnetizing effect of the off-diagonal fluctuations :
a stronger exchange than predicted by a simple Stoner picture is
necessary to have even a weak
nonzero ground-state magnetization {\it probability} (see
Fig. \ref{fig:probpf}), moreover
an even stronger one is necessary to achieve a significant probability.
All this is in qualitative agreement with equation (\ref{sigmax}). A direct
numerical check of this equation would however require a much larger
number of particles, beyond today's numerical capabilities.

\section{Real Space Models}\label{realsp}

  It is now evident from the results presented above that
fluctuations of IME introduce a new energy scale.
In addition
to the Stoner parameter $\lambda U/\Delta$, the ratio $\lambda$
between the exchange and the interaction fluctuations gives a second
relevant parameter for the emergence of a ferromagnetic phase.
We therefore turn our attention to the microscopic computation of
the magnetization parameter $\lambda$ for standard solid-state models.
This will allow us to estimate the strength of the demagnetizing effect
of off-diagonal fluctuations in more realistic situations.
We consider Anderson lattices whose one-body Hamiltonian
is given by

\begin{eqnarray}
H & = & V \sum_{\langle i;j \rangle} c^\dagger_{i,s} c_{j,s}
+ \sum_i W_i c^\dagger_{i,s} c_{i,s}
\end{eqnarray}

\noindent Here $\langle i;j \rangle$ restricts the sum to nearest neighbors,
and $W_i \in [-W/2;W/2]$ where $W$ is the disorder strength.
We study interaction potentials of the form

\begin{equation}
{\cal U}(i-j) = {\cal U}_0 \delta(i-j) +
{\cal U}_1/\left| \vec{r_i}-\vec{r_j} \right|
\end{equation}

\noindent i.e. for ${\cal U}_1=0$ we have a pure Hubbard interaction whereas
${\cal U}_1 \ne 0$ gives a long-range interaction.
Microscopically, $\lambda$ is given by the ratio of the average exchange term

\begin{equation}\label{exchange}
\langle U_{\alpha,\beta}^{\beta,\alpha} \rangle = \sum_{i,j}
{\cal U}(i-j) \overline{\psi_{\alpha}^*(i)  \psi_{\beta}^*(j)
   \psi_{\alpha}(j)  \psi_{\beta}(i)}
\end{equation}

\noindent and the r.m.s. of the distribution of IME (\ref{ime}).
By definition the average in equation (\ref{exchange}) is performed
over wavefunctions close to the Fermi level.
Fig. \ref{fig:lambda2d} and \ref{fig:lambda3d} show the disorder
dependence of $\lambda$,
for a pure Hubbard interaction ${\cal U}_1=0$ on two- and three-dimensional
lattices respectively and for different linear system sizes.
The data have been obtained
from averages over 30 wavefunctions in the middle of the Anderson band $E =0$
and for 10 ($L=80$ in 2D and $L=15$ in 3D) to 200
($L=10$ in 2D and $L=6$ in 3D) disorder realizations.
In both dimensionalities we can distinguish three
regimes : ($I$) At low disorder, the one-electron
dynamics undergoes a crossover from ballistic to diffusive regime
as the linear system size is increased beyond the elastic mean-free path
$l_e \sim (V/W)^2$. In the ballistic regime $l_e \gg L$,
wavefunctions are plane-waves. In this case, a Hubbard interaction
gives $\lambda \sim L^2$, since the
${\rm RMS} \left[U_{\alpha,\beta}^{\gamma,\delta}\right] \sim L^{-4}$ and
$\langle
U_{\alpha,\beta}^{\beta,\alpha} \rangle \sim L^{-2}$ \cite{remark3}, whereas
once the diffusive regime is reached, one expects
$\lambda \sim \Delta/(\Delta/g) \sim g$ \cite{blanter}.
In the crossover
between these two regimes, contributions from gaussian modes (those
corresponding to $|i-j| < l_e$ in (\ref{sigmaU})) may dominate
the fluctuations of the IME but eventually vanish as one increases $L$
as they are weighted by a factor $(l_e/L)^D$ \cite{blanter}.
Presumably these contributions still affect our data in region ($I$).
($II$) In the regime of intermediate disorder, both
off-diagonal fluctuations and exchange are increased by disorder, and
apparently they
compensate each other, resulting in a $L$-independent $\lambda \approx 4$,
in 2D. We expect that this behavior will hold as one
further increase the system size. We indeed numerically estimated
the elastic mean free path at $W/V=5$
from the distribution of inverse participation ratio \cite{prigodin}
and found a value $l_e \approx 4$. The gaussian modes are thus weighted by a
prefactor $1/400$ for $L=80$ and have therefore only a marginal influence on
the fluctuations of the IME, so that one may reasonably assume that finite-size
effects have only a marginal influence on the data presented in
Fig. \ref{fig:lambda2d}
in region ($II$). In the three-dimensional case, it even seems that
$\lambda$ decreases as the system size increases in the intermediate regime
$W/V \in [8,17]$, however this is due to the quite small linear
system sizes considered here, and once one reaches $L \gg l_e$, $\lambda$
should saturate at a finite, but quite small value.
It is interesting to note that the upper border
of this intermediate regime is quite close to the critical disorder
value for the Anderson localization transition.
($III$) In the regime of strong
disorder, one-particle wavefunctions are strongly localized on fewer
and fewer sites, the off-diagonal fluctuations are
sharply reduced (due to quasi selection rules discussed in
chapter \ref{model}) and again exchange dominates. Note that eventually, the
latter disappears also, but at a lower rate than the fluctuations.
These results indicate that at an intermediate disorder strength, off-diagonal
fluctuations may be strong enough to
play an important role for the magnetization properties of the ground-state.

  We next evaluate the influence of the long-range part
of the interaction. The average exchange interaction (\ref{exchange})
term is given by an average taken over one-particle
wavefunctions close to the Fermi level.
Due to their orthogonality,
taking this average over the full set
of wavefunctions gives a $\delta$-function and only on-site
contributions. This averaging procedure is however only justified
if the one-body dynamics
is described by Random Matrix Theory (RMT) for which the structure of the
eigenstates is homogeneous all through the spectrum. RMT however describes
systems which are of interest here only inside an energy window given
by the Thouless energy around the Fermi level \cite{altshklo} so that
the average over wavefunctions close to the Fermi level leads only to a
more or less sharply peaked function of $(\vec{r_i}-\vec{r_j})$.
There are also contributions to the exchange from
the long-range terms, but still we expect that the average damps
them with respect to their contribution to off-diagonal fluctuations
(This damping of course depends on the disorder strength.) which are
of the same order of magnitude as the short-range contribution up to
distances of the order of $l_e$ \cite{mirlin}. This means that we expect
a decrease of $\lambda$ upon increase of the interaction range. The
validity of this reasoning is illustrated for the two-dimensional
case on Fig. \ref{fig:screening} where we plot the evolution of
$\lambda$ for different disorders as the long-range part of the interaction
becomes more and more important. Clearly, $\lambda$ decreases as the range
of the electron-electron interaction increases, and therefore the Hubbard
results presented on Fig. \ref{fig:lambda2d} and \ref{fig:lambda3d}
give an upper bound for $\lambda$.
One thus expects the demagnetizing effect described in this paper
to be more efficient at low filling when the screening length
exceeds the elastic mean free path.

  In finite-sized systems like quantum dots where impurity
scattering is weak but wavefunction fluctuations are induced by
chaotic scattering at an irregular confining potential, standard
estimates give $\lambda \approx g$ for a short-range interaction,
whereas in the (unphysical) limit of an infinite range interaction
${\cal U}(\vec{r}-\vec{r}') = {\cal U}$ one gets $\lambda=1$ \cite{remark4}.
Therefore, and as
$g$ is not too large in such systems, it is a priori not justified
to neglect the effect of off-diagonal fluctuations, as they should at least
strongly suppress the probability of finding ground states of larger
spin beyond few ($\approx 2,3$) polarized electrons.
It has even been proposed by Blanter, Mirlin and Muzykantskii that
in confined systems the accumulation of charge at the surface of confinement
leads to stronger fluctuations of screened Coulomb interaction matrix
elements $\sim \Delta/\sqrt{g}$ which would give
$\lambda \sim \sqrt{g}$. As in quantum dots
$g$ is of the order of up to few tens, this would bring $\lambda$ down
to values where the demagnetizing effect of off-diagonal fluctuations
plays an important role. All this illustrates the relevance of
off-diagonal fluctuations for the magnetization properties of
the ground-state in regimes of intermediate disorder and for poor screening
of the electronic interactions - presumably, for low electronic densities
$\rho$ for which the distance between electron is smaller than the elastic
mean free path $\rho^{1/D} < l_e$.

Assuming still $U \sim \Delta/g$, the shift of the Stoner threshold 
is quite small,
of the order $O(\Delta/g)$. This is so, as the model we consider is valid
only in an energy window of the order of the Thouless
energy $E_c = g \Delta$ around the Fermi level, so that it is quite
natural to set $n,m \approx g$.
At larger magnetization however, the second term in (\ref{gapsigma})
takes over and induces a significant reduction of the ground-state
spin when the latter becomes comparable to $g$ with a prefactor depending on
the strength of the average exchange. This term strongly modifies
the Stoner scenario as it induces magnetization saturation above
the magnetization threshold and full polarization can be achieved only
once a second, significantly larger, threshold is reached.

We finally describe an experimental setup that allows to gain important
information on the ground-state spin of two-dimensional lateral
quantum dots in the Coulomb blockade regime. The
experiment was proposed in \cite{spings1} and consists
in applying an external 
magnetic field in the plane of a lateral, two-dimensional 
quantum dot. Because of the two-dimensional nature of the dynamics,
we assume that an in-plane field has no orbital effect
so that it introduces only a Zeemann coupling \cite{stern}.
The difference in ground-state spins for consecutive number of
electrons can then be determined experimentally by studying the motion
of Coulomb blockade conductance peaks at very low
temperature
$T \ll \Delta$ as the in-plane magnetic field is increased. 
The resonant gate voltage is given by a difference of two
many-body ground-state energies
$e V_g^n = {\cal E}^0_{n+1} - {\cal E}^0_{n}$, 
and it is always the difference of an even-odd pair (i.e. of ground state
energies corresponding to one even and one odd number of electrons on the 
dot). Upon application of an in-plane field, the peak position behaves like

\begin{eqnarray}
e V_g^n(B) & = &
{\cal E}^0_{n+1} - {\cal E}^0_{n} + g \mu_B B \delta \sigma_z(n)
\end{eqnarray}

\noindent $\delta \sigma_z(n)$ is the magnetization difference
between the two consecutive ground-states which can therefore be
extracted experimentally from the motion of conductance peaks in
an in-plane field. At minimal magnetization one has a sequence of
ground state spins $\sigma_z=0,$ 1/2, 0, 1/2, 0, 1/2... (For odd $n$,
and due to SRS,
the $\sigma_z=1/2$ and $\sigma_z=-1/2$ ground-states are degenerate so that
an arbitrarily weak in-plane field aligns the spins and
one always has $\sigma_z=1/2$), therefore
$\delta \sigma_z(n)= (-1)^n/2$ and one has
$|\partial V_g/\partial B| = g \mu_B/2$. As $B$ is increased the
ground-state
with even number of electrons is most likely to
magnetize first (this is because of both the even-odd effect mentioned 
in section \ref{pert} and the larger kinetic energy cost to flip
one spin for odd number of electrons), 
exactly reversing the slope of two consecutive peaks;
then as the field increases further the odd state will likely flip,
restoring the original slope.  As long as consecutive ground states
never differ by more than one unit of spin the absolute
value of the slope will remain constant as the system polarizes.
Correspondingly, if all slopes are constant 
$|\partial V_g/\partial B| = g \mu_B/2$, it is very likely that
no ground-state is magnetized (this would for instance require a sequence 
$\sigma_z=1,$ 1/2, 1, 1/2, 1, 1/2... We do not see any obvious
reason why {\it all} even $n$ ground-states should have $\sigma_z=1$ while
at the same time none of the odd $n$ ground-states are magnetized).

However if there exist many magnetized ground states, then the
probability to find pairs of consecutive ground-states with a larger 
difference in magnetization $|\delta \sigma_z(n)|> 1$ increases and
one expects a range of slopes to occur.
Then the corresponding peak heights may be strongly reduced
by the spin blockade mechanism \cite{weinmann},
which should be easily visible experimentally.
This argument neglects changes in the $g$-factor of the
electron with
changing $n$, which presumably are slow.
This is illustrated on 
Fig. \ref{fig:spindot} where
the peak positions are drawn as a function of the
Zeemann coupling for $\lambda=1$ and 5. It is clearly seen that at
weak $\lambda$, $|\partial V_g/\partial B|$ is constant and
corresponds to a minimal $\delta \sigma_z$,
while a larger $\lambda$ gives different slopes in agreement with
the above reasoning. Note also in this latter case,
that peaks evolve in parallel at weak
magnetic field, indicating the successive addition of two spins oriented
in the same direction. This feature is absent of the weak-exchange 
(right-hand) side of the graph for which the ground-states are
obtained by piling up electrons on the orbitals according to the Pauli
prescription. This results in a minimal ground-state spin, a sequence
$\delta \sigma_z(n)= (-1)^n/2$ of magnetization differences between consecutive
ground-states and a motion of neighboring peaks in opposite direction at low
field.

Recently, experiments in this direction have indeed been performed,
which have given serious evidences for the occurence of partially (but weakly)
magnetized ground-states with few polarized 
electrons \cite{marcus2,spindotexp}.
This means that the off-diagonal fluctuations are not dominant, in agreement
with the above considerations giving a large ratio $\lambda \approx g$ 
between the strength of the average exchange and the off-diagonal 
fluctuations. Therefore the perturbative treatment presented in section
\ref{pert} is expected to correctly describe 
semiconductor quantum dots either in the diffusive regime (large dots)
or in the ballistic regime with chaotic boundary scattering (smaller dots).
We note in this respect that from recent experiments on the distribution
of conductance peak spacings, the parameters $\lambda$ and $U$ have been 
tentatively extracted and seem to indicate a conductance $g \approx 6$
for which off-diagonal fluctuations should give a non-negligible 
contribution \cite{ensslin}.
Finally, we note another interesting experimental result which is the apparent
absence of suppression of the conductance peak found in some cases
for larger spin
difference between consecutive ground-states \cite{spindotexp}. This is
in major disagreement with the spin-blockade mechanism proposed by
Weinmann, Ha\"usler and Kramer \cite{weinmann} and is yet to be understood.

\section{Conclusions}\label{concl}

In this article we have illustrated how fluctuations of the interaction
matrix elements tend to reduce the ground-state magnetization, both when
they can be treated perturbatively (the regime which is relevant
for condensed matter physics) and in the asymptotic regime where they
give the dominant terms in the Hamiltonian (which is relevant for nuclear
physics). The mechanism behind this effect is in a way similar to the
Stoner picture where itinerant ferromagnetism occurs due to a larger
number of diagonal interactions at low magnetization. As in a mean-field
or self-consistent approach, each of these terms
gives a positive contribution (for a repulsive interaction),
this directly favors spin polarization.
Similarly, we have shown that interactions induce more transitions
in the low spin sectors. Each of these transitions gives one contribution
in second order perturbation theory which this time is however negative
({\it both for attractive and repulsive interactions}) if one considers the
lowest level in each sector, and this therefore favors a low spin
ground-state. 
In the perturbative regime, we have seen that these fluctuations
induce two terms in an effective Hamiltonian formalism : a
$\vec{S} \cdot \vec{S}$ term which simply induces a small shift of the
Stoner threshold and a
second $|\vec{S}|^3$-term which is switched on at larger magnetization
where it
results in a saturation of $\sigma$. This is a major qualitative
modification of the Stoner scenario : even neglecting discrepancies
in the one-body density of states, full polarization is not achieved once
ground-state magnetization has been triggered by the exchange interaction.
The latter must indeed also overcome the $\sim |\vec{S}|^3$ term,
which requires an even larger exchange. The strongest effect of off-diagonal
fluctuations occurs in the large spin regime $\sigma = O(g)$
where a mean-field picture overestimates the value of the ground-state spin.

 From the point of view of nuclear physics, our analysis of the regime
of large fluctuations, based on a study of the many-body density
of states, clearly indicated a strong bias toward low angular momentum
ground-state. We have not explained however why numerical results indicate
an almost 100\% predominance of $\sigma=0$ ground-state for models similar
to the one we have studied \cite{spings1,nuphys},
and this question is still open.

Our findings should finally
be put in perspective with the Renormalization
Group (RG) treatment for disordered interacting electronic systems of
Finkelstein \cite{finkel}. In his approach, one indeed finds that the RG flow
renormalizes the ferromagnetic spin-spin coupling to larger and larger
values, possibly indicating the occurence of a ferromagnetic phase due to
the combined effect of disorder and interaction. The perturbative
treatment we presented in section \ref{pert} did not allow us to find
any contribution favoring a higher spin and this apparent disagreement
between the RG approaches and ours is at present not
understood. We note however that it has been suggested that
the divergence of the exchange coupling induced by the RG flow could
indicate a crossover to the singlet-only universality
class \cite{finkel,castellani}. In this respect it is worth noticing that the
scattering processes in the singlet and triplet channels as defined in
the present work have coupling constants corresponding
to the sum and the difference of the couplings
$\Gamma$ and $\Gamma_2$ as defined in \cite{finkel} respectively. It can be
checked that the
ratio $\sigma(W)/\sigma(V)=(\Gamma+\Gamma_2)/(\Gamma-\Gamma_2)$ 
satisfies the same RG equation as
the exchange coupling ($\gamma_2 \equiv \Gamma_2/z$ in \cite{castellani})
so that 
the triplet channel vanishes at the same rate as the (ferromagnetic)
exchange flows to strong coupling, which may indicate a cancellation of the
ferromagnetic instability by the effect studied in the present paper.

\acknowledgments

It is our pleasure to acknowledge interesting discussions with I. Aleiner,
D. Goldhaber-Gordon, B. Narozhny and S. Sachdev. This work has been 
supported by the Swiss National Science Foundation
and the NSF grant No PHY9612200. Numerical simulations were performed at the
Centro Svizzero di Calcolo Scientifico in Manno, Switzerland.


\appendix
\section{}\label{appendix1}

Under a rotation in spin space, the triplet operators in
equation (\ref{triplet0}) are brought into one another, whereas the
singlet operators (\ref{singlet}) are left invariant. 
SRS on the other hand implies a number of interaction-induced 
two-body transitions which is invariant under such a rotation.
SRS can be easily checked for initial states without double occupancy,
and it is equally easy to convince oneself that the singlet operators
(\ref{singlet}) is spin conserving.
For initial states with double occupancies however,
the triplet operators (first three terms between brackets
in (\ref{usrs})) are not individually SRS but must be considered
as one single spin conserving triplet operator. 
To check this one acts on a four-particle state with two double 
occupancies (which has thus $\sigma=\sigma_z=0$) 
with the full triplet operator of (\ref{usrs})

\begin{eqnarray}\label{tripamplitude}
& & \left[ 
c^{\dagger}_{\alpha,\uparrow} c^{\dagger}_{\beta,\uparrow} c_{\gamma,\uparrow}
c_{\delta,\uparrow} + 
c^{\dagger}_{\alpha,\downarrow} c^{\dagger}_{\beta,\downarrow} 
c_{\gamma,\downarrow} c_{\delta,\downarrow} + 
\frac{1}{2}
(c^{\dagger}_{\alpha,\uparrow} c^{\dagger}_{\beta,\downarrow}+
c^{\dagger}_{\alpha,\downarrow} c^{\dagger}_{\beta,\uparrow})
(c_{\gamma,\downarrow} c_{\delta,\uparrow} +
c_{\gamma,\uparrow} c_{\delta,\downarrow}) \right]
c^{\dagger}_{\gamma,\downarrow} c^{\dagger}_{\gamma,\uparrow}
c^{\dagger}_{\delta,\downarrow} c^{\dagger}_{\delta,\uparrow} |0 \rangle 
\nonumber \\
& = & \left[(c^{\dagger}_{\alpha,\uparrow} c^{\dagger}_{\delta,\downarrow}
- c^{\dagger}_{\alpha,\downarrow} c^{\dagger}_{\delta,\uparrow})
(c^{\dagger}_{\beta,\uparrow} c^{\dagger}_{\gamma,\downarrow}
- c^{\dagger}_{\beta,\downarrow} c^{\dagger}_{\gamma,\uparrow}) +
\frac{1}{2} (c^{\dagger}_{\alpha,\uparrow} c^{\dagger}_{\beta,\downarrow}
- c^{\dagger}_{\alpha,\downarrow} c^{\dagger}_{\beta,\uparrow})
(c^{\dagger}_{\delta,\uparrow} c^{\dagger}_{\gamma,\downarrow}
- c^{\dagger}_{\delta,\downarrow} c^{\dagger}_{\gamma,\uparrow})
 \right] |0 \rangle \equiv 2 |\Psi \rangle+|\Psi' \rangle 
\end{eqnarray}

\noindent This is the sum of two products of two singlets, and it is 
obviously spin conserving. Moreover, acting on the same initial state with
a singlet interaction operator gives

\begin{eqnarray}\label{a2}
& & \frac{1}{2}
(c^{\dagger}_{\alpha,\uparrow} c^{\dagger}_{\beta,\downarrow}-
c^{\dagger}_{\alpha,\downarrow} c^{\dagger}_{\beta,\uparrow})
(c_{\gamma,\downarrow} c_{\delta,\uparrow} -
c_{\gamma,\uparrow} c_{\delta,\downarrow})
c^{\dagger}_{\gamma,\downarrow} c^{\dagger}_{\gamma,\uparrow}
c^{\dagger}_{\delta,\downarrow} c^{\dagger}_{\delta,\uparrow} 
|0 \rangle \nonumber \\
& = & \frac{1}{2} (c^{\dagger}_{\alpha,\uparrow} c^{\dagger}_{\beta,\downarrow}
- c^{\dagger}_{\alpha,\downarrow} c^{\dagger}_{\beta,\uparrow})
(c^{\dagger}_{\delta,\uparrow} c^{\dagger}_{\gamma,\downarrow}
- c^{\dagger}_{\delta,\downarrow} c^{\dagger}_{\gamma,\uparrow}) |0 \rangle
\end{eqnarray}

\noindent The two final states in (\ref{tripamplitude})

\begin{eqnarray}
|\Psi \rangle & = & \frac{1}{2}(c^{\dagger}_{\alpha,\uparrow} c^{\dagger}_{\delta,\downarrow}
- c^{\dagger}_{\alpha,\downarrow} c^{\dagger}_{\delta,\uparrow})
(c^{\dagger}_{\beta,\uparrow} c^{\dagger}_{\gamma,\downarrow}
- c^{\dagger}_{\beta,\downarrow} c^{\dagger}_{\gamma,\uparrow})|0 \rangle
\nonumber \\
|\Psi' \rangle & = &\frac{1}{2} (c^{\dagger}_{\alpha,\uparrow} 
c^{\dagger}_{\beta,\downarrow}
- c^{\dagger}_{\alpha,\downarrow} c^{\dagger}_{\beta,\uparrow})
(c^{\dagger}_{\delta,\uparrow} c^{\dagger}_{\gamma,\downarrow}
- c^{\dagger}_{\delta,\downarrow} c^{\dagger}_{\gamma,\uparrow})|0 \rangle
\end{eqnarray}

\noindent are not orthogonal to each other and it can easily be checked that 
the sum $2 |\Psi \rangle+|\Psi' \rangle=\sqrt{3} |\Phi \rangle$ is a
singlet. $|\Phi \rangle$ is normalized and the factor $\sqrt{3}$ appears
in the case of an initial state with double occupancies
for which the number of triplet transitions
is reduced by a factor three (without double occupancies, the three triplet
operators in (\ref{usrs}) are individually SRS). Thus the total transition
{\it probability} is kept constant. Finally, $|\Phi \rangle$ is orthogonal 
to $|\Psi' \rangle$, i.e. singlet and triplet
channels give transitions onto orthogonal
states. Their contribution to second order perturbation theory will 
therefore add incoherently and give a transition probability

\begin{eqnarray}\label{res22}
3 \overline{(V_{\alpha,\beta}^{\gamma,\delta})^2}+ 
\overline{(W_{\alpha,\beta}^{\gamma,\delta})^2} = 16 U^2
\end{eqnarray}

\noindent A calculation going along similar lines shows that if one of the 
final orbitals (e.g. $\alpha$ or $\beta$) is partially occupied,
the transition probability is reduced by a factor $1/2$.

We also calculate the transition probability for one singly, one
doubly occupied initial orbitals. The initial state to consider is

\begin{eqnarray}
|\Psi_{\pm} \rangle & = & \frac{1}{\sqrt{2}}
        (c^{\dagger}_{\alpha,\uparrow} c^{\dagger}_{\beta,\downarrow}
\pm c^{\dagger}_{\alpha,\downarrow} c^{\dagger}_{\beta,\uparrow})
c^{\dagger}_{\delta,\downarrow} c^{\dagger}_{\delta,\uparrow}|0 \rangle
\end{eqnarray}

\noindent The label $\pm$ refers to either a triplet or a singlet 
$\sigma_z=0$ two-particle
state on the orbitals $\alpha$ and $\beta$. The transition amplitudes
can be calculated in the same way for both cases, and we restrict ourselves
below to the triplet case with $|\Psi_{+} \rangle$.
Note that this latter state can be brought via a rotation in spin space onto 
the following $\sigma_z=1$ state

\begin{eqnarray}
|\Psi_{z} \rangle & = & 
c^{\dagger}_{\alpha,\uparrow} c^{\dagger}_{\beta,\uparrow}
c^{\dagger}_{\delta,\downarrow} c^{\dagger}_{\delta,\uparrow}|0 \rangle
\end{eqnarray}

\noindent and that the calculations to be presented below give the same 
transition amplitude
for both $|\Psi_{+} \rangle$ and $|\Psi_{z} \rangle$ and are thus fully
SRS. Acting on $|\Psi_{+} \rangle$ with a singlet interaction operator gives

\begin{eqnarray}\label{a4}
& & \frac{1}{2}
(c^{\dagger}_{\gamma,\uparrow} c^{\dagger}_{\mu,\downarrow}-
c^{\dagger}_{\gamma,\downarrow} c^{\dagger}_{\mu,\uparrow})
(c_{\beta,\downarrow} c_{\delta,\uparrow} -
c_{\beta,\uparrow} c_{\delta,\downarrow}) |\Psi_{+} \rangle \nonumber \\
& = & \frac{1}{2 \sqrt{2}}
(c^{\dagger}_{\gamma,\uparrow} c^{\dagger}_{\mu,\downarrow}-
c^{\dagger}_{\gamma,\downarrow} c^{\dagger}_{\mu,\uparrow})
(c^{\dagger}_{\alpha,\uparrow} c^{\dagger}_{\delta,\downarrow}
+c^{\dagger}_{\alpha,\downarrow} c^{\dagger}_{\delta,\uparrow})|0 \rangle
\end{eqnarray}

\noindent which once again is SRS. Two remarks are in order here. First,
the above transition amplitude has picked up a factor $1/\sqrt{2}$
with respect to the case where the initial state has two double occupancies.
This is due to the vanishing of one transition ``channel'', as the
$\beta$ orbital is only singly occupied, and will result in a factor
$1/2$ for the transition amplitude. Note that this factor is counterbalanced
by a twice larger number of transitions for the case considered here, as
one has the freedom to 
destroy (or create) a particle on the $\alpha^{\rm th}$ or
the $\beta^{\rm th}$ orbital. Secondly, doing the same calculation
with a triplet operator acting on the singlet initial state 
$|\Psi_{+} \rangle$ is not SRS per se, but once again requires to 
consider the $\sigma_z=\pm 1$ triplet operators, as we did above for the
case of two double occupancies.

Only in the situation where both initial and final states correspond 
to partially occupied orbitals does one get an uncompensated reduction
of the transition amplitude with respect to the above
case of doubly occupied initial and empty final orbitals. As we are now
going to show, this results from the coherent addition of the triplet
and singlet transitions which lead to the same final state. The initial
state is e.g.

\begin{eqnarray}
|\Psi_{in} \rangle & = & \frac{1}{\sqrt{2}}
(c^{\dagger}_{\mu,\uparrow} c^{\dagger}_{\beta,\downarrow} +
c^{\dagger}_{\mu,\downarrow} c^{\dagger}_{\beta,\uparrow})
c^{\dagger}_{\delta,\uparrow} c^{\dagger}_{\delta,\downarrow} | 0 \rangle
\end{eqnarray}

\noindent and one acts on it with the operator

\begin{eqnarray}\label{op}
O_{\pm} & = & \frac{1}{2}
(c^{\dagger}_{\gamma,\uparrow} c^{\dagger}_{\mu,\downarrow} \pm
c^{\dagger}_{\gamma,\downarrow} c^{\dagger}_{\mu,\uparrow})
(c_{\beta,\downarrow} c_{\delta,\uparrow} \pm
c_{\beta,\uparrow} c_{\delta,\downarrow})
\end{eqnarray}

\noindent A straightforward calculation gives the same result {\it for both
singlet and triplet operator}

\begin{eqnarray}\label{opst}
O_{\pm} |\Psi_{in} \rangle & = & \frac{1}{2\sqrt{2}} 
(c^{\dagger}_{\gamma,\uparrow} c^{\dagger}_{\delta,\downarrow} +
c^{\dagger}_{\gamma,\downarrow} c^{\dagger}_{\delta,\uparrow})
c^{\dagger}_{\mu,\uparrow} c^{\dagger}_{\mu,\downarrow} | 0 \rangle
\end{eqnarray}

\noindent The result is SRS, i.e. gives a four-particle 
final state with $\sigma=1$, $\sigma_z=0$. The key point here is that
both singlet and triplet channels go to the same final state. Thus one gets
the corresponding second order transition probability by adding their
amplitude coherently. The average transition probability reads then

\begin{eqnarray}\label{tramplitude}
\frac{1}{4} \overline{(V_{\gamma,\mu}^{\beta,\delta}+
W_{\gamma,\mu}^{\beta,\delta})^2} = \overline{(U_{\gamma,\mu}^{\beta,\delta}+
U_{\mu,\gamma}^{\delta,\beta})^2} = 2 U^2
\end{eqnarray}
 
The same contribution arises from the interchange $\beta \leftrightarrow \mu$
in the operator (\ref{op}). The above transition probability comes therefore
with a factor two. This is because once the particles are triplet (or singlet)
paired on different orbitals, transitions become distinguishable.
In addition, one has to consider the four $\sigma=1$ 
transitions induced by ($s=\uparrow,\downarrow$)

\begin{eqnarray}
O_{s} & = & c^{\dagger}_{\gamma,s} c^{\dagger}_{\mu,s} 
c_{\beta,s} c_{\delta,s} \ \ \ \ {\rm or} \ \ \ \ 
c^{\dagger}_{\gamma,s} c^{\dagger}_{\beta,s} 
c_{\mu,s} c_{\delta,s}
\end{eqnarray}

\noindent which together give a transition 
probability $ 2 \overline{V^2}=8 U^2$.
The total transition probability is then $12 U^2$ instead of $16 U^2$ in the
case of doubly occupied initial and empty final orbitals (\ref{res22}). 
The same result is obtained 
in the case of a singlet initial state

\begin{eqnarray}
|\Psi_{in} \rangle & = & \frac{1}{\sqrt{2}}
(c^{\dagger}_{\mu,\uparrow} c^{\dagger}_{\beta,\downarrow} -
c^{\dagger}_{\mu,\downarrow} c^{\dagger}_{\beta,\uparrow})
c^{\dagger}_{\delta,\uparrow} c^{\dagger}_{\delta,\downarrow} | 0 \rangle
\end{eqnarray}

\noindent for which singlet and triplet transitions (\ref{op}) and
(\ref{opst}) also add coherently,
resulting in the same reduction of the transition 
amplitude (\ref{tramplitude}).
Consequently, and for $n>3$, the splitting between the $\sigma=1$ yrast and 
the first $\sigma=0$ excited state is much smaller than the spin gap
between the two lowest yrasts.

Note that these calculations must be modified in realistic systems for which
$\overline{V^2} = \overline{W^2}$ does not necessary hold. For a purely
local (Hubbard) interaction  with time-reversal symetry for instance, one has
$\overline{V^2}=0 \ne \overline{W^2}$ as the antisymetrized matrix elements 
vanish exactly.
Then, the reduction in transition probability occurs due to the vanishing
of singlet transitions as one goes to larger magnetizations.

\section{}\label{appendix2}

We first calculate each sector's connectivity $K$ which is
the number of basis states directly connected to an arbitrary
initial many-body state of a given sector, alternatively the number
of nonzero matrix elements per row (or column)
of the Hamiltonian matrix.
We saw in Appendix A that some transitions have increased weights,
in particular triplet transitions involving two doubly occupied
orbitals pick up a factor $\sqrt{3}$ that is absent of all other
transitions. In absence of double occupancies however, these transitions
are replaced by three times as many triplet transitions so that 
the total transition probability is conserved. The latter quantity is in fact 
the physically relevant one as it appears in second order perturbation
theory and determines the scaling of the MBDOS in the regime of dominant
fluctuations. We therefore calculate the weighted connectivity, where
the number of transitions are multiplied by the square of their 
relative amplitude. With this definition 
and for the case we are considering of 
spin $1/2$ particles, the connectivity
is constant within one sector. From (\ref{usrs}), $K$ is the sum of
a singlet and a triplet channel contribution which differ only in that
the former allows transition from and to double occupancies. 
For $\sigma=0$ we may consider the $U=0$ ground-state as our initial state.
It is easily seen then that the number of directly connected states
can be expressed as a sum over four contributions $K=K_0+K_1+K_s+K_t$ 
which for $\sigma=0$ are given by 

\begin{eqnarray}\label{connecst}
K_0(\sigma=0) & = & 1 \nonumber \\
K_1(\sigma=0) & = & n/2 (m/2-n/2) \nonumber \\
K_s(\sigma=0) & = & n/2(n/2+1)(m/2-n/2)(m/2-n/2+1)/4 \\
K_t(\sigma=0) & = & 3 n/2(n/2-1)(m/2-n/2)(m/2-n/2-1)/4 \nonumber
\end{eqnarray}

\noindent The first term corresponds to trivial diagonal transitions,
the second one to partially off-diagonal transitions changing a single 
one-body occupation, while the third and fourth ones correspond to generic
two-body transitions induced respectively 
by the singlet and triplet interaction operators in (\ref{usrs}). Note
the discrepancy in the prefactor $3$ between $K_s$ and $K_t$ due to
the enhancement of triplet transition amplitude discussed in Appendix A.
As $\sigma$ increases, some singlet transitions are replaced by additional
triplet transitions but some other disappear which we are going to identify.
The connectivity at full polarization is also easily calculated as
there are no long singlet transitions and particles may be considered
spinless. One has $K=K_0+K_1+K_t$ 

\begin{eqnarray}
K_0(\sigma=n/2) & = & 1 \nonumber \\
K_1(\sigma=n/2) & = & n (m/2-n) \\
K_t(\sigma=n/2) & = & n(n-1)(m/2-n)(m/2-n-1)/4 \nonumber
\end{eqnarray}

\noindent The connectivity difference between minimal and maximal
polarization is therefore given by

\begin{equation}\label{test}
K(\sigma=0)-K(\sigma=n/2) = \frac{n^3 m}{8} - \frac{3 n^4}{16} 
+O(n^3,m^3,n^2m,nm^2)
\end{equation}

\begin{figure}
\epsfxsize=3.3in
\epsfysize=2.5in
\epsffile{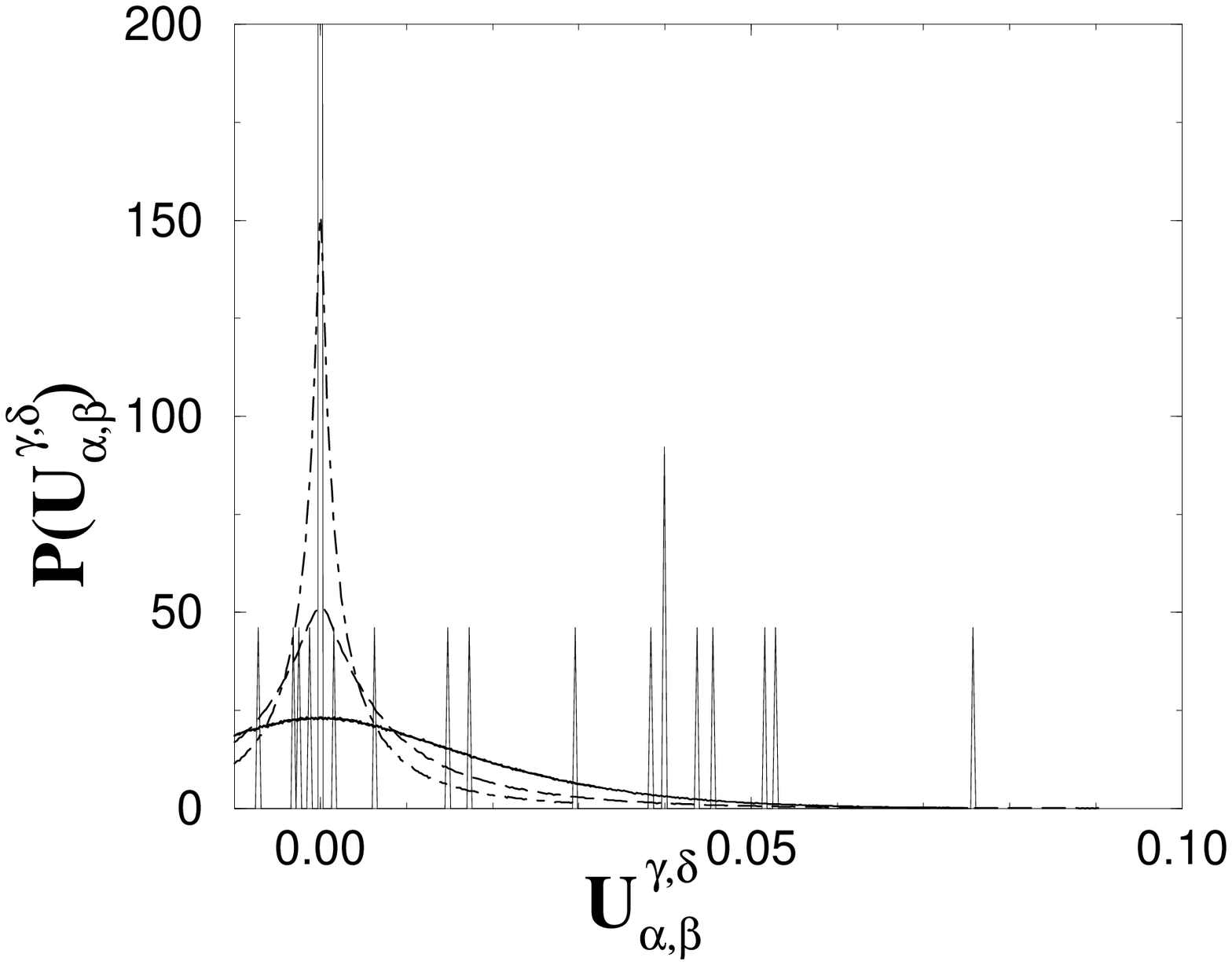}

\vspace{2 mm}

\caption
{Distribution of off-diagonal interaction matrix elements (\ref{ime})
for a one-dimensional lattice model with nearest and next-nearest
neighbor hopping and a Hubbard interaction (Details of the model can be
found in reference \protect\cite{daul}) and for different strength $W/V$ of the
disordered potential $W/V=0$ (thin solid line), 1 (thick solid line), 
2 (dashed line) and 3 (dotted-dashed line).}
\label{fig:puodt1t2}

\vspace{5 mm}

\epsfxsize=1.8in
\epsfysize=1.55in
\epsffile{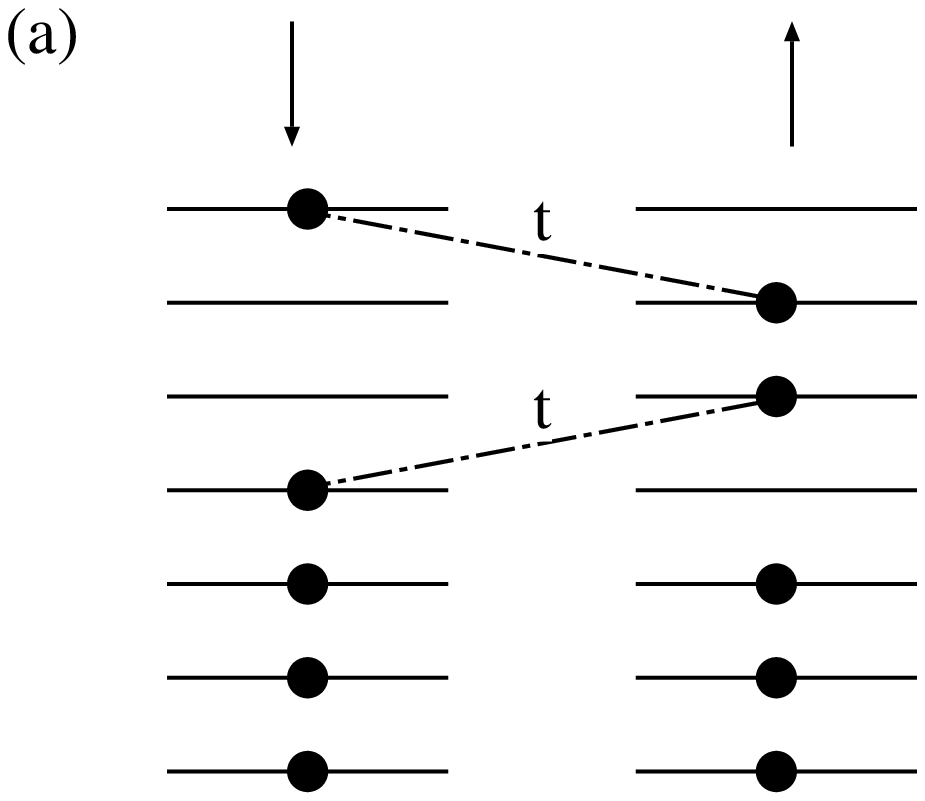} 
\vspace{-1.55in}
\hspace{5cm}
\epsfxsize=1.8in
\epsfysize=1.55in
\epsffile{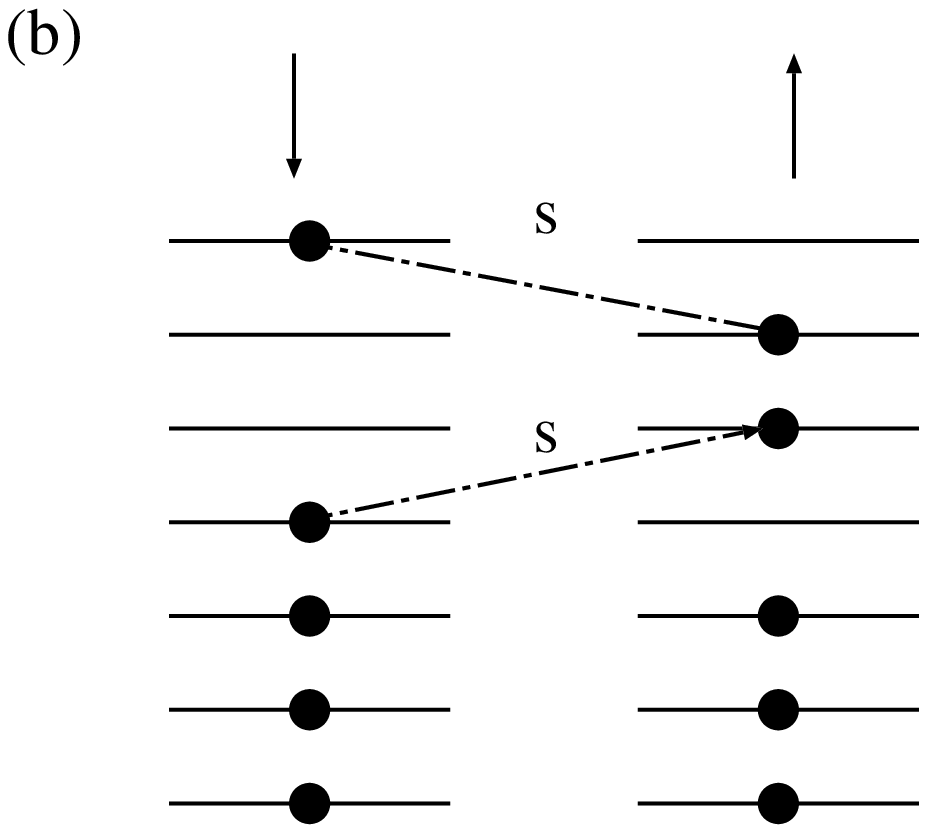} 
\hspace{1cm}
\epsfxsize=1.8in
\epsfysize=1.55in
\epsffile{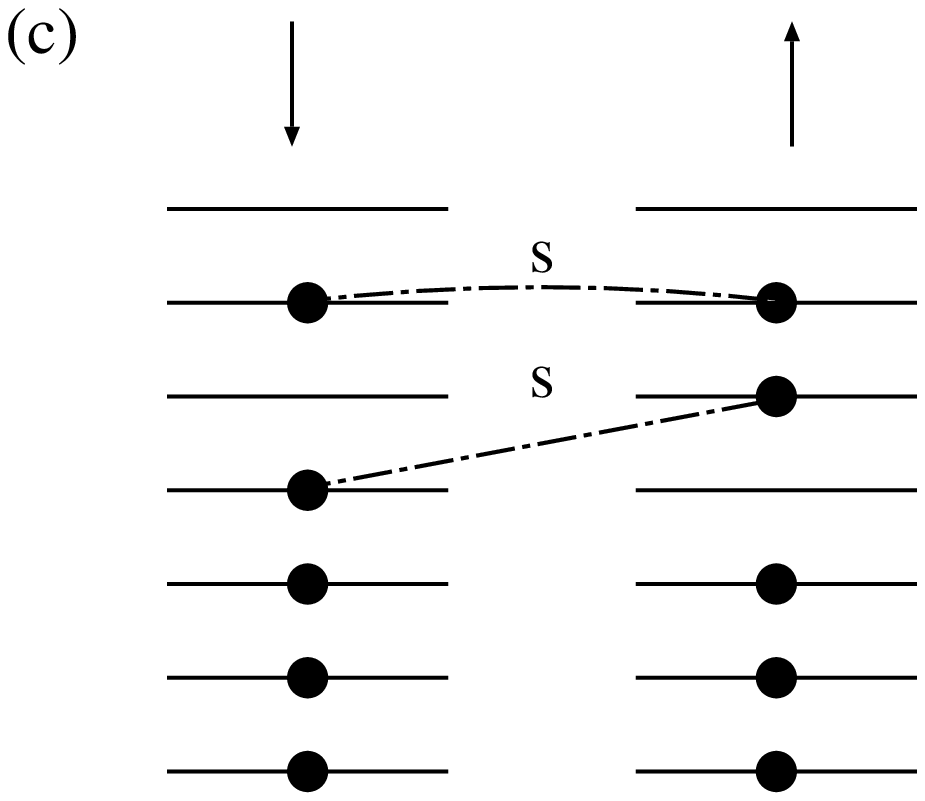}

\vspace{5 mm}

\caption
{Representation of $\sigma_z=0$ many-body states with $\sigma=2$ (a) and
0 (b). These two states differ only by the nature of the
two-particles bonds connecting
pairs of fermions on partly filled orbitals which are either 
singlets or triplets (the nature of the bonds is indicated
by the letter $s$ or $t$).
As fermions on doubly 
occupied orbitals can only be singlet paired, they cannot provide 
for a nonzero spin. Together with SRS, this forbids the scattering from a
triplet bond configuration onto a double occupancy, so that the rightmost
state (c) can only be coupled to the $\sigma=0$ state (b).}
\label{fig:sketch}

\vspace{5 mm}

\epsfxsize=3.3in
\epsfysize=2.5in
\epsffile{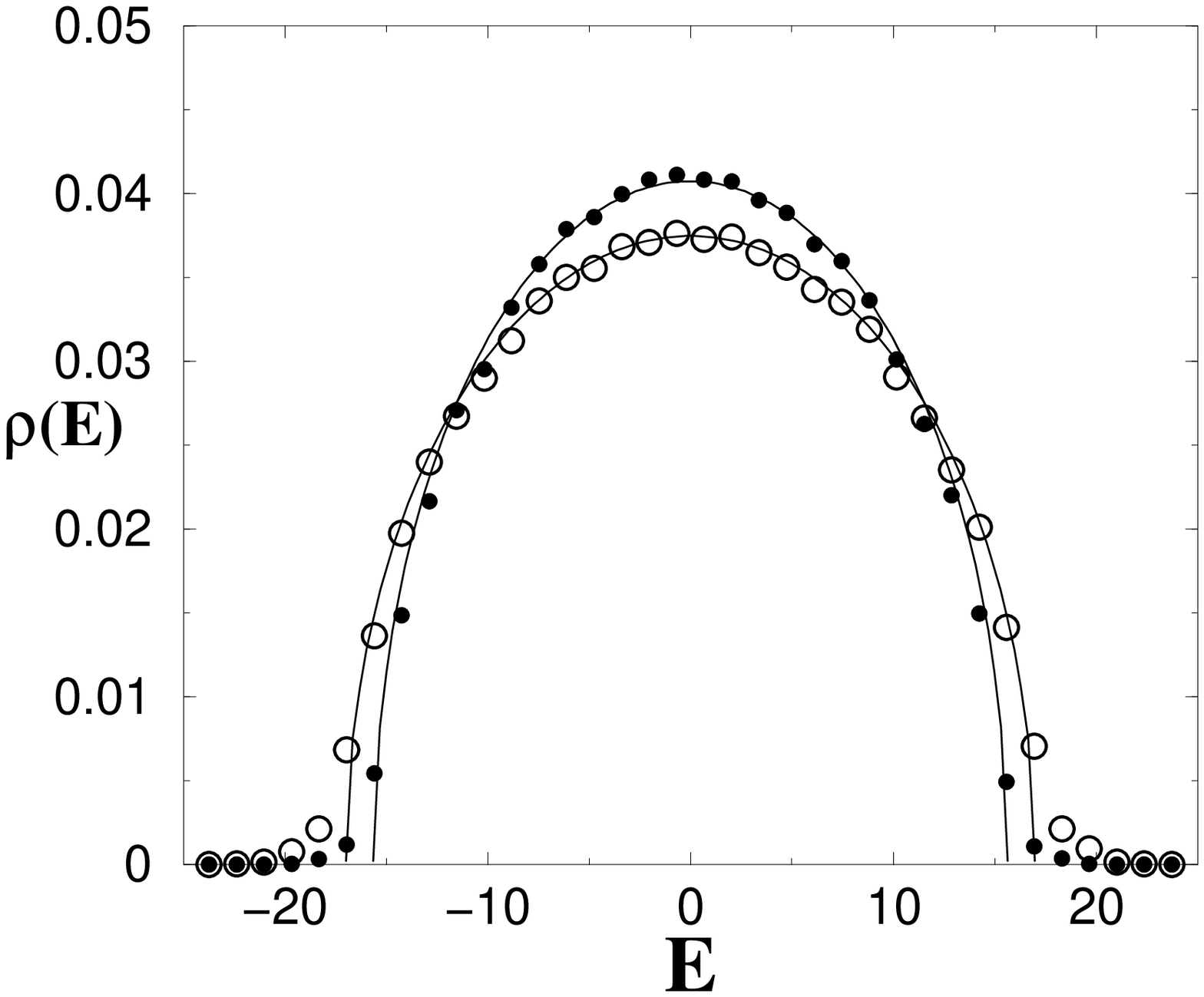}

\vspace{2 mm}

\caption
{Density of states for $n=2$, $m=12$ $\sigma=0$ (empty circles)
and $\sigma=1$ (full circles) computed from 5000 realizations
of ${\cal U}_f$. The full lines give the corresponding
semicircle law (\ref{halfcircle}). Tails develop due to the 
finiteness of the Hilbert
space size ($N(0)=78$ and $N(1)=66$).}
\label{fig:dostip}

\epsfxsize=3.3in
\epsfysize=2.5in
\epsffile{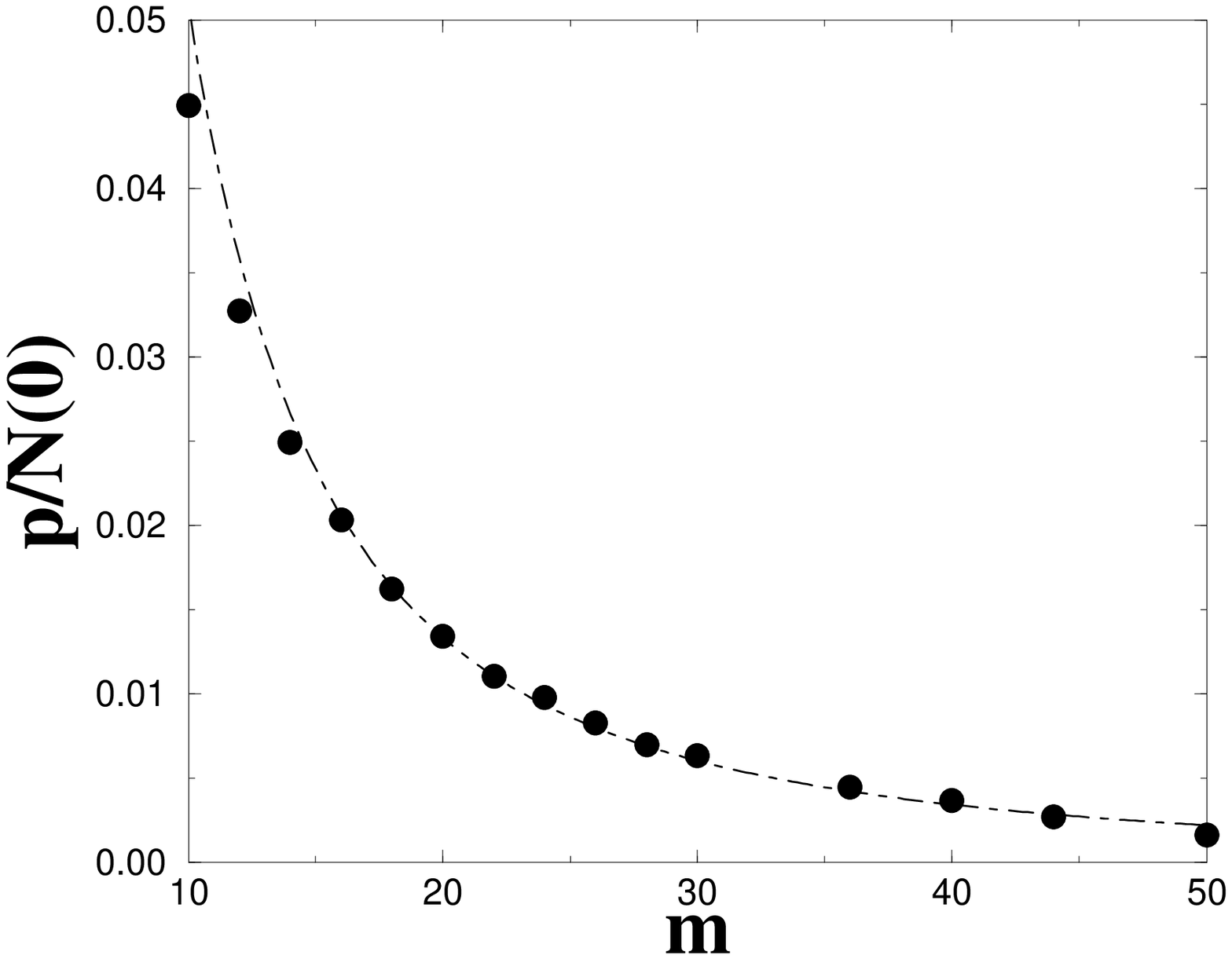}

\vspace{1 mm}

\caption
{Average number $p$ of levels in the spin gap between the two yrasts
for $n=2$, divided by the total number of levels $N(0)=m(m+1)/2$
in the $\sigma=0$
sector as a function of the number $m$ of one-particle orbitals.
The dashed line shows the dependence $p/N \sim 1/N$ in agreement with
a $m$-independent number of levels in the gap.}
\label{fig:numgaptip}

\vspace{4 mm}

\epsfxsize=3.1in
\epsfysize=2.5in
\epsffile{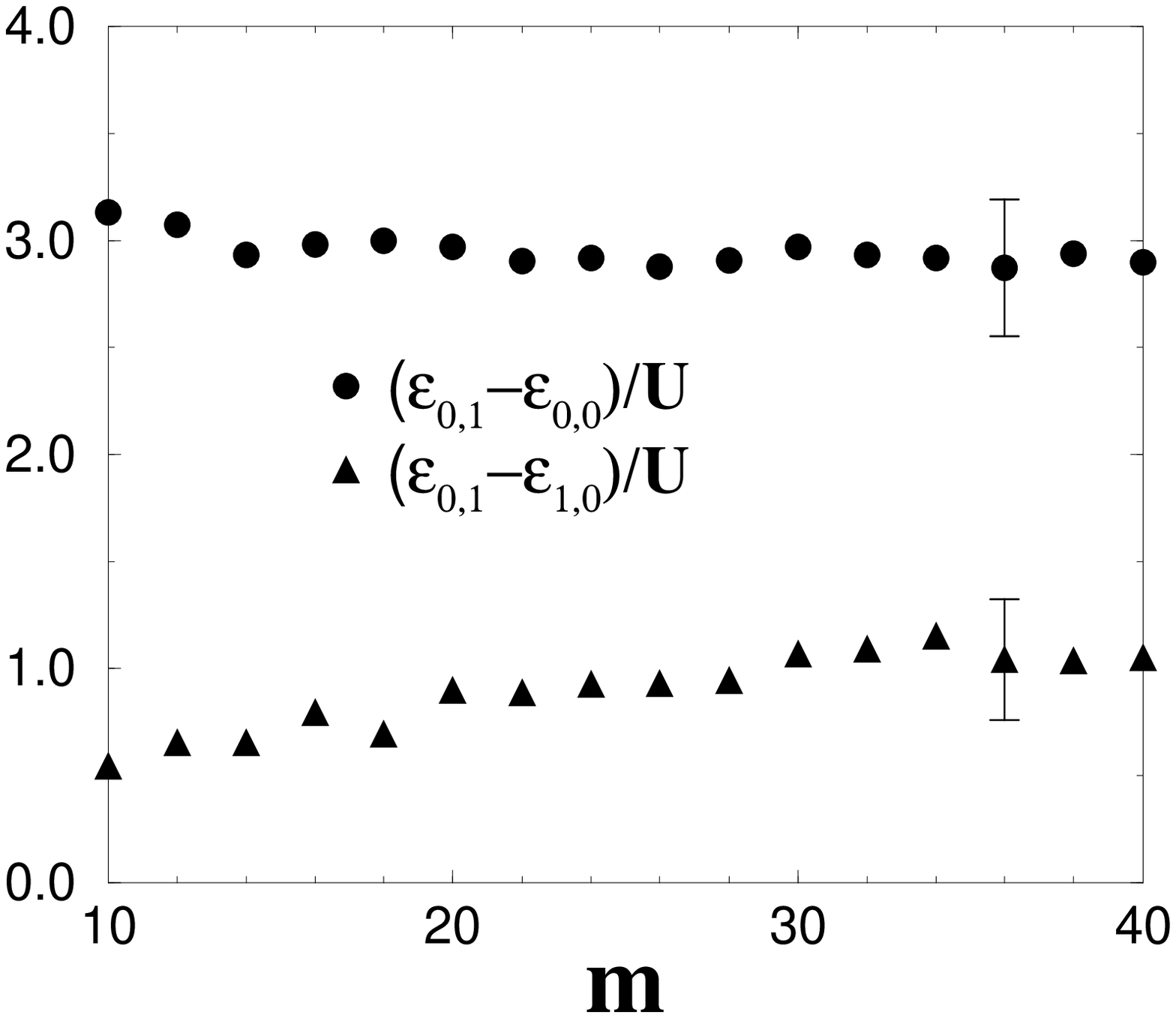}

\vspace{1 mm}

\caption
{Average spin gap between the two lowest yrast levels (circles)
and average splitting
between the first excited $\sigma=0$ level and the $\sigma=1$ yrast level 
(triangles) of ${\cal U}_f$ for $n=2$ as a function of the number 
of one-particle
orbitals $m$. The data show almost no $m$-dependence in agreement 
with (\ref{gapinftip}).}
\label{fig:gapinftip}

\vspace{4 mm}

\epsfxsize=3.1in
\epsfysize=2.4in
\epsffile{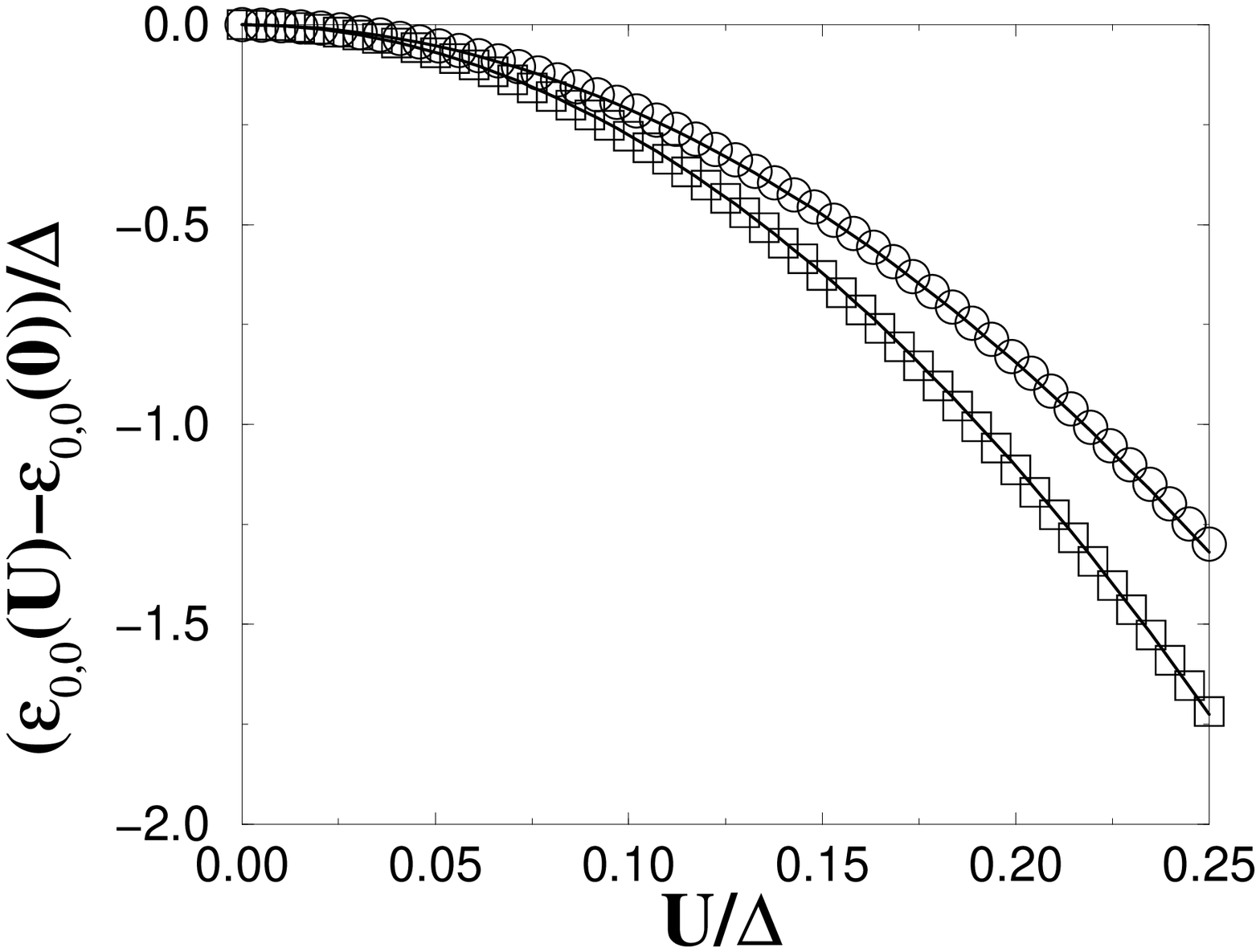}

\vspace{1 mm}

\caption
{Ground-state energy for the Hamiltonian (\ref{hamfin}) with $\lambda=0$
at $n=2$, $\sigma=0$, $m=12$ (circles) and $m=16$ (squares) 
as a function of the strength of off-diagonal fluctuations
$U/\Delta$. The solid lines indicate the perturbative result
$(\epsilon_{0,0}(U)-\epsilon_{0,0}(0))/\Delta = 
A (U/\Delta)^2$ with a numerical coefficient determined by (\ref{2ndf})
$A=-21.12$ and $-27.56$ 
respectively.}
\label{fig:e0tip}

\vspace{4 mm}

\epsfxsize=3.1in
\epsfysize=2.4in
\epsffile{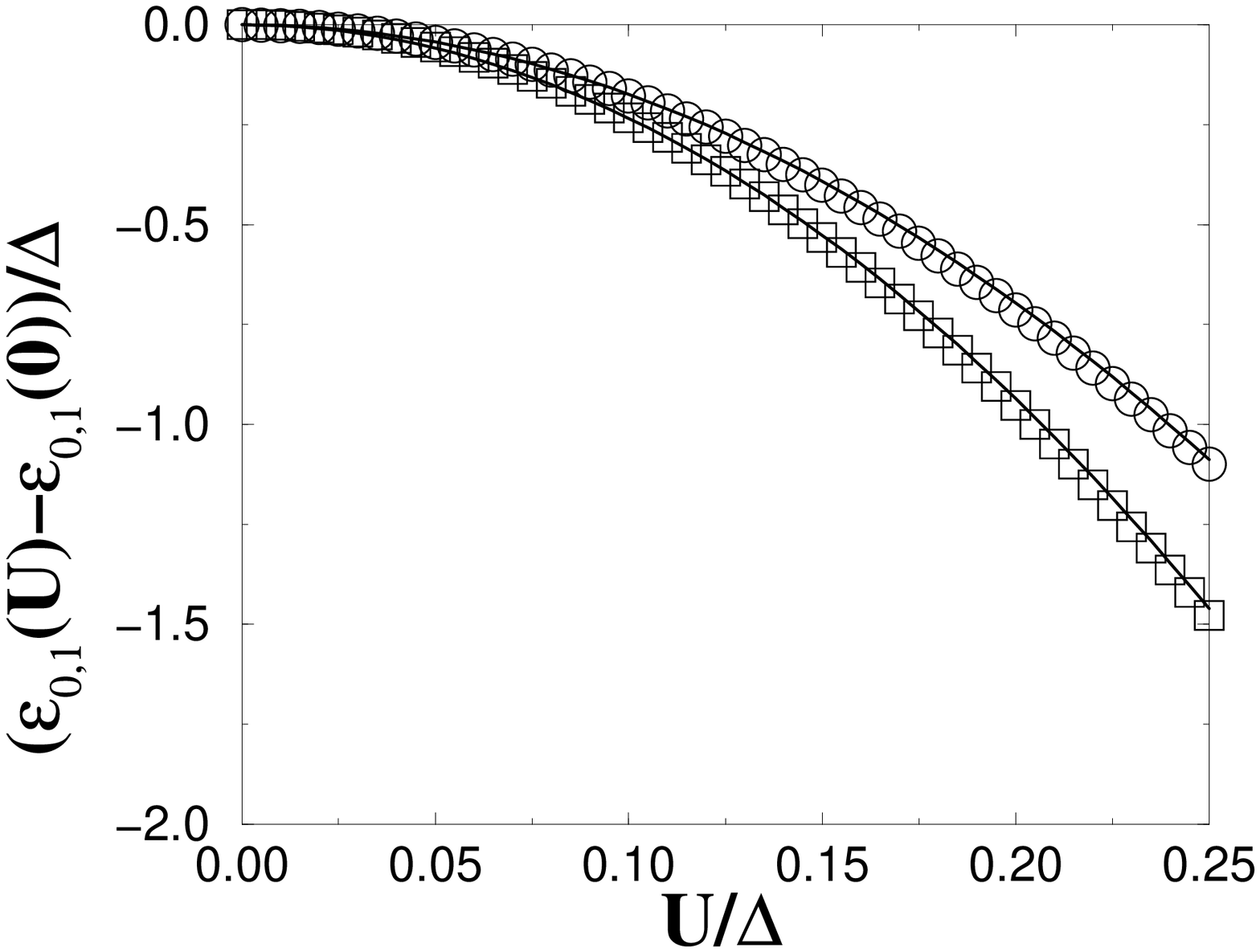}

\vspace{4 mm}

\caption
{Ground-state energy for the Hamiltonian (\ref{hamfin}) 
with $n=2$, $\sigma=1$, $m=12$ (circles) and $m=16$
(squares) as a function of the strength of off-diagonal fluctuations
$U/\Delta$ in absence of exchange interaction. 
The solid lines indicate the perturbative result
$(\epsilon_{0,1}(U)-\epsilon_{0,1}(0))/\Delta = 
A (U/\Delta)^2$ with a numerical coefficient 
determined by (\ref{2ndf})$A=-17.5$ and $-23.91$ 
respectively.}
\label{fig:e1tip}

\vspace{8 mm}

\epsfxsize=3.1in
\epsfysize=2.4in
\epsffile{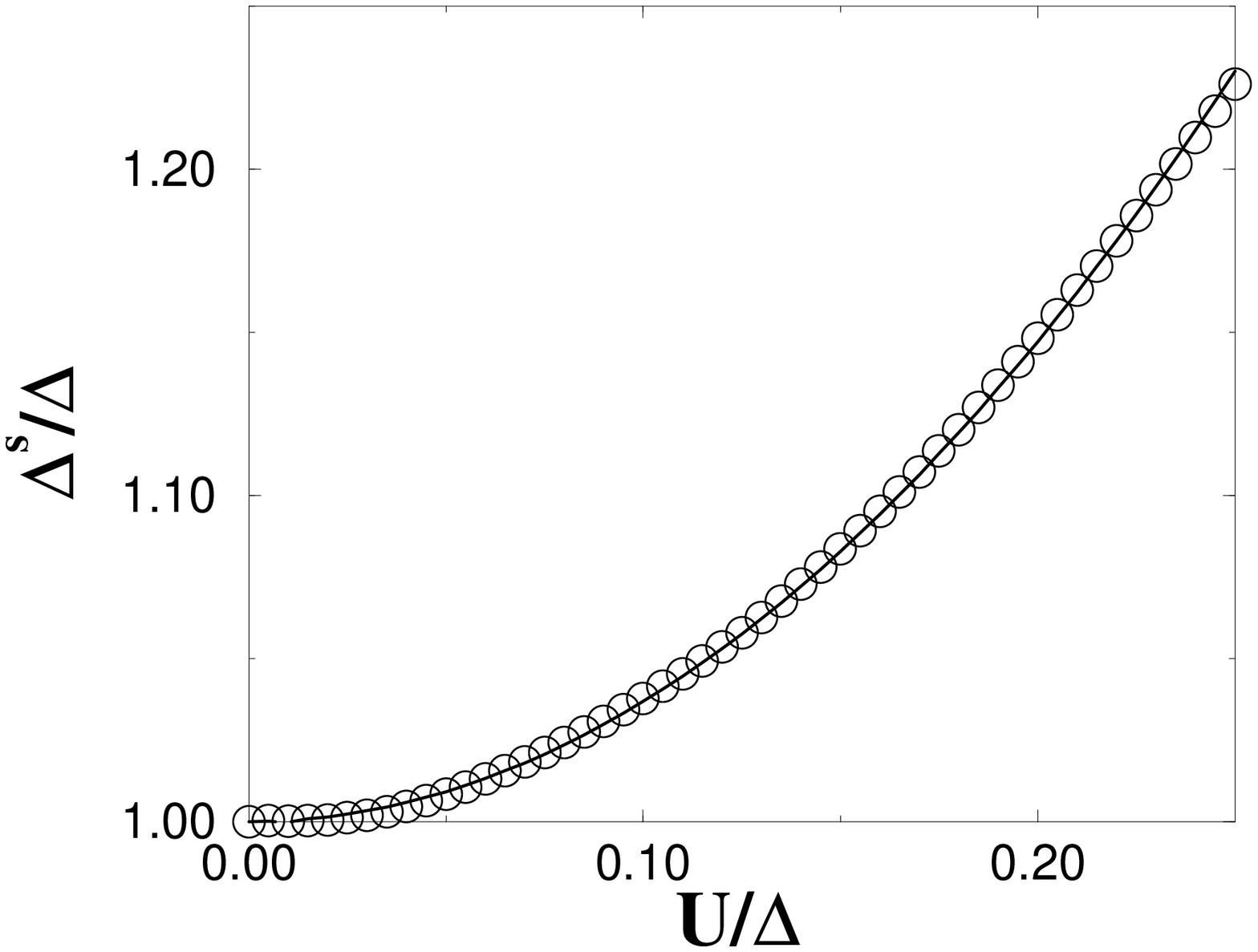}

\vspace{4 mm}

\caption
{Spin gap between the two lowest yrast states for $n=2$ and $m=16$
as a function of the strength of off-diagonal fluctuations $U/\Delta$. 
The solid line gives the perturbative result from (\ref{tip2nd}), giving
$(\epsilon_{0,1}(U)-\epsilon_{0,0}(U))/\Delta = 1 + 
A (U/\Delta)^2$ with a numerical coefficient  
determined by (\ref{tip2nd}), $A=3.66$.}
\label{fig:gaptip}

\newpage

\epsfxsize=3.1in
\epsfysize=2.4in
\epsffile{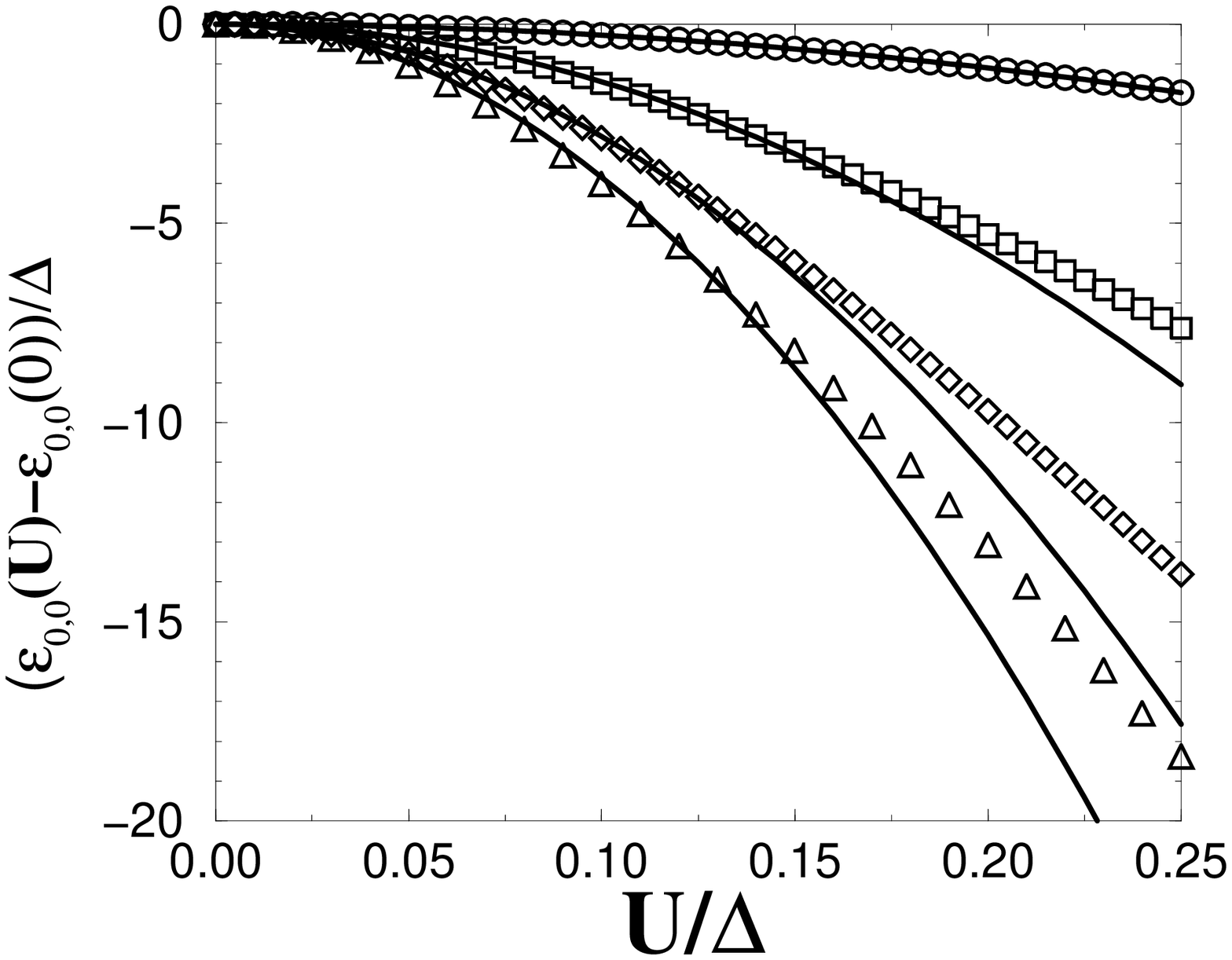}

\vspace{2 mm}

\caption
{Energy of the $\sigma=0$ yrast state for $m=16$, $n=2$ (circles),
4 (squares), 6 (diamonds) and 8 (triangles),
as a function of the strength of off-diagonal fluctuations $U/\Delta$. 
The solid lines give the perturbative results extracted 
from (\ref{pertnip0}), giving
$(\epsilon_{0,0}(U)-\epsilon_{0,0}(0))/\Delta = 
A (U/\Delta)^2$ with a numerical coefficient 
determined by (\ref{pertnip0})$A=-27.56$ $(n=2)$, 
-144.75 $(n=4)$, -281.09 $(n=6)$ and -373.57 $(n=8)$. Note that the breakdown
of the perturbative expression coincides with the emergence of the
large $U/\Delta$ linear regime.}
\label{fig:e0nip}

\vspace{6 mm}

\epsfxsize=2.in
\epsfysize=1.8 in
\epsffile{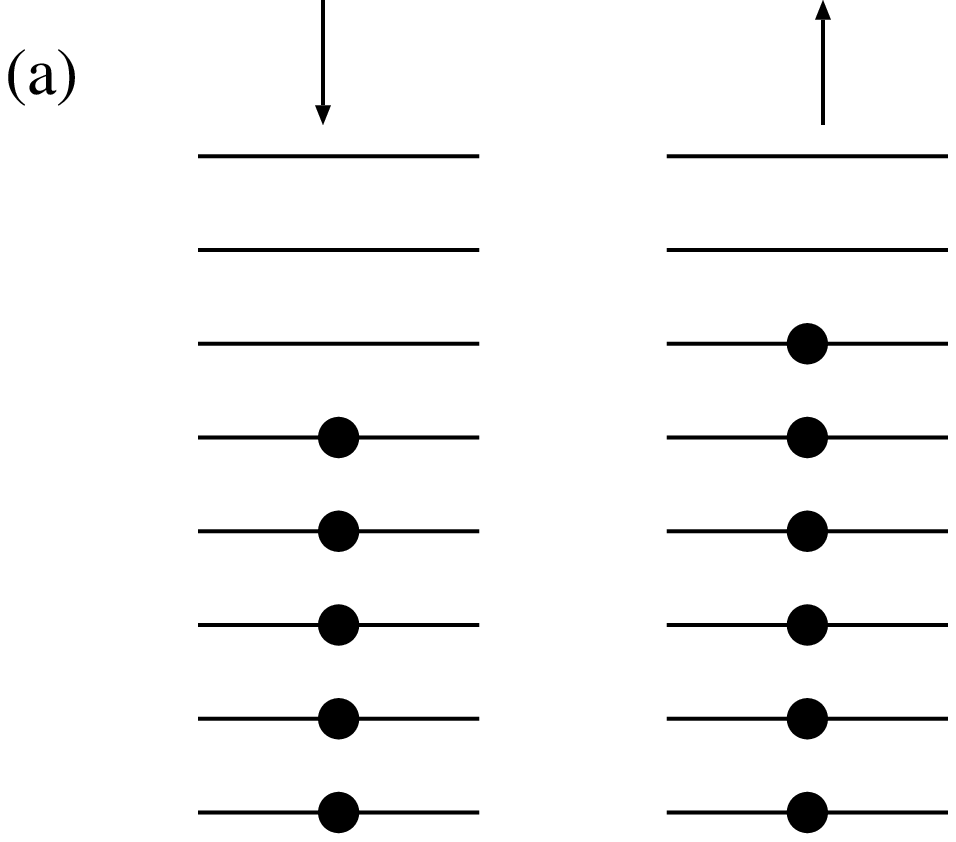} 
\vspace{-1.8in}
\hspace{7.cm}
\epsfxsize=2.in
\epsfysize=1.8in
\epsffile{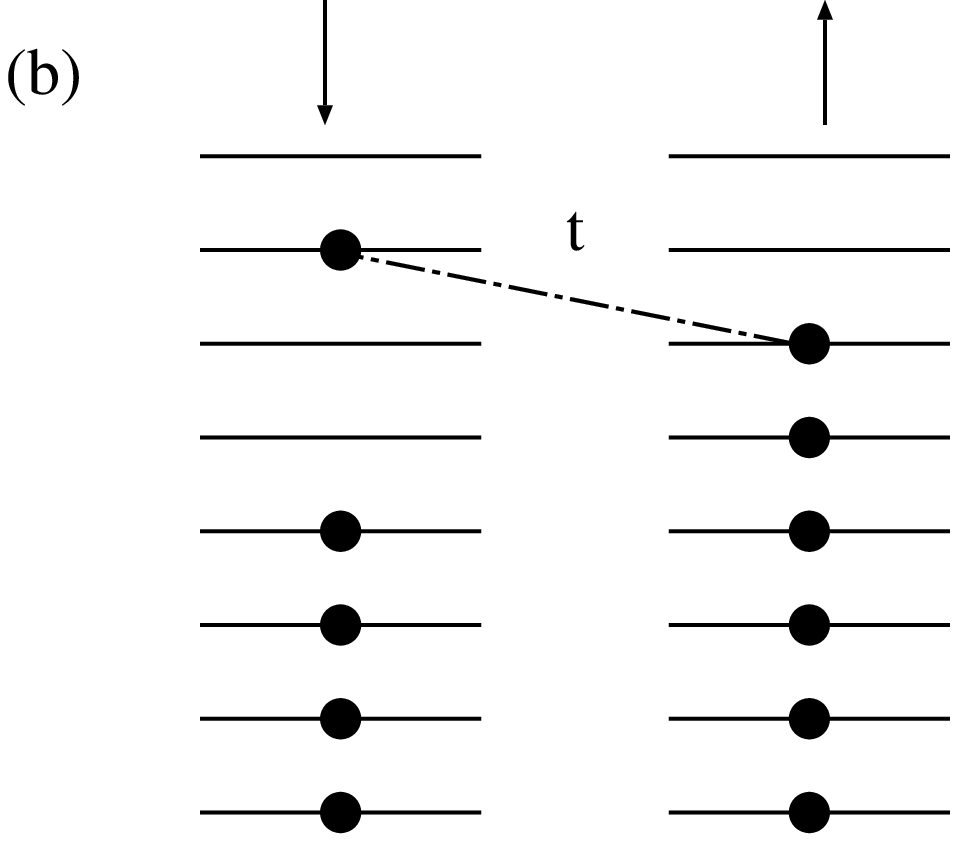} 

\vspace{6 mm}

\caption
{The two lowest yrast states $\sigma=1/2$ (a) and 3/2 (b) for odd number
of fermions.}
\label{fig:sketchodd}

\vspace{6 mm}

\epsfxsize=2.in
\epsfysize=1.8 in
\epsffile{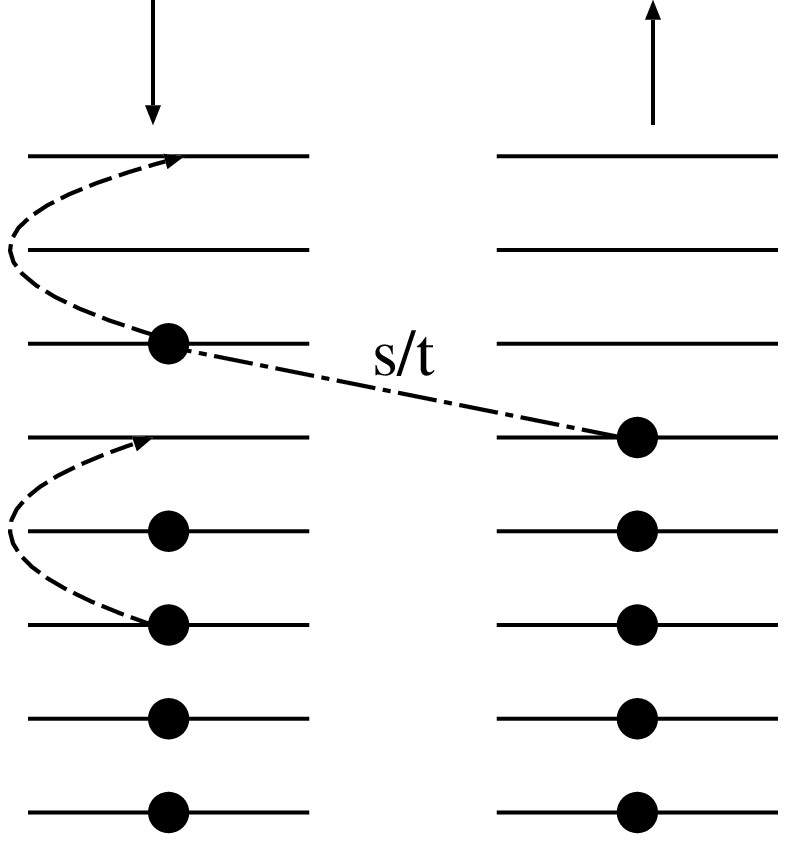} 
\vspace{-1.8in}
\hspace{7.cm}
\epsfxsize=2.6in
\epsfysize=1.8in
\epsffile{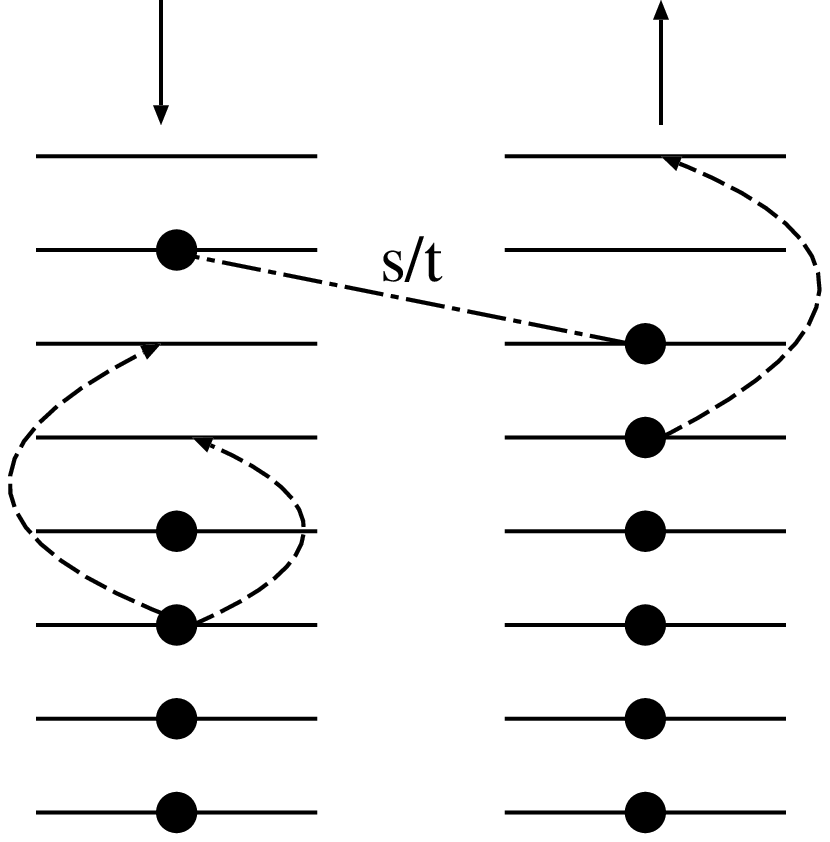} 

\vspace{6 mm}

\caption
{Left: Transitions that have a different transition amplitude for
$\sigma=0$ and $\sigma=1$ and thereby
give the dominant contribution to the spin gap
between the two lowest yrast levels for even number of
fermions. Right: Corresponding transitions for odd number of
fermions giving different transition amplitudes for
$\sigma=1/2$ and $\sigma=3/2$.}
\label{fig:sketchgap}

\vspace{6 mm}

\epsfxsize=3.3in
\epsfysize=2.5in
\epsffile{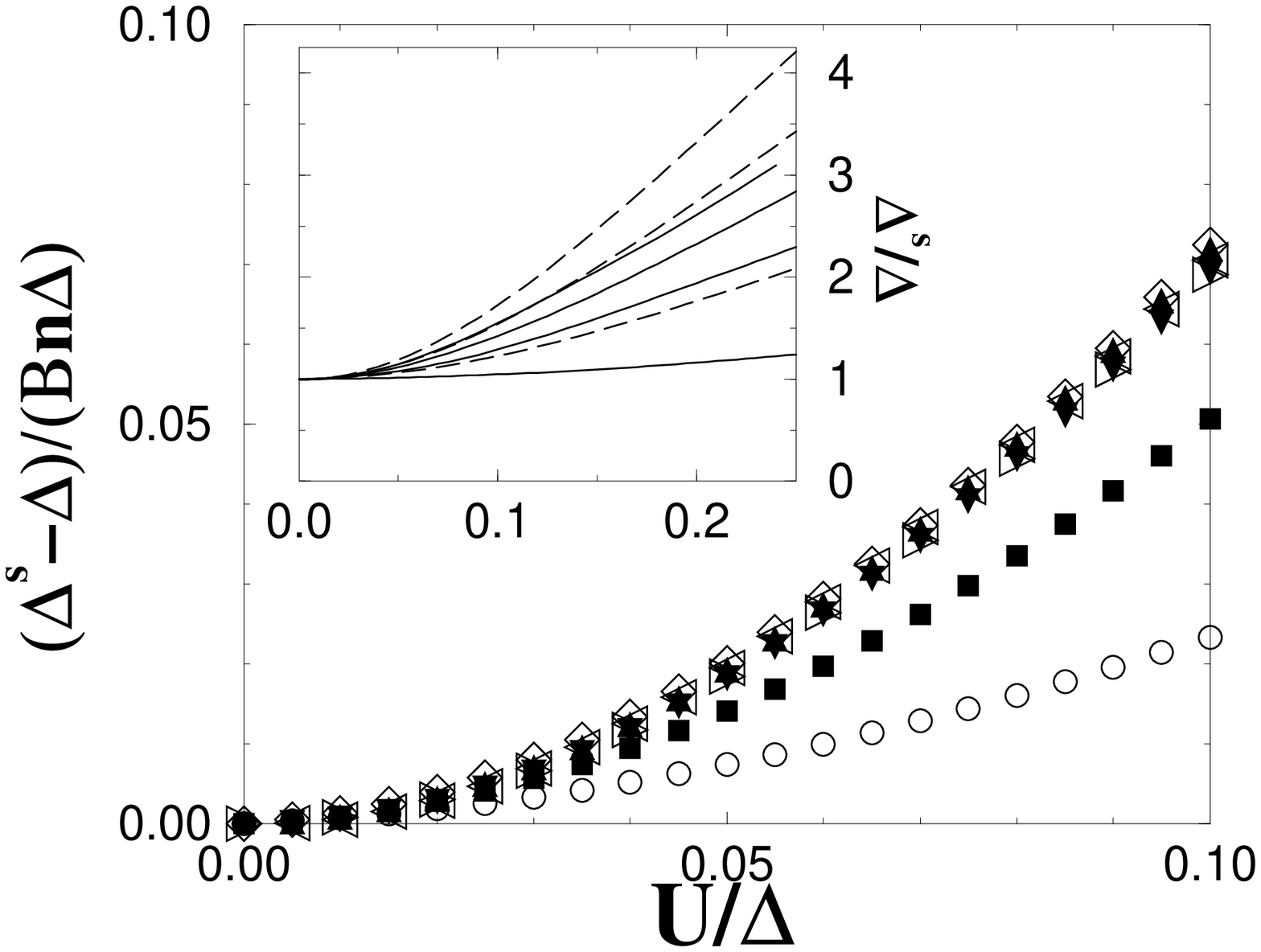}

\vspace{2 mm}

\caption
{Rescaled spin gap between the two lowest yrast states for $m=16$
and $n=2$ (circles), 3 (squares), 4 (diamonds), 5 (triangles up),
6 (triangles left), 7 (triangles down) and 8 (triangles right). 
Symbols corresponding to odd (even) $n$ are filled (empty). The scaling
parameter satisfies $\alpha=1$ (1.5) for even (odd) number of particles
(see text). The scaling holds quite well already for a small number
of particles $n \ge 4$ and the even-odd dependence of the gap confirms
the theory presented in the text.
Inset : spin gap before rescaling for the same cases as above.
Lines corresponding to odd (even) $n$ are dashed (full) to stress 
the even-odd dependence of the gap. Note that for larger $n$, the
gap starts to have a linear dependence above $U/\Delta \approx 0.1$ 
indicating
the border between perturbative and asymptotic regimes.}
\label{fig:gapnip}

\vspace{6 mm}

\epsfxsize=3.1in
\epsfysize=2.4in
\epsffile{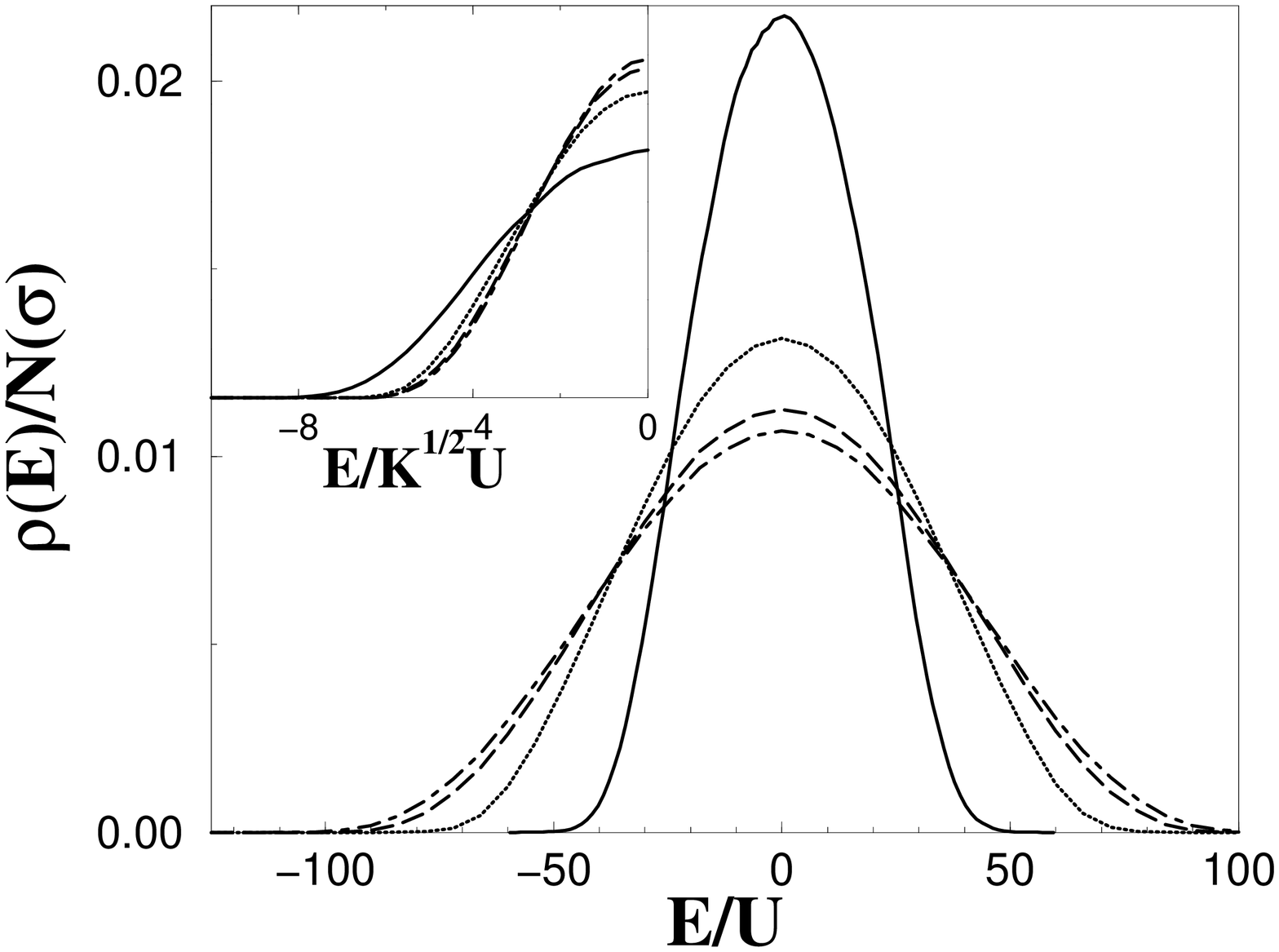}

\vspace{2 mm}

\caption
{Density of states for the Hamiltonian ${\cal U}_f$
with $n=6$ particles and $m=16$ orbitals,
corresponding to the magnetization blocks $\sigma=-3$ (solid line),
-2 (dotted line), -1 (dashed line) and 0 (dotted-dashed line).
Inset: rescaled density of states showing the approximate scaling in
$E/K^{1/2}U$.}
\label{fig:mbdos}

\newpage

\epsfxsize=3.1in
\epsfysize=2.4in
\epsffile{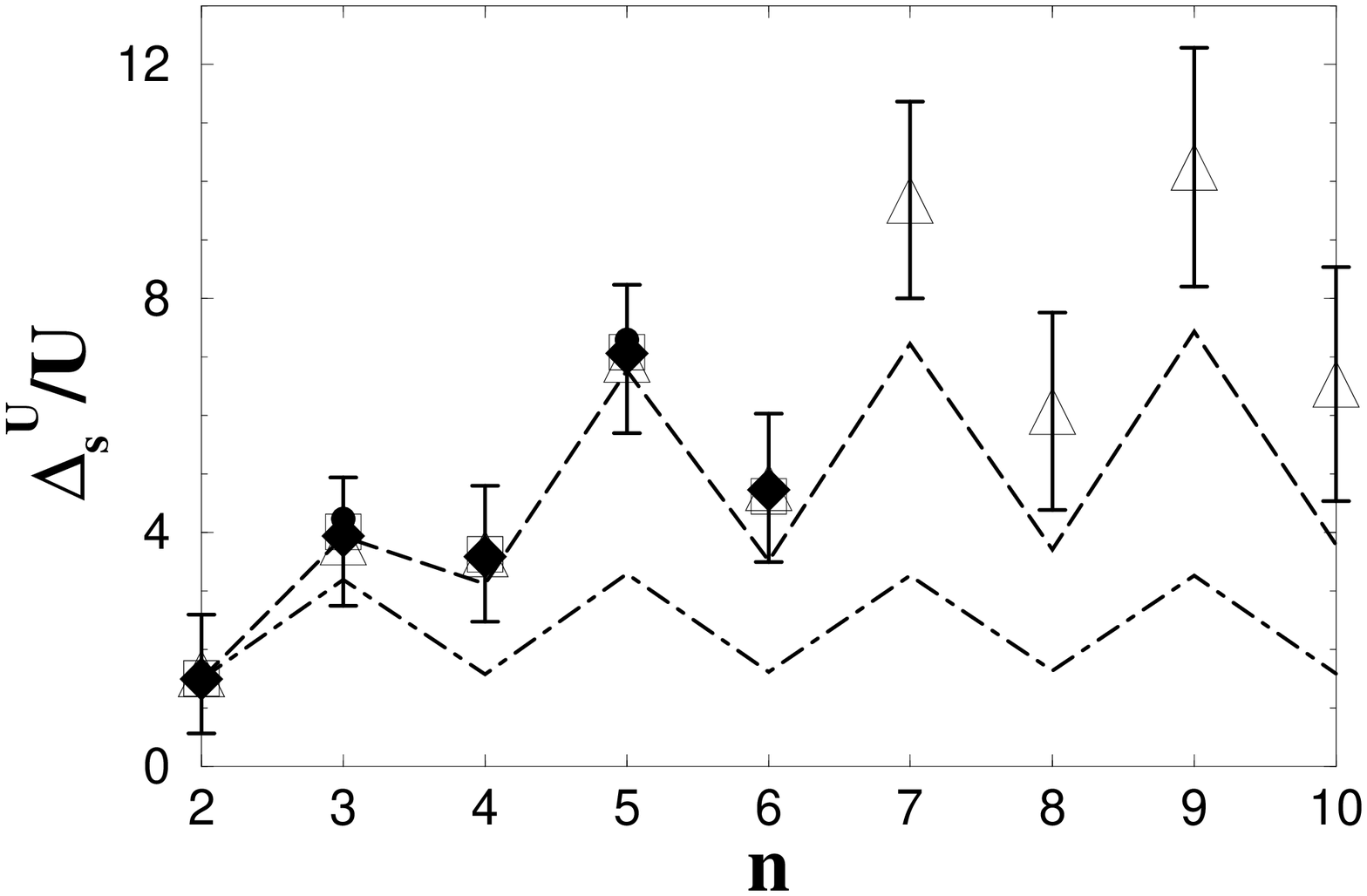}

\vspace{2 mm}

\caption
{Dependence of the finite-size spin gap in the number $n$ of
particles in the regime of dominating fluctuations $U/\Delta \gg 1$
and $\lambda = 0$.
Points correspond to numerical results for $m=10$ (full circles), 12
(empty
squares), 14 (full diamonds) and 16 (empty triangles).
For the case $m=16$ and 1000 Hamiltonian realizations,
the error bars indicate the r.m.s. of the gap distribution while the
dashed 
and dotted-dashed lines show the numerically computed variances 
(Left-hand side of eq. (5.4)) for the full Hamiltonian and after setting
to zero non-generic interaction matrix elements respectively.}
\label{fig:gapinf}

\vspace{1 mm}

\epsfxsize=3.3in
\epsfysize=2.5in
\epsffile{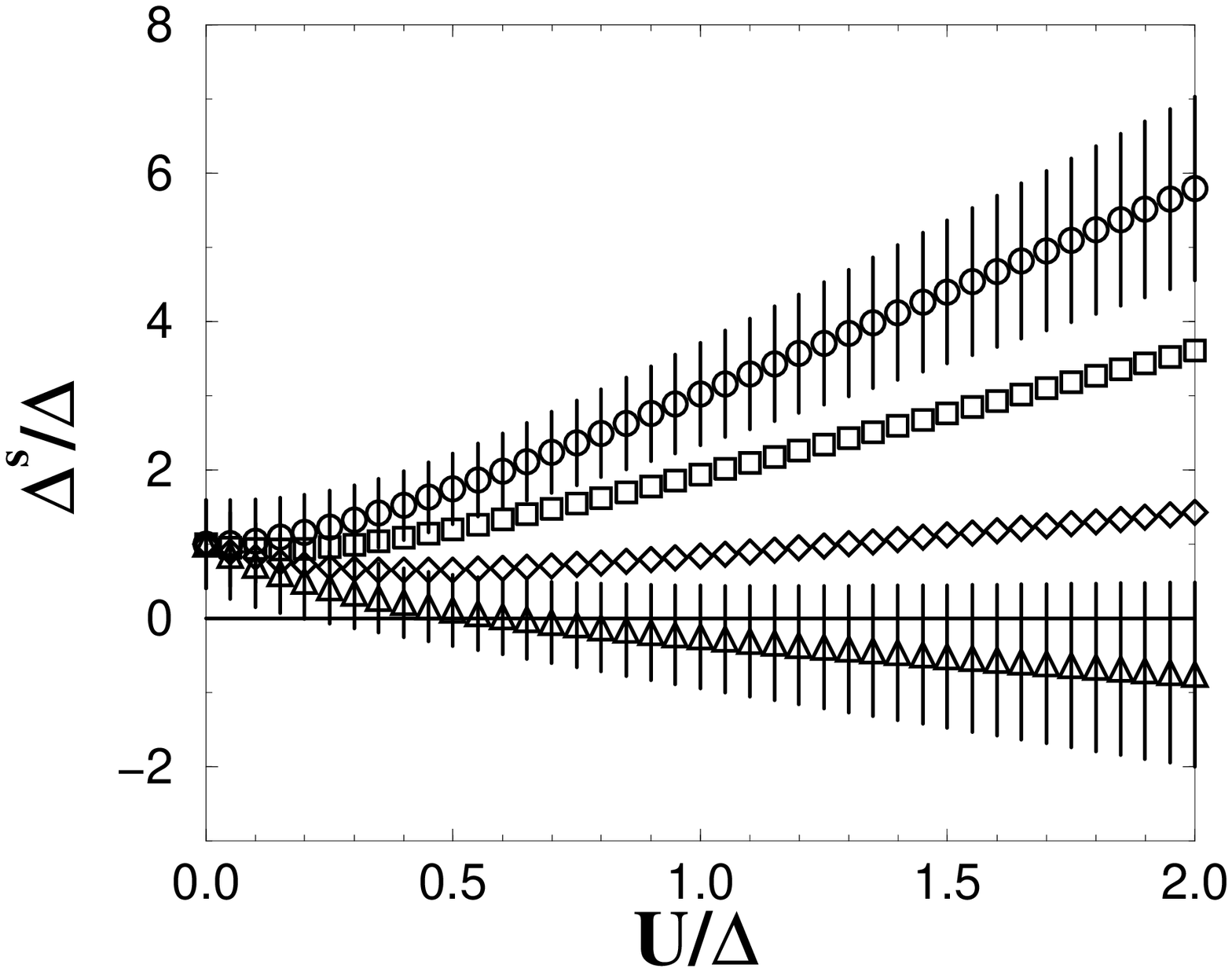}

\vspace{2 mm}

\caption
{Evolution of the energetic distance between the two lowest yrast
for $n=5$ and $m=12$ as a function of the strength of interaction
fluctuations $U/\Delta$. Shown are curves corresponding to an exchange
$\lambda U/\Delta=0.$ (circles), 2 (squares), 4 (diamonds) and 6 (triangles).
The error bar indicate the r.m.s of the distribution of $\Delta^s/\Delta$.}
\label{fig:gapxover}

\vspace{5 mm}

\epsfxsize=2.2in
\epsfysize=1.75in
\epsffile{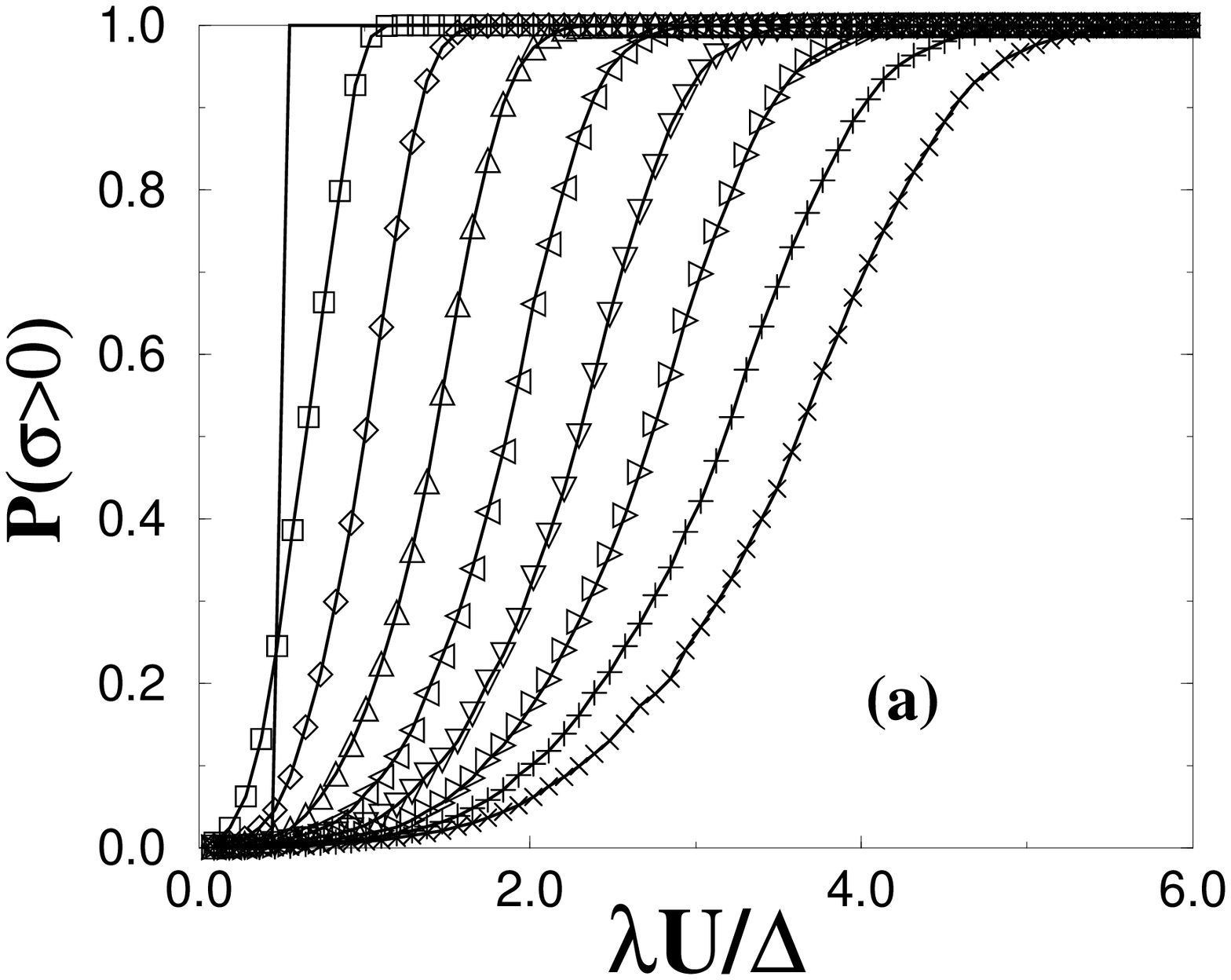}
\vspace{-1.75in}
\hspace{5.5cm}
\epsfxsize=2.2in
\epsfysize=1.75in
\epsffile{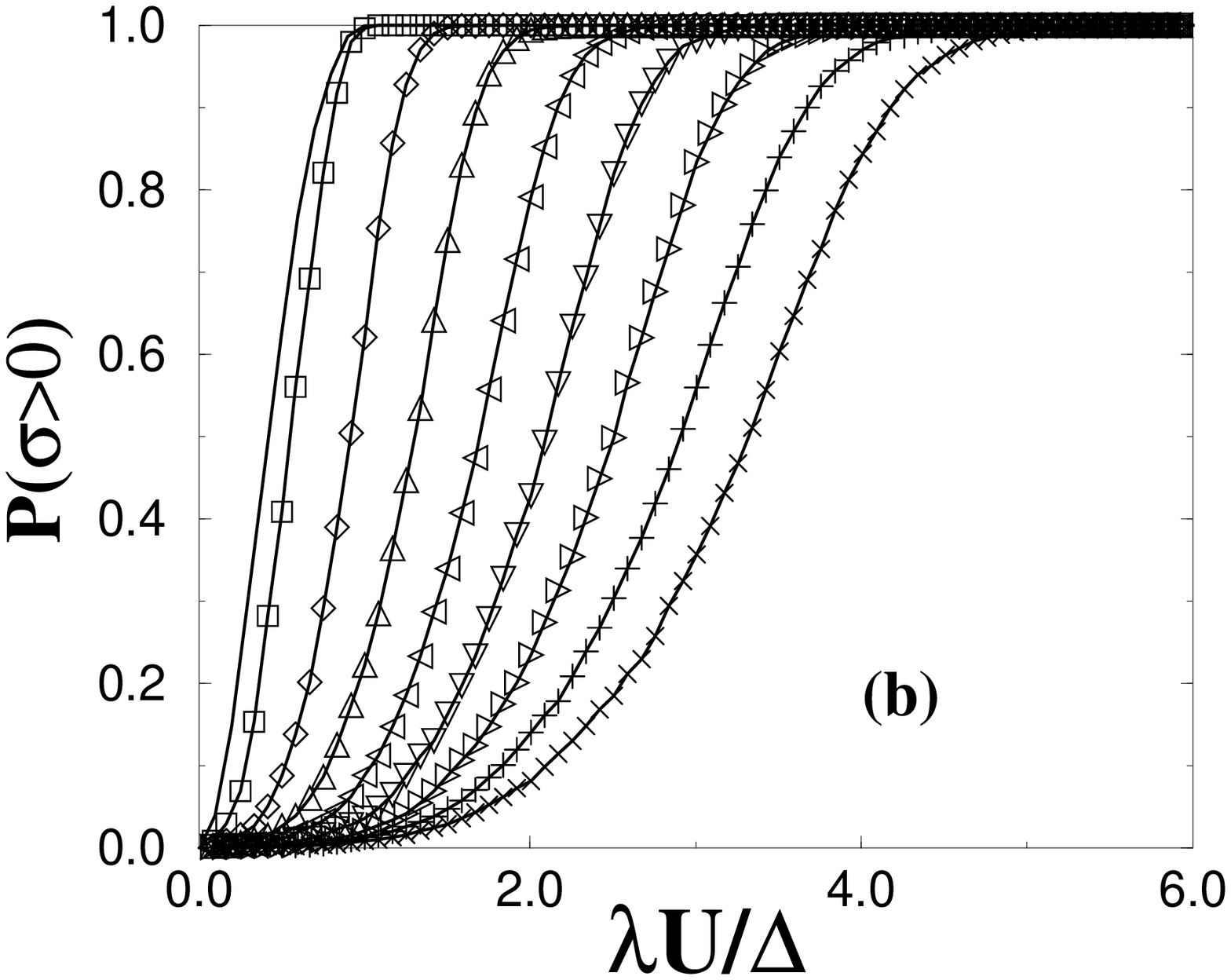}
\hspace{0.3cm}
\epsfxsize=2.2in
\epsfysize=1.75in
\epsffile{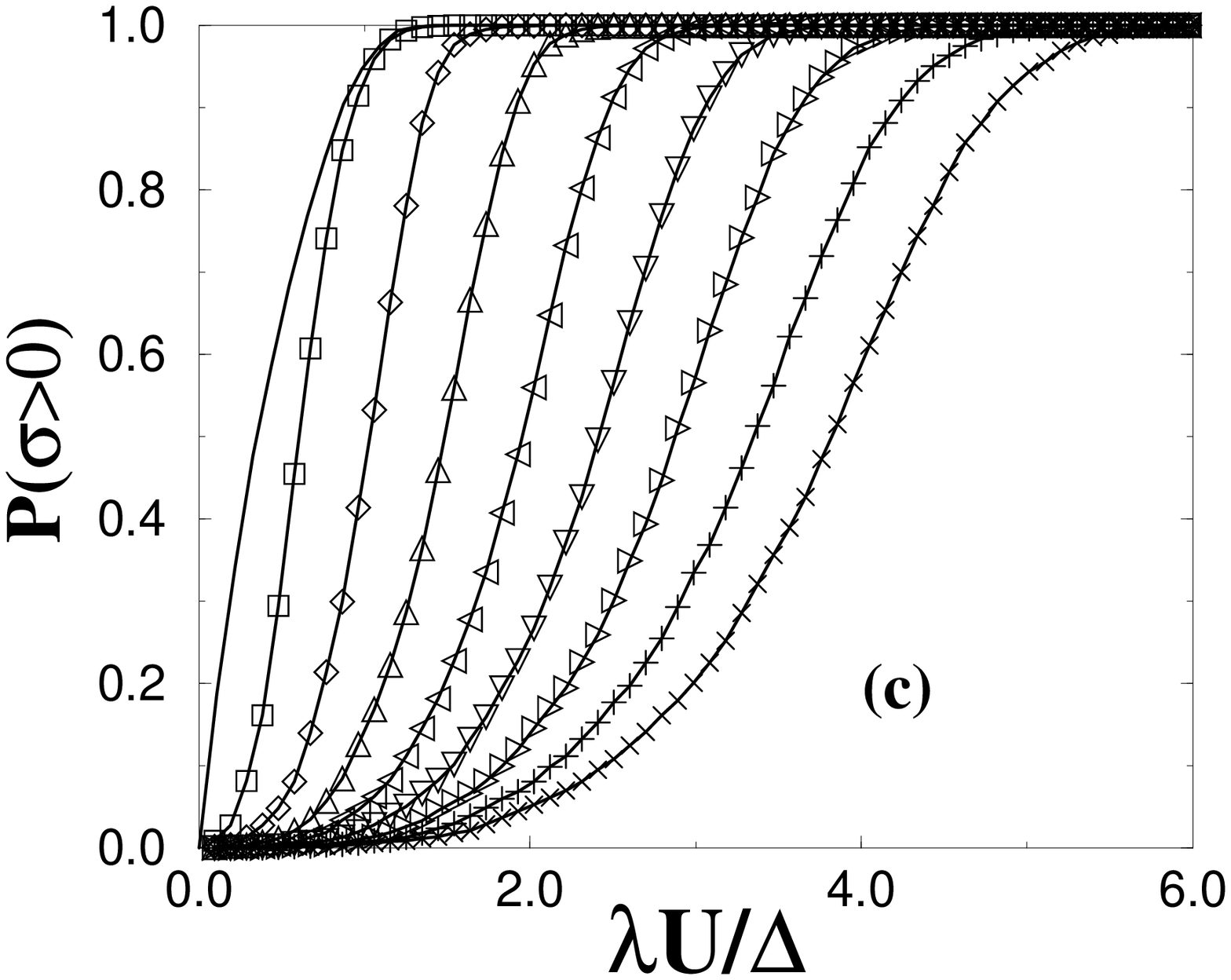}

\vspace{2 mm}

\caption
{Probability for a magnetized ground-state as a function of
the exchange $\lambda U/\Delta$ for 5000 realizations of
Hamiltonian (\ref{hamfin}) with $n=6$, $m=12$.
Three cases with equidistant (a), GOE (b) and random (c)
one-body spectra are shown. Different curves correspond to
different fluctuations of IME :
$U/\Delta = 0$ (solid line), 0.1,0.2,0.3...0.8, (symbols, from left to
right). }
\label{fig:probpf}

\vspace{4 mm}

\epsfxsize=2.8in
\epsfysize=2.1in
\epsffile{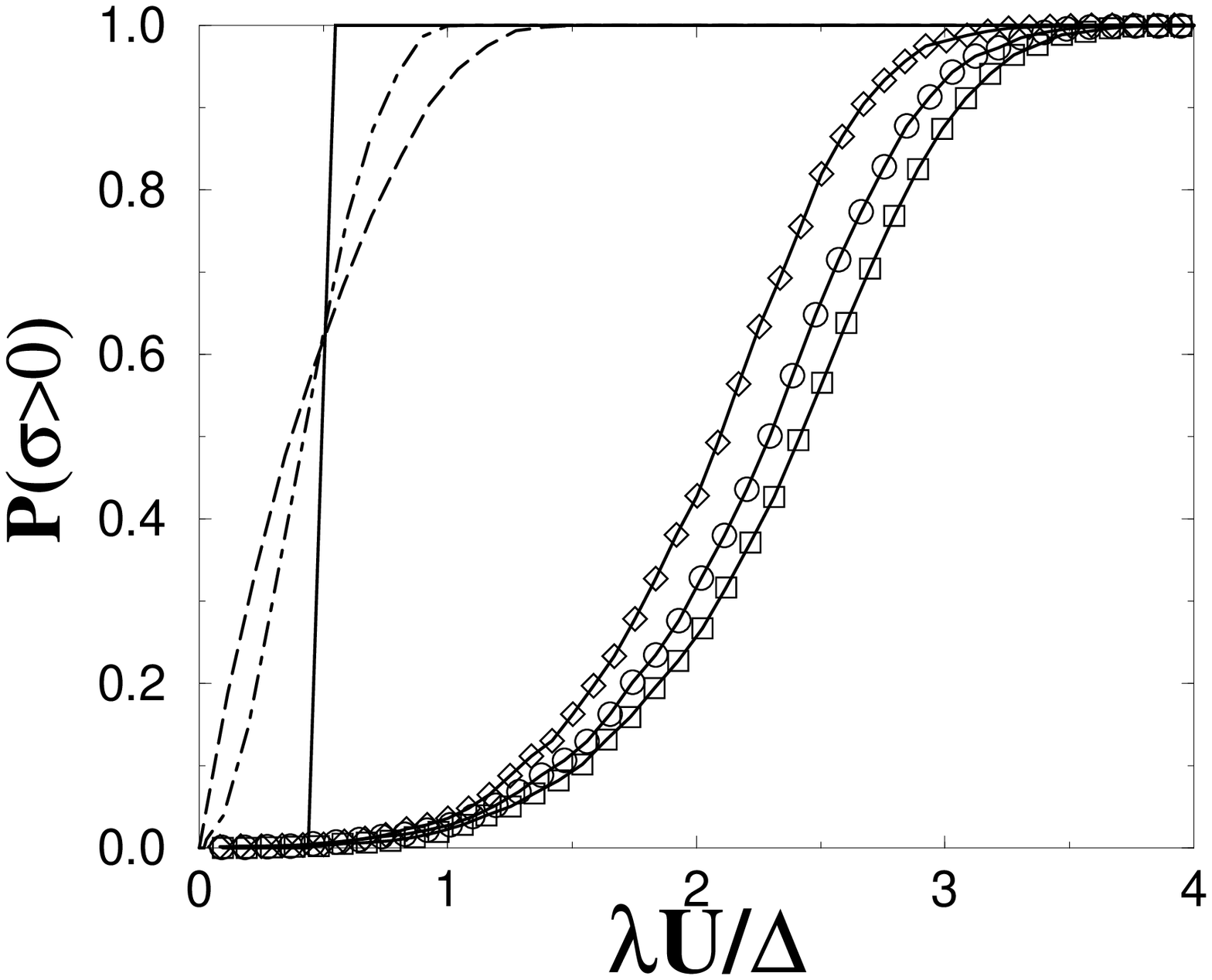}

\vspace{2 mm}

\caption
{Comparison of the ground-state magnetization probability for $n=6$
and $m=12$ at $U/\Delta=0$
(lines) and 0.5 (symbols) for equidistant (solid line and circles), 
Poissonian (dashed line and squares)
and Wigner-Dyson (dotted-dashed line and diamonds) orbitals distribution.}
\label{fig:probcomp}

\epsfxsize=3.1in
\epsfysize=2.2in
\epsffile{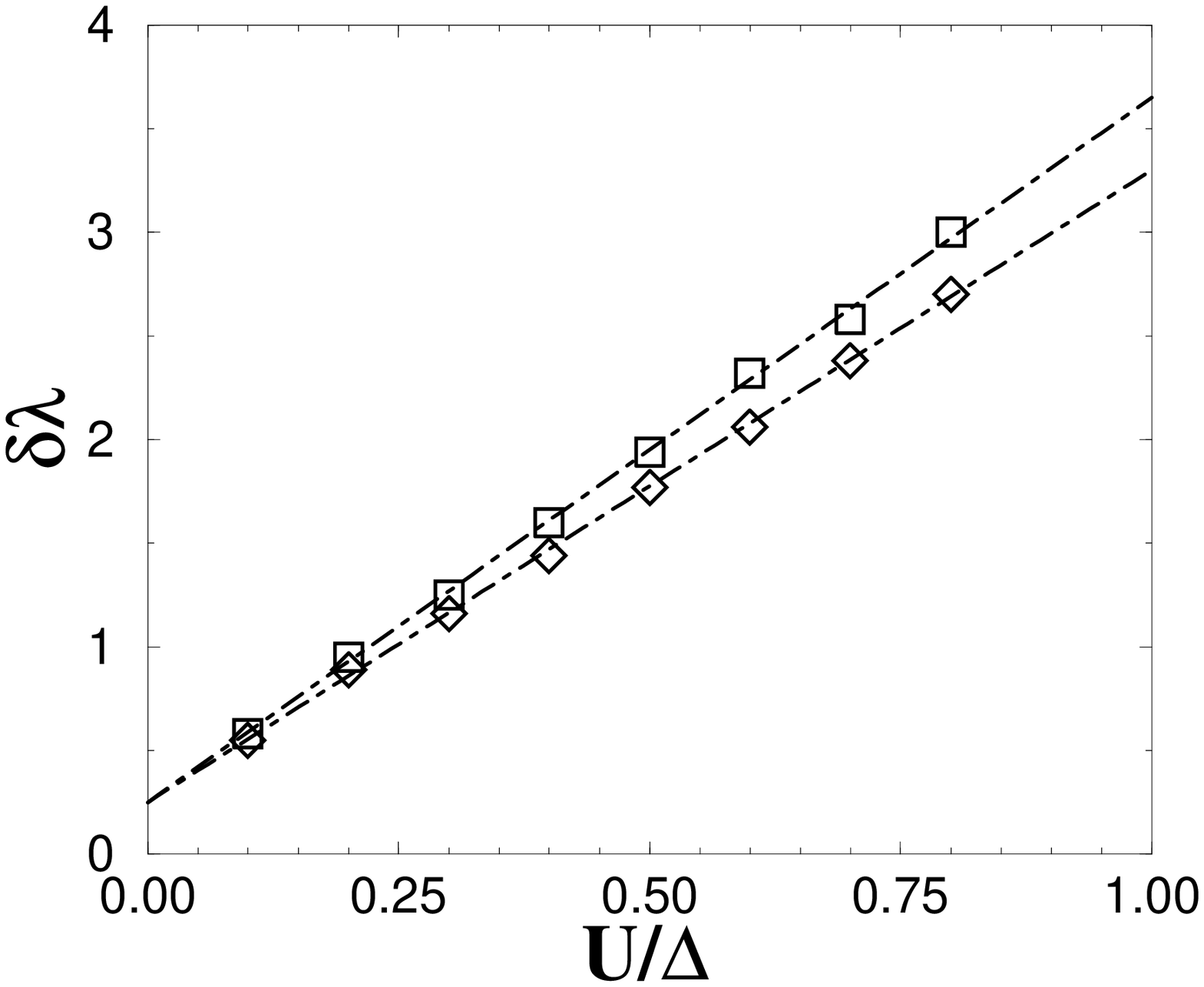}

\vspace{2 mm}

\caption
{Dependence of the average distance $\delta \lambda$ from the magnetization
threshold $\lambda_c(U)$ for a Poissonian (squares) and Wigner-Dyson
(diamonds) orbital distributions. $\lambda_c(U)$ is extracted from
Fig. \ref{fig:probpf} as the value at 
which $P(\sigma>0)=0.01$
and $\lambda_c(U)+\delta \lambda$ corresponds to $P(\sigma>0)=0.5$.
The linear fits do not extrapolate
to zero since $P(\sigma>0)=0.5$ requires a finite $\delta \lambda$
at $U/\Delta=0$ (see Figs. \ref{fig:probpf}).}
\label{fig:dlambda}

\epsfxsize=3.1in
\epsfysize=2.2in
\epsffile{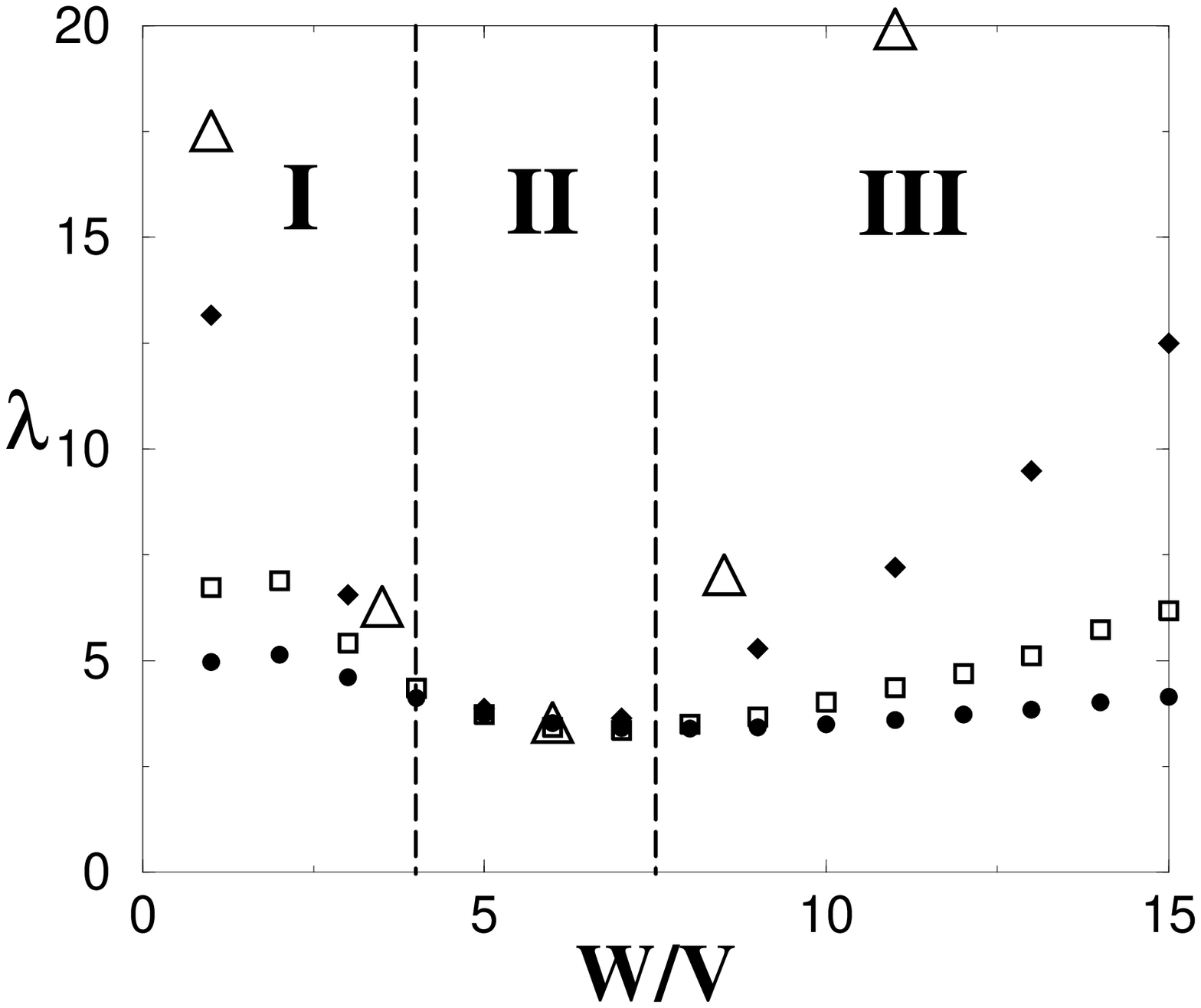}

\vspace{2 mm}

\caption
{Magnetization parameter $\lambda$ vs. disorder strength $W/V$,
for a Hubbard interaction
and a two-dimensional 10$ \times $10 (full circles),
20$ \times $20 (empty squares), 50$ \times $50 (full diamonds) and
80$ \times $80 (empty triangles) Anderson lattice. One clearly
differentiates three regimes : (I) At small disorder, $\lambda$ increases
due to a crossover from ballistic to diffusive behavior (see text).
(II) At intermediate disorder, exchange
and fluctuations compensate each other so that $\lambda$ is size-independent.
(III) At large
disorder one-body states are strongly localized over very few sites,
which kills the off-diagonal fluctuations faster than
the exchange and the latter dominates again.}

\label{fig:lambda2d}

\epsfxsize=3.1in
\epsfysize=2.2in
\epsffile{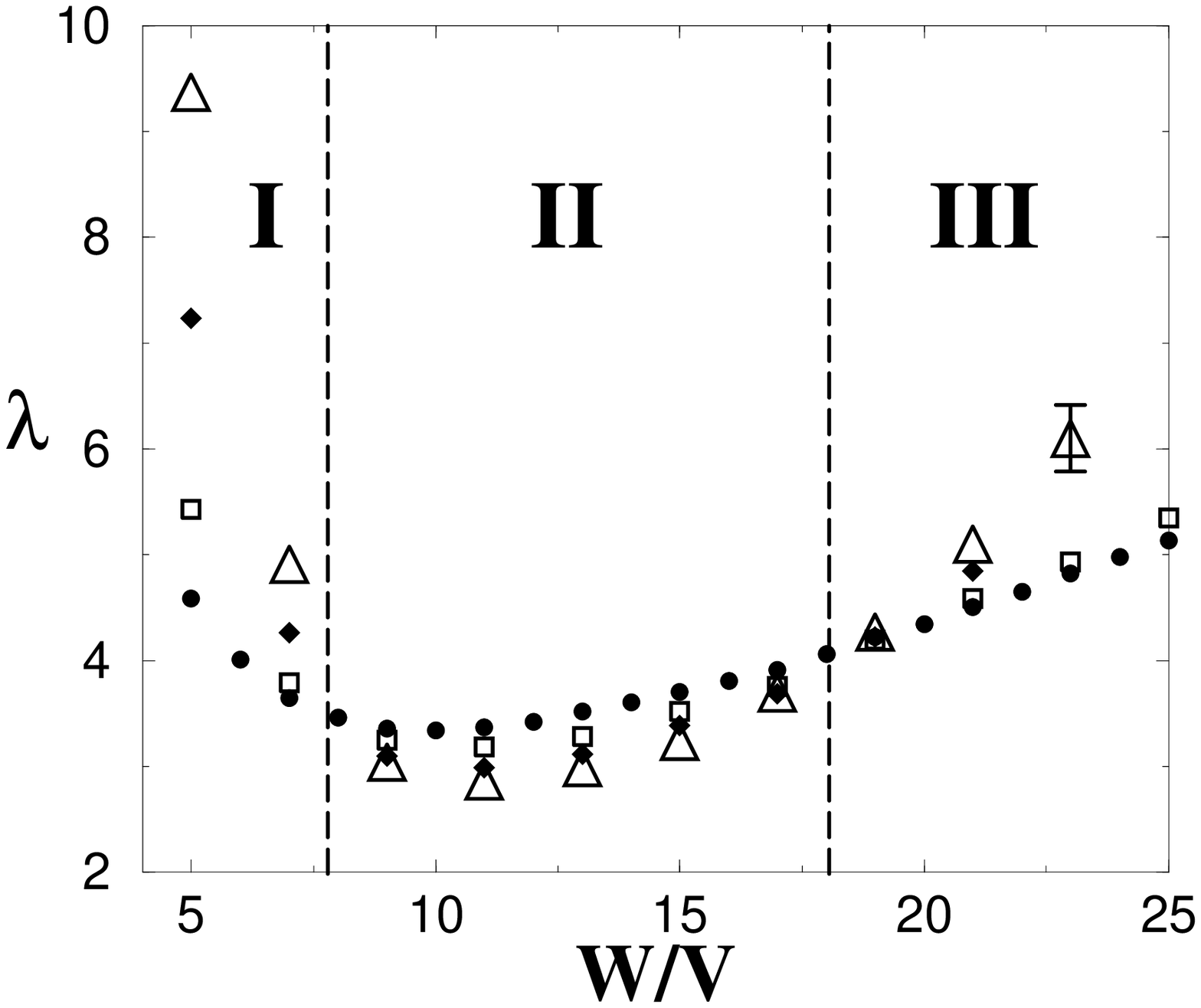}

\vspace{1 mm}

\caption
{Magnetization parameter $\lambda$ vs. disorder strength $W/V$,
for a Hubbard interaction
and a three-dimensional 6$ \times $6$ \times $6 (full circles),
8$ \times $8$ \times $8 (empty squares), 12$ \times $12$ \times $12 
(full diamonds) and 15$ \times $15$ \times $15
(empty triangles) Anderson lattice. One clearly
differentiates three regimes : (I) At small disorder, $\lambda$ increases
due to a crossover from ballistic to diffusive behavior (see text).
(II) At intermediate disorder, fluctuations seem to take over
and $\lambda$ decreases with system size.
(III) At large
disorder one-body states are strongly localized over very few sites,
which kills the off-diagonal fluctuations faster than
the exchange and the latter dominates again.}
\label{fig:lambda3d}

\vspace{4 mm}

\epsfxsize=3.3in
\epsfysize=2.5in
\epsffile{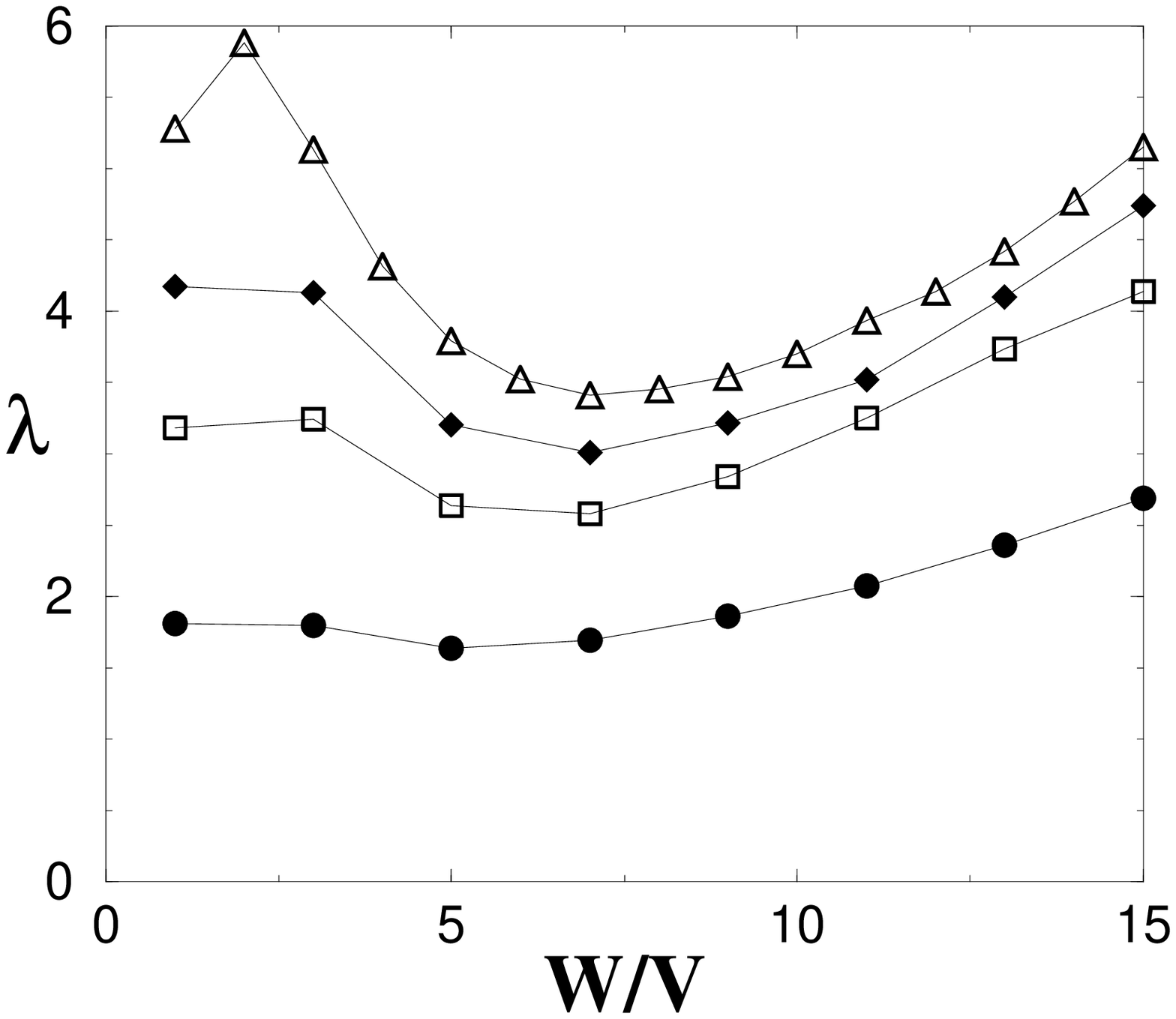}

\vspace{2 mm}

\caption
{Effect of the interaction range on the magnetization
parameter $\lambda$ for a two-dimensional 15$ \times $15
Anderson lattice and ratio ${\cal U}_0/{\cal U}_1=1$ 
(full circles), 4 (empty squares),
9 (full diamonds) and $\infty$ (Hubbard interaction - empty triangles).
Increasing the interaction range leads to a stronger
increase of the fluctuations than of the exchange, resulting in a lowering
of $\lambda$.}
\label{fig:screening}

\vspace{20 mm}

\epsfxsize=3.9in
\epsfysize=2.7in
\epsffile{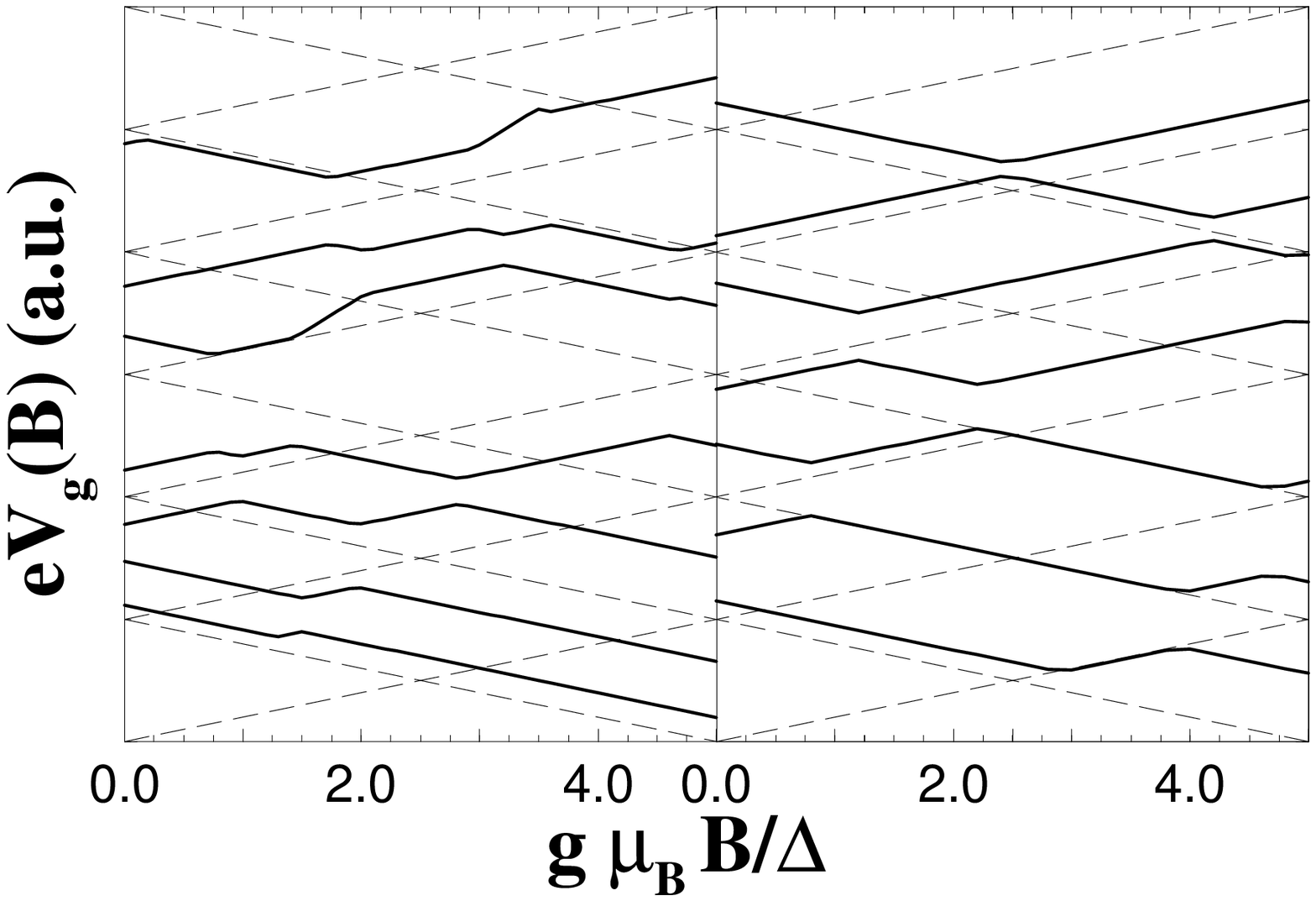}

\vspace{2 mm}

\caption{Schematic of the conductance peaks in a 2D lateral
quantum dot as a function of an in-plane magnetic field $g \mu_B B/\Delta$
for $m=14$, 
$\lambda U/\Delta = 2$ (left) and 0.4 (right) and $U/\Delta=0.4$ corresponding
to the addition of the
$n=3,4,...9^{\rm th}$ electrons (from bottom to top). 
The dashed lines indicate slopes of $\pm g \mu_B/2$ and serve as a guide
to the eye. At weak exchange (right), electrons
are piled up on the orbitals according to the Pauli prescription, so that
the spin is always minimized and the peaks move in parallel to the dashed 
lines. The sequence of magnetization difference between consecutive
peaks $\delta \sigma_z(n)=(-1)^n/2$ results in peaks
moving in opposite direction at low field.
Larger slopes appear at larger exchange strength (left) and
the successive addition of electrons
of same spin results in conductance peaks evolving in parallel at
low field. Note that due to the subtraction of the average
charge-charge interaction, the model does not reproduce the charging
energy so that the vertical distance between consecutive peaks is
arbitrary.}
\label{fig:spindot}

\vspace{5mm}

\epsfxsize=3.8in
\epsfysize=2.8in
\epsffile{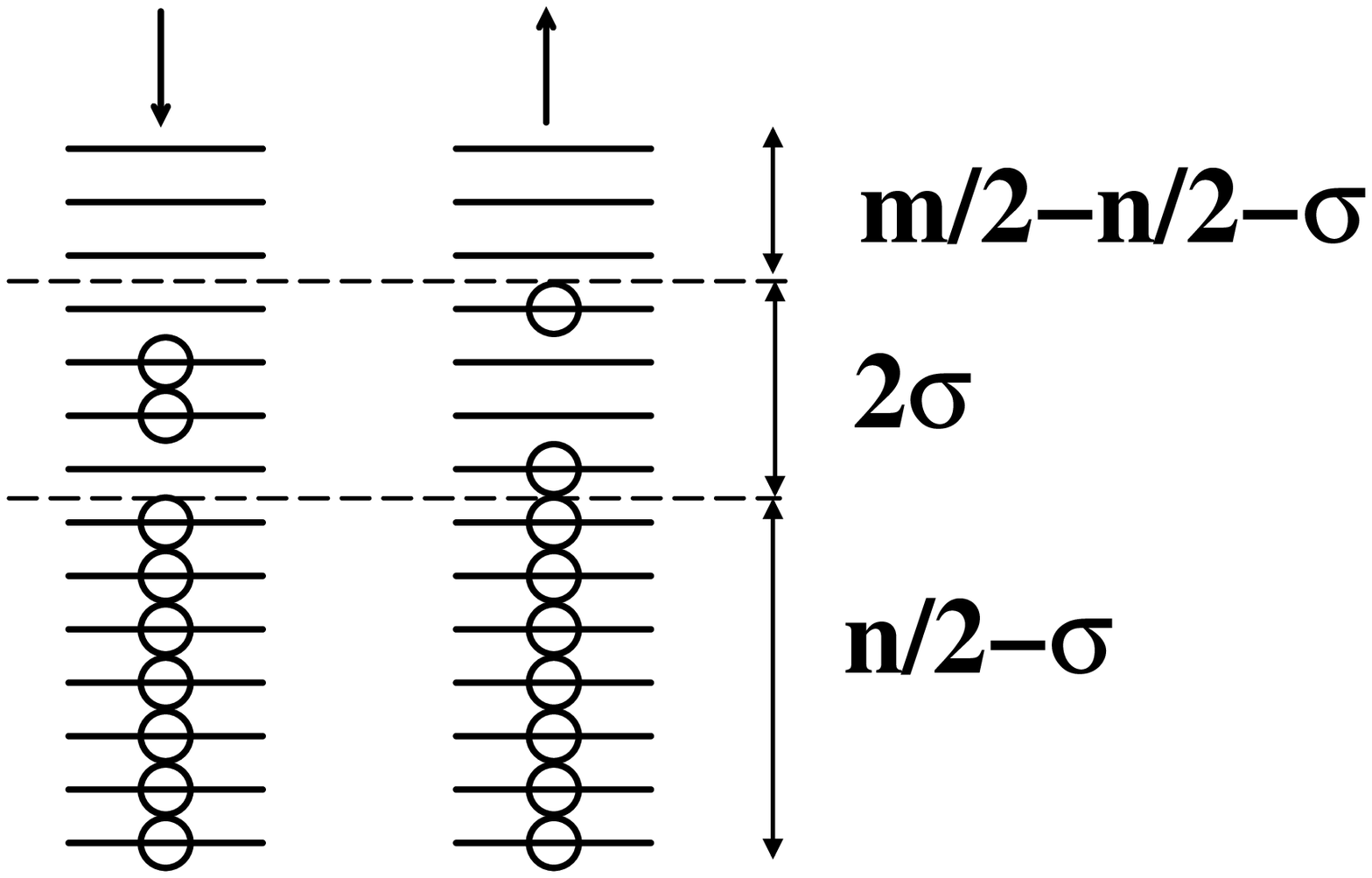}

\caption
{Representation of the $\sigma_z=0$ many-body yrast state with good total 
spin $\sigma$ at $U/\Delta=0$. The state consists of a filled fermi
sea with $n/2-\sigma$ doubly occupied orbital and a layer of $2 \sigma$
orbitals where particles are paired tripletwise.}
\label{fig:stlayer}

\vspace{6 mm}

\epsfxsize=2.45in
\epsfysize=2.1in
\epsffile{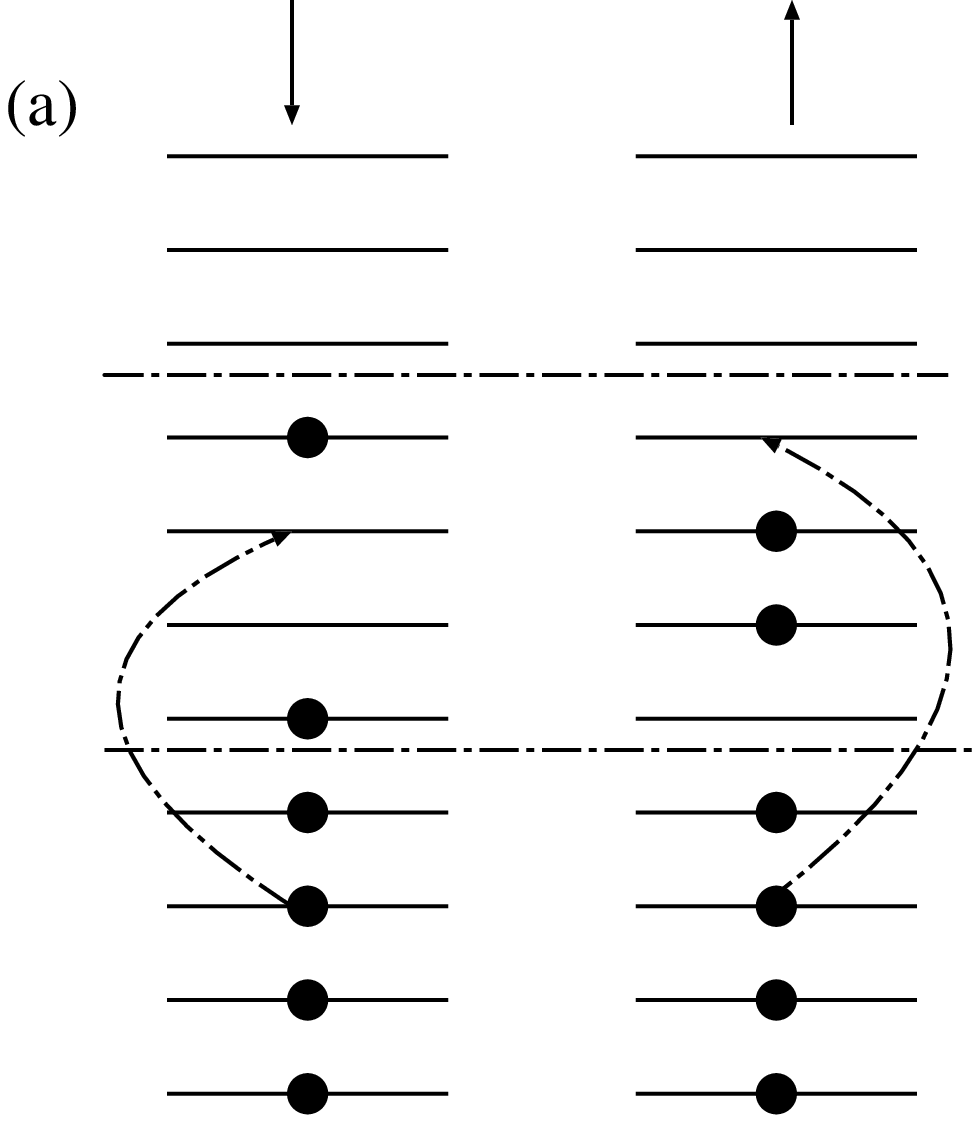} 
\vspace{-2.1in}
\hspace{3.in}
\epsfxsize=2.45in
\epsfysize=2.1in
\epsffile{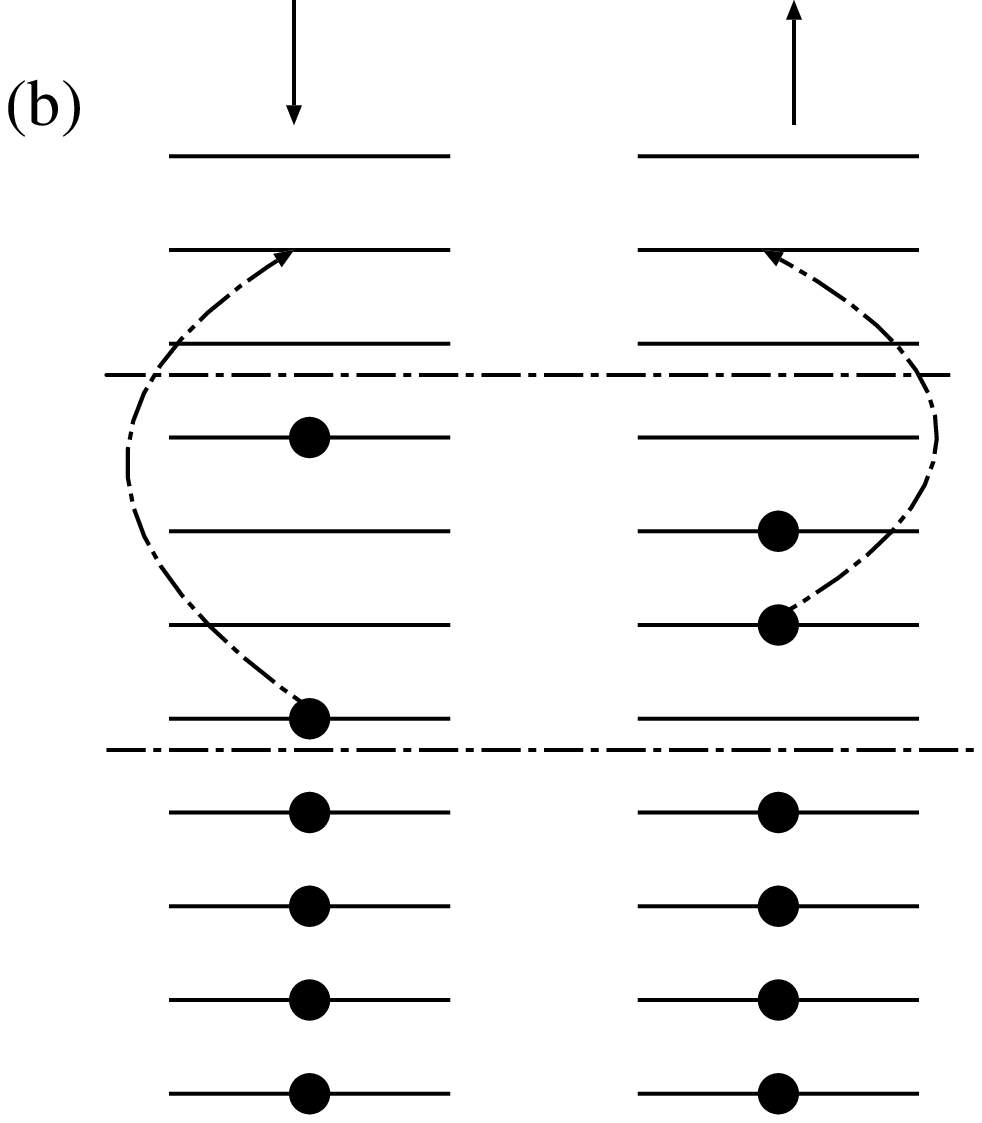} \\[1cm]
\epsfxsize=2.25in
\epsfysize=2.in
\epsffile{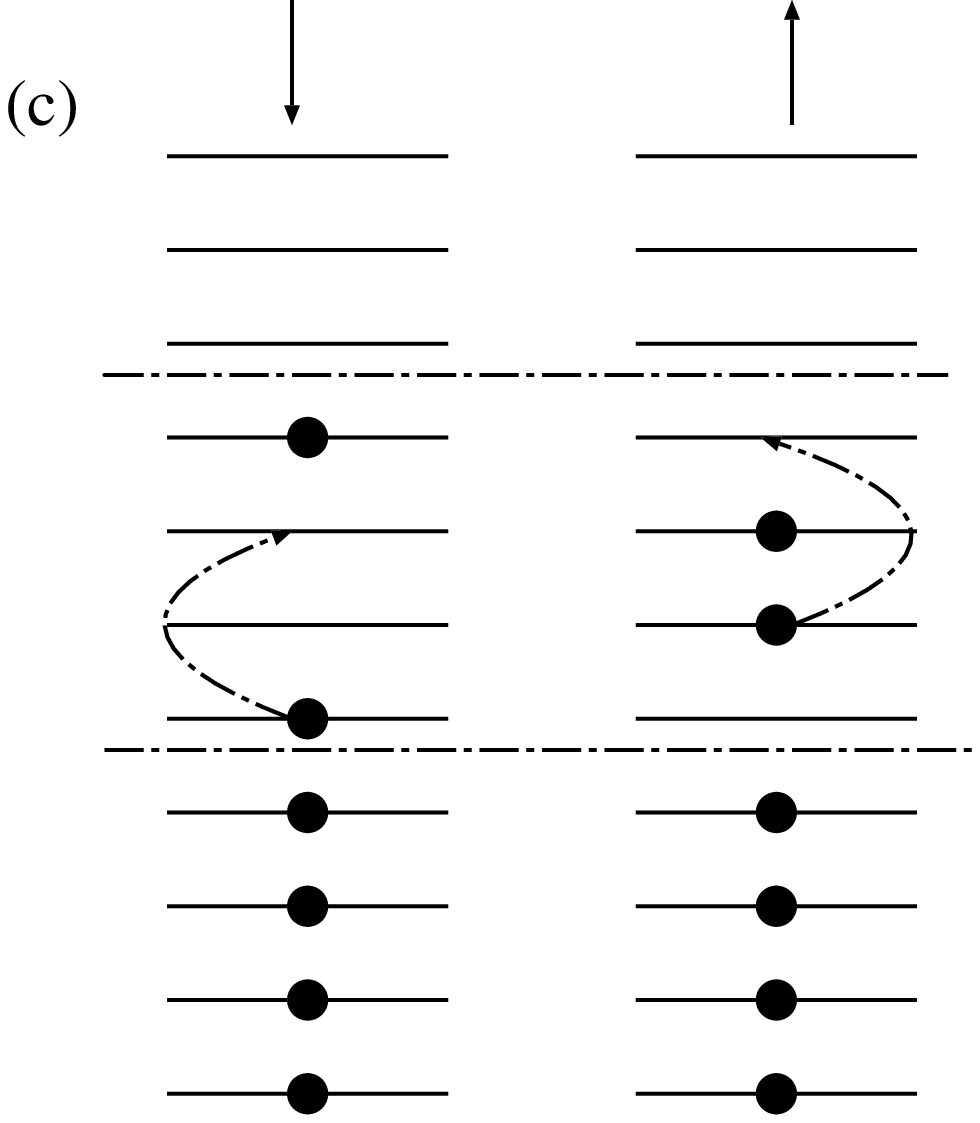} 
\hspace{2.5cm}
\epsfxsize=2.4in
\epsfysize=2.in
\epsffile{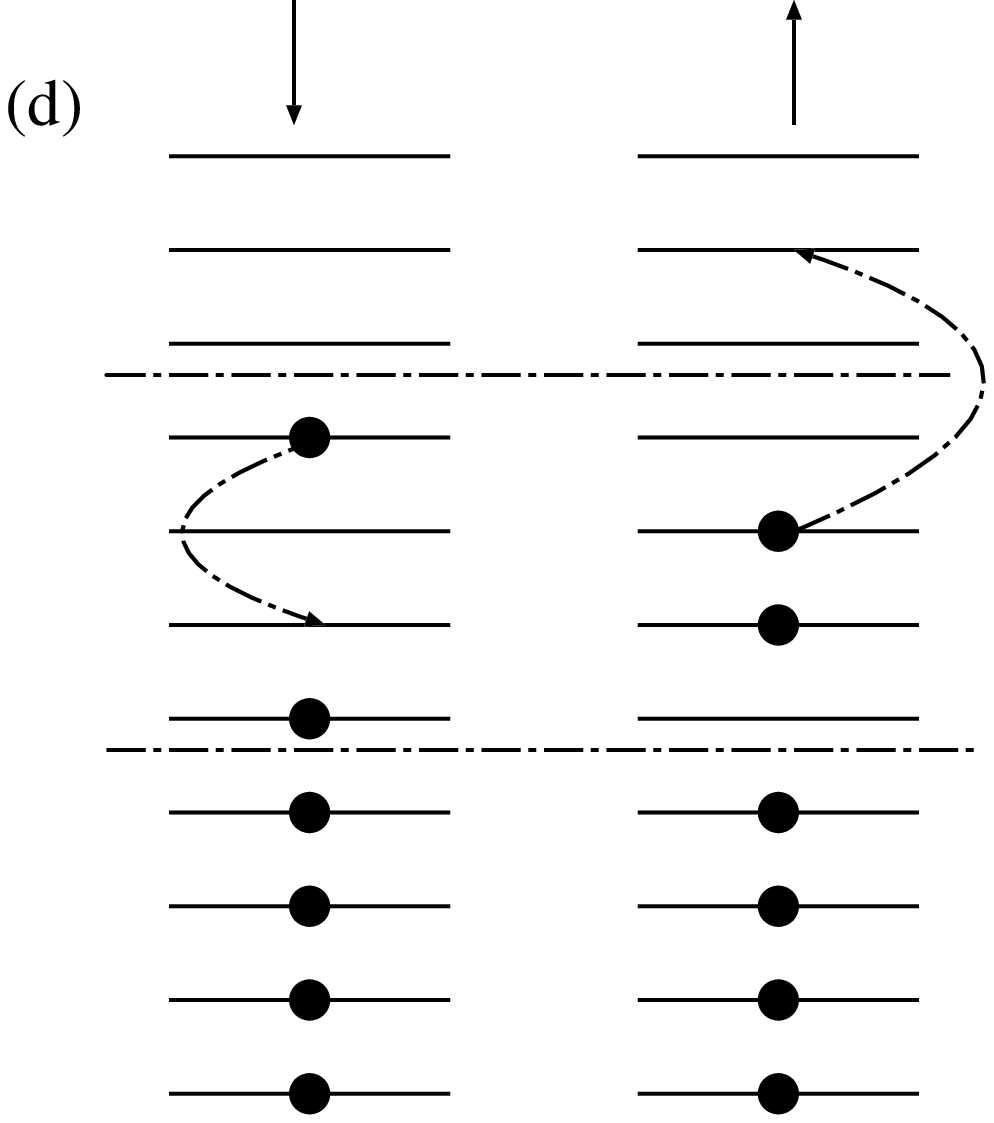}

\vspace{6 mm}

\caption
{Transitions that do not exist for the $\sigma \ne 0$ yrast state.}
\label{fig:scat}

\end{figure}

\noindent For a finite magnetization, 
we may represent the lowest level in the sector
as a sea of $n/2-\sigma$ double occupancies separated from $m/2-n/2-\sigma$
empty levels by a layer of $2 \sigma$ singly occupied levels as
depicted on Fig.\ref{fig:stlayer}. The 
transitions included in (\ref{connecst}) and that are now forbidden
correspond to singlet transitions involving either initial
or final states with at least one scattering particle
in the $2 \sigma$ layer. These transitions can be classified as : \\
1) transitions from a double occupancy in the Fermi sea onto the
$2 \sigma$-layer (Fig.\ref{fig:scat} (a))\\
2)  transitions from the $2 \sigma$-layer onto 
a double occupancy in one of the $m/2-n/2-\sigma$
empty levels (Fig.\ref{fig:scat} (b))\\
3) One- and two-body transitions within the $2 \sigma$-layer 
(Fig.\ref{fig:scat} (c)) and \\
4) Two-body transitions from the $2 \sigma$-layer, one of the particles
being transferred to a new orbital in the $2 \sigma$-layer, the other
one onto one of the $m/2-n/2-\sigma$ empty levels 
(Fig.\ref{fig:scat} (d)) \\
A simple counting of the number of these transitions finally gives

\begin{eqnarray}\label{connec}
K(\sigma) & = & K(0)-[\sigma (2 \sigma-1) (n/2-\sigma)
+ \sigma (2 \sigma-1) (m/2-n/2-\sigma) \nonumber \\
& + & (2 \sigma (2 \sigma-1) + \sigma (\sigma-1)^2(2\sigma-3)/2)
+ \sigma (2 \sigma-1) (\sigma-1) (m/2-n/2-\sigma)]
\end{eqnarray}

\noindent which in particular correctly reproduces 
the difference (\ref{test}). It is easily checked (e.g.
from (\ref{test}) that
the ratio $K(\sigma=n/2)/K(0)=1-A \nu + B \nu^2$ is a function of the filling
factor $\nu = n/m$ only. Also it is remarkable that the connectivity difference
between $\sigma=0$ and $\sigma=1$ is $m/2$ for any number of particles.

\newpage

\begin{references}
\bibitem{remspin} We will use the notations
$\vec{S}$ and $S_z$ for spin operators and
$\sigma$ and $\sigma_z$ for the corresponding eigenvalues.
\bibitem{caveat} To avoid degeneracies in the $U=0$ many-body spectrum,
we added in our numerics a small random part to the equidistant spectrum,
$\epsilon_\alpha = (\alpha -1 + R_{\alpha}) \Delta$, $|R_\alpha| \ll 1$.
\bibitem{remark5} Note that this is a necessary but not sufficient
condition. For example combining two $\sigma=1$, $\sigma_z=0$ objects gives a
nonzero $\sigma=0$ component.
\bibitem{remark} This is 
already the case in the perturbative regime where this width is 
essentially determined by first order corrections. However the
dependence in $n$ and $m$ is different.
\bibitem{remark2} We use $P(\sigma>0)$ indifferently for both even and odd $n$
even though in the latter case one should write $P(\sigma>1/2)$.
\bibitem{remark3} These results should be considered as a limit
$W/V \rightarrow 0$ as fluctuations are killed
in a clean system.
\bibitem{remark4} This is not exactly true as the infinite-ranged interaction
kills both fluctuations and exchange exactly. This result should be 
considered as a limit ${\cal U} = U/|\vec{r}-\vec{r}'|^{\alpha}$, 
$\alpha \rightarrow 0$.
\bibitem{remark6} Note that also in absence of fluctuations
it is easier to magnetize an even number than an odd number of
fermions as the doubling of the spin-gap at $U/\Delta=0$ from 
$\Delta$ to $2 \Delta$ is only
partially compensated by a 1.5-fold increase of the gain in 
exchange energy.
\end{references}
\end{document}